\pdfoutput=1
\documentclass[PAPER, atlasdraft=false, cernpreprint, UKenglish, texmf, orcidlogo]{atlasdoc}
\usepackage{atlaspackage}
\usepackage{atlasbiblatex}
\usepackage{multirow}
\usepackage{array}
\usepackage{xcolor}
\usepackage{bbding}
\usepackage{pifont}
\usepackage{wasysym}
\usepackage{lscape}
\usepackage{booktabs}
\usepackage{dcolumn}
\newcolumntype{~}{>{\global\let\currentrowstyle\relax}}
\newcolumntype{^}{>{\currentrowstyle}}
\newcommand{\rowstyle}[1]{\gdef\currentrowstyle{#1}%
#1\ignorespaces
}
\usepackage{atlasphysics}
\graphicspath{{logos/}{figures/}}

\usepackage{ANA-EXOT-2021-31-PAPER-defs}

\AtlasTitle{Search for vector-like leptons coupling to first- and second-generation Standard Model leptons in $pp$ collisions at $\sqrt{s}=13~\tev$ with the ATLAS detector}

\AtlasAbstract{
A search for pair production of vector-like leptons coupling to first- and second-generation Standard Model leptons is presented. The search is based on a dataset of proton--proton collisions at $\sqrt{s}=13~\tev$ recorded with the ATLAS detector during Run 2 of the Large Hadron Collider, corresponding to an integrated luminosity of 140~fb$^{-1}$. Events are categorised depending on the flavour and multiplicity of leptons (electrons or muons), as well as on the scores of a deep neural network targeting particular signal topologies according to the decay modes of the vector-like leptons. In each of the signal regions, the scalar sum of the transverse momentum of the leptons and the missing transverse momentum is analysed.  The main background processes are estimated using dedicated control regions in a simultaneous fit with the signal regions to data. No significant excess above the Standard Model background expectation is observed and limits are set at 95\% confidence level on the production cross-sections of vector-like electrons and muons as a function of the vector-like lepton mass, separately for SU(2) doublet and singlet scenarios. The resulting  mass lower limits are $1220~\gev$ ($1270~\gev$) and $320~\gev$ ($400~\gev$) for vector-like electrons (muons) in the doublet and singlet scenarios, respectively.
}

\AtlasRefCode{EXOT-2021-31}

\PreprintIdNumber{CERN-EP-2024-296}

\AtlasJournalRef{\JHEP 05 (2025) 075}
\AtlasDOI{https://doi.org/10.1007/JHEP05(2025)075}

\hypersetup{pdftitle={ATLAS document},pdfauthor={The ATLAS Collaboration}}

\begin{document}

\maketitle

\tableofcontents

\section{Introduction}
\label{sec:intro}


The Standard Model (SM) of particle physics is the most successful and tested theory of the known fundamental particles and their interactions. The last missing piece of the SM puzzle, the Higgs boson, was discovered in 2012 by the ATLAS and CMS Collaborations~\cite{HIGG-2012-27,CMS-HIG-12-028} at the Large Hadron Collider (LHC)~\cite{Evans:2008zzb}. Despite its many achievements, the SM remains an incomplete theory, as it does not provide answers for open questions such as the structure of masses and mixings of elementary fermions (also known as the `flavour puzzle'), the hierarchy and fine-tuning problems, the observed baryon asymmetry in the universe, and the nature of dark matter and dark energy. Many beyond-the-SM (BSM) theories, typically introducing new particles and interactions, or a new space-time structure, have been proposed to address these and other shortcomings of the SM.

Vector-like fermions are hypothetical particles whose left- and right-handed chiral components have the same transformation properties under the weak-isospin SU(2) gauge group~\cite{delAguila:1982fs,PhysRevD.33.2632,Fishbane:1987tx,Montvay:1988av,Fujikawa:1994we}. Consequently, they have Dirac masses, without a Yukawa coupling proportional to their mass, becoming less constrained by Higgs boson measurements~\cite{DELAGUILA19901}. Vector-like fermions arise in many BSM scenarios, such as Composite Higgs models~\cite{Kaplan:1983sm,Agashe:2004rs}, models with extra spatial dimensions~\cite{Kong_2010,Huang:2012kz}, supersymmetric models~\cite{PhysRevD.81.035004,Zheng:2019kqu}, and grand unified theories~\cite{Nevzorov:2012hs,PhysRevD.89.115004,Joglekar:2016yap}. In particular, vector-like fermions can provide an explanation to the flavour puzzle via their mixings with SM fermions~\cite{PhysRevLett.102.101801,Redi:2013pga}, or even provide a dark matter candidate~\cite{Schwaller:2013hqa,Halverson:2014nwa,Bahrami:2016has,Bhattacharya:2018fus}. Naturalness arguments~\cite{Susskind:1978ms} require that quadratic divergences that arise from the radiative corrections to the Higgs boson mass are cancelled out by some new mechanism to avoid fine-tuning, and vector-like quarks play such role, e.g in Composite Higgs models. On the other hand, the observed tensions between the measured and predicted values of the muon~\cite{PhysRevD.73.072003,Muong-2:2023cdq} and electron~\cite{Parker:2018vye,Morel:2020dww,Fan:2022eto} anomalous magnetic moments or the so-called `Cabibbo angle anomaly'~\cite{Belfatto:2019swo,Grossman:2019bzp,Crivellin:2020ebi} can be explained by BSM models including vector-like leptons (VLL)~\cite{PhysRevD.84.075017,PhysRevD.88.013017,Megias:2017dzd,Hiller:2020fbu,Dehghani:2024lgz}. VLLs mixing with first-, second-, or third-generation SM leptons ($\ell = e, \mu, \tau$) are referred to as vector-like electrons, muons, or $\tau$-leptons, respectively.

At the LHC, VLLs are predominantly produced in pairs via the electroweak interaction, and consequently have a considerably lower production cross-section than vector-like quarks,
for which a broad search programme has been developed~\cite{ATLAS:2024fdw,CMS:2024bni}. In contrast, only few LHC searches for VLLs exist, which are summarised below.
The production and decay modes for VLLs depend on the assumed SU(2) representation~\cite{Kumar:2015tna,Bhattiprolu:2019vdu}. In the doublet scenario, two mass-degenerate VLLs at tree level, one electrically charged ($L^\pm$) and one electrically neutral ($N^0$), form an SU(2) doublet ($L$, $N^0$). They can be produced in association via the exchange of a virtual $W$ boson in the $s$-channel, $pp \to W^* \to L^+ \bar{N}^0$,\footnote{The charge-conjugate process, $pp \to W^* \to L^- N^0$, is also implied.} or in pairs via a virtual $Z$-boson or photon exchange, $pp \to Z^*/\gamma^* \to L^+ L^-$ and $pp \to Z^* \to N^0 \bar{N}^0$. In the doublet scenario, the charged VLL decay modes are $L \to \ell Z$ and $\ell H$, where $H$ is the SM Higgs boson, with branching ratios that depend on the VLL mass $m_L$ and asymptotically reach 50\% each for $m_L \gg m_H$, in accordance with the Goldstone boson equivalence theorem~\cite{Chanowitz:1985hj}; at lower masses, the branching ratio to $\ell H$ decreases as it becomes kinematically disfavoured. In contrast, the neutral VLL decay mode is $N^0 \to \ell W$ with 100\% branching ratio. In the singlet scenario, only the charged VLL is present and is also produced in pairs, $pp \to Z^*/\gamma^* \to L^+ L^-$; its decay modes are $L \to \nu W$, $\ell Z$ and $\ell H$, with branching ratios asymptotically reaching 50\%, 25\%, and 25\%, respectively.

Searches by the L3 Collaboration at the LEP Collider excluded vector-like electrons, muons and $\tau$-leptons with masses less than $\sim$$100~\gev$ at the 95\% confidence level (CL)~\cite{L3:2001xsz}; the limits are similar for both the doublet and singlet scenarios. At the LHC, using a data sample of proton--proton ($pp$) collisions at a centre-of-mass energy of $\sqrt{s}=8~\tev$ corresponding to an integrated luminosity of 20.3~fb$^{-1}$, the ATLAS Collaboration excluded vector-like electrons (muons) in the mass range 129--176~$\gev$ (114--168~$\gev$), except for the interval 144--163~$\gev$ (153--160~$\gev$), in the singlet scenario~\cite{EXOT-2014-08}. Both the ATLAS and CMS Collaborations searched for vector-like $\tau$-leptons using the full LHC Run 2 dataset at $\sqrt{s}=13~\tev$, corresponding to about 140~fb$^{-1}$. Assuming the doublet scenario, the ATLAS search excluded vector-like $\tau$-leptons in the mass range of 130--900~$\gev$~\cite{EXOT-2020-07}. The CMS search considered both the doublet and singlet scenarios, excluding vector-like $\tau$-leptons in the mass ranges of 100--1045~$\gev$ and 125--150~$\gev$, respectively~\cite{CMS-EXO-21-002}.

\begin{figure}[tb]
\centering
\subfloat[]{\includegraphics[width=0.3\textwidth]{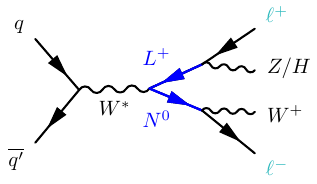}\label{fig:prod_decay_a}}
\subfloat[]{\includegraphics[width=0.3\textwidth]{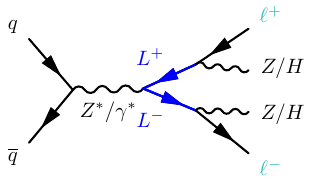}\label{fig:prod_decay_b}}
\subfloat[]{\includegraphics[width=0.3\textwidth]{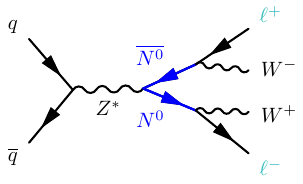}\label{fig:prod_decay_c}}\\
\subfloat[]{\includegraphics[width=0.3\textwidth]{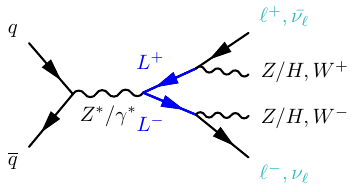}\label{fig:prod_decay_d}}
\caption{Feynman diagrams illustrating the pair production and decay of vector-like leptons: (a-c) refer to the doublet scenario, and (d) to the singlet scenario.}
\label{fig:prod_decay}
\end{figure}

This paper presents a search for doublet or singlet VLLs coupling to first- and second-generation SM leptons. Final states with two opposite-sign, three and four leptons are considered in the signal regions, where the leptons can originate from the VLL decays or the sub-sequent decays of the $Z$, $H$, or $W$ bosons. Figure~\ref{fig:prod_decay} illustrates the signal processes targeted in this analysis. The search is based on a dataset of $pp$ collisions at $\sqrt{s}=13~\tev$ recorded with the ATLAS detector during Run 2 of the LHC, corresponding to an integrated luminosity of 140~fb$^{-1}$. A mass range between $150~\gev$ and $1600~\gev$ is considered for vector-like electrons (\vlle) and vector-like muons (\vllmu). Vector-like leptons in the doublet scenario with mass values lower than $150~\gev$ would have already been seen by differential cross-section measurements in four-lepton events~\cite{STDM-2018-30}, and are therefore not considered in this analysis. A categorisation based on a deep neural network (DNN) is performed to enhance the purity of the various signal types and to discriminate signal against the SM background. Control regions (CRs), orthogonal to the signal regions (SRs), are defined to constrain the normalisation of the main backgrounds: \ttbar, \Zjets, \ttW, \ttZ, \VV and non-prompt lepton backgrounds. A maximum-likelihood fit is performed across event categories to search for the signal and constrain several leading background processes simultaneously.

%
%
%
%
%
%
%
%
%
%
%
%
%
%
%
%
%
%
%
%
%
%
%
%
%
%
%
%
%
%
%
%
%
%
%
%
%
%
%
%
%
%
%
%
%
%
%
%


%

%
\section{ATLAS detector}
\label{sec:detector}

%
\newcommand{\AtlasCoordFootnote}{%
ATLAS uses a right-handed coordinate system with its origin at the nominal interaction point (IP)
in the centre of the detector and the \(z\)-axis along the beam pipe.
The \(x\)-axis points from the IP to the centre of the LHC ring,
and the \(y\)-axis points upwards.
Polar coordinates \((r,\phi)\) are used in the transverse plane,
\(\phi\) being the azimuthal angle around the \(z\)-axis.
The pseudorapidity is defined in terms of the polar angle \(\theta\) as \(\eta = -\ln \tan(\theta/2)\) and is equal to the rapidity
$ y = \frac{1}{2} \ln \left( \frac{E + p_z}{E - p_z} \right) $ in the relativistic limit.
Angular distance is measured in units of \(\Delta R \equiv \sqrt{(\Delta y)^{2} + (\Delta\phi)^{2}}\).}

The ATLAS detector~\cite{PERF-2007-01} at the LHC covers nearly the entire solid angle around the collision point.\footnote{\AtlasCoordFootnote}
It consists of an inner tracking detector surrounded by a thin superconducting solenoid, electromagnetic and hadronic calorimeters,
and a muon spectrometer incorporating three large superconducting air-core toroidal magnets.

The inner-detector system (ID) is immersed in a \qty{2}{\tesla} axial magnetic field
and provides charged-particle tracking in the range \(|\eta| < 2.5\).
The high-granularity silicon pixel detector covers the vertex region and typically provides four measurements per track,
the first hit generally being in the insertable B-layer (IBL) installed before Run~2~\cite{PIX-2018-001}.
It is followed by the SemiConductor Tracker (SCT), which usually provides eight measurements per track.
These silicon detectors are complemented by the transition radiation tracker (TRT),
which enables radially extended track reconstruction up to \(|\eta| = 2.0\).
The TRT also provides electron identification information
based on the fraction of hits (typically 30 in total) above a higher energy-deposit threshold corresponding to transition radiation.

The calorimeter system covers the pseudorapidity range \(|\eta| < 4.9\).
Within the region \(|\eta|< 3.2\), electromagnetic calorimetry is provided by barrel and
endcap high-granularity lead/liquid-argon (LAr) calorimeters,
with an additional thin LAr presampler covering \(|\eta| < 1.8\)
to correct for energy loss in material upstream of the calorimeters.
Hadronic calorimetry is provided by the steel/scintillator-tile calorimeter,
segmented into three barrel structures within \(|\eta| < 1.7\), and two copper/LAr hadronic endcap calorimeters.
The solid angle coverage is completed with forward copper/LAr and tungsten/LAr calorimeter modules
optimised for electromagnetic and hadronic energy measurements respectively.

The muon spectrometer (MS) comprises separate trigger and
high-precision tracking chambers measuring the deflection of muons in a magnetic field generated by the superconducting air-core toroidal magnets.
The field integral of the toroids ranges between \num{2.0} and \qty{6.0}{\tesla\metre}
across most of the detector.
Three layers of precision chambers, each consisting of layers of monitored drift tubes, cover the region \(|\eta| < 2.7\),
complemented by cathode-strip chambers in the forward region, where the background is highest.
The muon trigger system covers the range \(|\eta| < 2.4\) with resistive-plate chambers in the barrel, and thin-gap chambers in the endcap regions.

The luminosity is measured mainly by the LUCID--2~\cite{LUCID2} detector that records Cherenkov light produced in the quartz windows of photomultipliers located close to the beampipe.

Events are selected by the first-level trigger system implemented in custom hardware,
followed by selections made by algorithms implemented in software in the high-level trigger~\cite{TRIG-2016-01}.
The first-level trigger accepts events from the \qty{40}{\MHz} bunch crossings at a rate below \qty{100}{\kHz},
which the high-level trigger further reduces in order to record complete events to disk at about \qty{1}{\kHz}.

A software suite~\cite{SOFT-2022-02} is used in data simulation, in the reconstruction
and analysis of real and simulated data, in detector operations, and in the trigger and data acquisition
systems of the experiment.


%

%
\section{Data and simulated event samples}
\label{sec:datamc}

This analysis uses data from $pp$ collisions at $\sqrt{s}=13~\tev$ collected by the ATLAS experiment from 2015 to 2018. After the application of data-quality requirements~\cite{DAPR-2018-01} to ensure that all parts of the detector are operational during data-taking, the data sample corresponds to an integrated luminosity of 140~\ifb.
The number of additional $pp$ interactions per bunch crossing (pile-up) in this sample ranges from about 8 to 70, with an average of 34.
The trigger requirements are discussed in Section~\ref{sec:events}.

Monte Carlo (MC) simulation samples are produced for the different signal and background processes. Table~\ref{tab:mcconfig} gives a detailed summary of all signal and SM background samples used in this analysis.
All samples showered with \textsc{Pythia} use the A14 set of tuned parameters~\cite{ATL-PHYS-PUB-2014-021} (referred to as `tune'), whereas those showered with \textsc{Herwig} use the H7-UE tune~\cite{Bellm:2015jjp}.
In all samples simulated with \Sherpa~\cite{Gleisberg:2008ta}, the matrix elements (MEs) are calculated with the Comix~\cite{Gleisberg:2008fv} and \OPENLOOPS~\cite{Buccioni:2019sur,Cascioli:2011va,Denner:2016kdg} libraries. They are matched with the \SHERPA parton shower (PS)~\cite{Schumann:2007mg} using the \MEPSatNLO prescription~\cite{Hoeche:2011fd,Hoeche:2012yf,Catani:2001cc,Hoeche:2009rj} with the set of tuned parameters developed by the \SHERPA authors.
Pile-up is modelled using events from minimum-bias interactions generated with \PYTHIA[8.186]~\cite{Sjostrand:2007gs} with the A3 tune~\cite{ATL-PHYS-PUB-2011-014}, and overlaid onto the simulated hard-scatter events according to the luminosity profile of the recorded data.
All samples include leading-logarithm photon emission, either modelled by the PS generator or by \textsc{Photos}~\cite{Golonka:2005pn}.
The mass of the top quark and SM Higgs boson are set to $172.5~\gev$ and $125~\gev$, respectively.
The generated events are processed through either a full simulation of the ATLAS detector geometry and response using \textsc{Geant4}~\cite{Agostinelli:2002hh}, or a faster simulation where the full \textsc{Geant4} simulation of the calorimeter response is replaced by a detailed parameterisation of the shower shapes (ATLAS Fast Simulation)~\cite{SOFT-2010-01}. Both types of simulated events are processed through the same reconstruction software used for the $pp$ collision data.
Corrections are applied to the simulated events so that the particle candidates' selection efficiencies, energy scales and energy resolutions match those determined from data control samples.
The simulated samples are normalised to their theoretical cross-sections, most of which are computed to the highest order available in perturbation theory.

Signal samples for \vlle and \vllmu from $SU(2)$ singlet (\vlles, \vllmus) and doublet (\vlled, \vllmud) models are simulated using \MGNLO[2.9.5]~\cite{Alwall:2014hca} at next-to-leading-order (NLO) in QCD with the \NNPDF[3.0nlo] PDF set~\cite{Ball:2014uwa} and \PYTHIA[8.245]~\cite{Sjostrand:2014zea}, and processed through the ATLAS Fast Simulation.
The NLO cross-section obtained from \textsc{Madgraph} is used for the normalisation of the signals.

The production of \ttbar events is modelled using the \POWHEGBOX[v2]~\cite{Frixione:2007nw,Nason:2004rx,Frixione:2007vw,Alioli:2010xd,Re:2010bp,Frederix:2012dh,Alioli:2009je}~
generator at NLO with the \NNPDF[3.0nlo] PDF set. The events are interfaced to \PYTHIA[8.230] to model the PS, hadronisation, and underlying event, using the \NNPDF[2.3lo] set of PDFs. The decays of bottom and charm hadrons are performed by \EVTGEN[1.6.0]~\cite{Lange:2001uf}. The \ttbar process is modelled with the \hdamp parameter~\footnote{The \hdamp parameter is a resummation damping factor and one of the parameters that controls the matching of \POWHEG MEs to the PS, thus effectively regulating the high-\pT radiation against which the hard-process system recoils.} set to 1.5\,$m_t$~\cite{ATL-PHYS-PUB-2016-020}. The \ttbar sample is normalised to the \xs prediction at next-to-next-to-leading-order (NNLO) in QCD including the resummation of next-to-next-to-leading logarithmic (NNLL) soft-gluon terms calculated using \TOPpp[2.0]~\cite{Beneke:2011mq,Cacciari:2011hy,Baernreuther:2012ws,Czakon:2012zr,Czakon:2012pz,Czakon:2013goa-fixed,Czakon:2011xx}. This \xs is \(\sigma(\ttbar)_{\text{NNLO+NNLL}} = 832 \pm 51\,\si{\pb}\).

The $Z/\gamma^* \to  \ell \ell$ process (with $\ell=e,\mu,\tau$) is simulated with \Sherpa[2.2.11]~\cite{Gleisberg:2008ta} using the \NNPDF[3.0nnlo] PDF set~\cite{Ball:2014uwa}.
For strong production of $Z/\gamma^* + \mathrm{jets}$, where a QCD coupling facilitates the production of the additional jets, processes with up to two coloured partons are modelled at NLO in the strong coupling, while processes with up to five additional partons are modelled at leading-order (LO) accuracy.
The $Z$+jets sample is normalised to the theoretical cross-section calculated at NLO accuracy in QCD~\cite{Catani:2009sm}.
The samples used to model the \ttW and the \ttll backgrounds are simulated using \Sherpa[2.2.10]~\cite{Bothmann:2019yzt} and \Sherpa[2.2.11], where the MEs are calculated for up to one and zero additional partons at NLO in QCD, respectively, and up to two partons at LO in QCD.
These samples are simulated using the \NNPDF[3.0nnlo] PDF set.
The invariant mass of the lepton pair ($\mll$) in the \ttll sample is set to be greater than $1~\gev$.
Both the factorisation and renormalisation scales are set to $\mu_r = \mu_f = m_T / 2$ in the \ttW sample, where $m_T$ is defined as the scalar sum of the transverse masses $\sqrt{m^2 + p_{T}^2}$ of the particles generated from the ME calculation.
In addition to this \ttW prediction at NLO in QCD, higher-order corrections related to electroweak (EW) contributions are also included. First, event-by-event correction factors are applied that provide virtual NLO EW corrections of the order $\alpha^{2}\alphas^{2}$ derived using the formalism described in Ref.~\cite{Kallweit:2015dum} along with LO corrections of order $\alpha^3$. Second, real emission contributions from the sub-leading EW corrections at order $\alpha^{3}\alphas$\,\cite{Frederix:2017wme} are simulated with \Sherpa[2.2.10] produced at LO in QCD and included as a separate sample.
Diboson ($\VV$) background processes are simulated with \Sherpa[2.2.2]~\cite{Bothmann:2019yzt} and include $\WZ$, $\ZZ$, and $\WW$ processes.
The ME is calculated with NLO accuracy in QCD for up to one additional parton and at LO accuracy for up to three additional partons.
The \NNPDF[3.0nnlo] set of PDFs is used.
The simulation includes off-shell effects and Higgs boson contributions, where appropriate.
Samples for the loop-induced processes $gg \to VV$ are simulated using LO-accurate MEs for up to one additional parton emission.
Samples for $\tth$, and single top production are simulated using the NLO generator \POWHEGBOX[v2] and interfaced with \PYTHIA[8] for the PS and fragmentation. These samples used the \NNPDF[3.0nlo] PDF set. The decays of bottom and charm hadrons are performed by \EVTGEN[1.6.0]. The production of a top quark in association with a \(W\) boson (\(tW\)) is modelled using the five-flavour scheme. The diagram removal scheme~\cite{Frixione:2008yi} is used to remove interference and overlap with \ttbar production. Single-top \(s\)- and \(t\)-channel production is modelled using the five- and four-flavour schemes, respectively.

A dedicated $\ttbar$ sample including rare $t \to Wb \gamma^*(\to \ell^{+}\ell^{-})$ radiative decays, $\ttbar \to W^+bW^-\bar{b}\ell^{+}\ell^{-}$, is simulated using a ME calculated at LO in QCD and requiring $\mll > 1~\gev$. In this sample the photon can be radiated from the top quark, the $W$ boson, or the $b$-quark. Both the \ttll and $\ttbar \to W^+bW^-\bar{b}\ell^{+}\ell^{-}$ samples are combined and together form the `$\ttdy$' sample.
The contribution from internal photon conversions ($\gamma^* \to \ell^{+}\ell^{-}$, referred to as `IntC') with $\mll < 1~\gev$ is modelled by QED multi-photon radiation via the PS in an inclusive $\ttbar$ sample.
Dedicated $Z$+jets samples containing electrons from material photon conversion ($\gamma \to e^+e^-$, referred to as `MatC') or internal photon conversion are generated with \POWHEGBOX and interfaced with \PYTHIA[8] for the PS and fragmentation. These samples are used to model the data in control regions enriched in material and internal conversion electrons, as explained in Section~\ref{sec:events}.

The remaining rare background contributions listed in Table~\ref{tab:mcconfig} are normalised using their LO theoretical cross-sections, except for the \tttt, \tZ, single top, \Wjets and $VH$ processes, for which a NLO cross-section is used.

\begin{table}[htb]
\begin{center}
\caption{
Simulated signal and background event samples, with the corresponding ME generator, ME order (which is the order in the strong coupling constant of the perturbative calculation), PS generator, the generator PDF sets and the underlying set of tuned parameters of the PS generator used.
The samples used to estimate the systematic uncertainties are indicated in parentheses and grey.
$V$ refers to production of an electroweak boson ($W$ or $Z/\gamma^*$).
The `$\ttbar W$ (EW)' sample also includes next-to-leading-order electroweak corrections.
MG5\_aMC refers to \textsc{MadGraph5\_aMC@NLO} 2.2, 2.3, or 2.6;
\textsc{Pythia} 8 refers to version 8.2;
\textsc{MePs@Nlo} is the method used in \textsc{Sherpa} to match the ME to the PS.
}
\label{tab:mcconfig}
\vspace{0.25cm}
\small
\setlength\tabcolsep{1.5pt}
\begin{tabular}{~l^l^l^l^l^l}
\hline\hline
Process & Generator & ME order & PS & PDF & Tune  \\
\hline
VLL signal             & \textsc{MG5\_aMC}  & NLO & \textsc{Pythia} 8\ & \NNPDF[3.0nlo] & A14 \\
$\ttbar$               & \textsc{Powheg-BOX} & NLO & \textsc{Pythia} 8 & \NNPDF[3.0nlo] & A14  \\
\rowstyle{\color{gray}}
& (\textsc{Powheg-BOX}) & (NLO) & (\textsc{Herwig7.1.3}) & (\NNPDF[3.0nlo]) &  (H7-UE-MMHT) \\
$Z \to \ell^{+}\ell^{-}$         & \textsc{Sherpa} 2.2.11 & \textsc{MePs@Nlo} & \textsc{Sherpa} & \NNPDF[3.0nlo] & \textsc{Sherpa} default \\
$Z \to \ell^{+}\ell^{-} (\gamma \to e^{+}e^{-})$ & \textsc{Powheg-BOX} & NLO & \textsc{Pythia} 8 & \CTEQ[6L1nlo]~\cite{Pumplin_2002} & A14 \\
$Z \to \ell^{+}\ell^{-} (\gamma* \to e^{+}e^{-})$ & \textsc{Powheg-BOX} & NLO & \textsc{Pythia} 8 & \CTEQ[6L1nlo] & A14 \\
$\ttbar W$             & \textsc{Sherpa 2.2.10} & \textsc{MePs@Nlo} & \textsc{Sherpa} &  \NNPDF[3.0nnlo] & \textsc{Sherpa} default \\
\rowstyle{\color{gray}}
& (MG5\_aMC)  & (NLO) & (\textsc{Pythia} 8) & (\NNPDF[3.0nlo]) & (A14)   \\
$\ttbar W$ (EW)        & \textsc{Sherpa 2.2.10} & LO & \textsc{Sherpa} &  \NNPDF[3.0nnlo] & \textsc{Sherpa} default \\ %
\rowstyle{\color{gray}}
& (MG5\_aMC)  & (LO) & (\textsc{Pythia} 8) & (\NNPDF[3.0nlo]) & (A14)   \\
\ttll                  & \textsc{Sherpa 2.2.11} & \textsc{MePs@Nlo} & \textsc{Sherpa} &  \NNPDF[3.0nnlo] & \textsc{Sherpa} default \\ %
\rowstyle{\color{gray}}
& (MG5\_aMC) & (NLO) & (\textsc{Pythia} 8) & (\NNPDF[3.0nlo]) & (A14)  \\
$VV$, $VVV$    		   & \textsc{Sherpa} 2.2.2  & \textsc{MePs@Nlo} & \textsc{Sherpa} & \NNPDF[3.0nnlo] & \textsc{Sherpa} default  \\
\ttH                   & \textsc{Powheg-BOX}  & NLO & \textsc{Pythia} 8\ & \NNPDF[3.0nlo] & A14 \\
\rowstyle{\color{gray}}
& (\textsc{Powheg-BOX}) & (NLO) & (\textsc{Herwig7.0.4}) &  (\NNPDF[3.0nlo]) &  (H7-UE-MMHT) \\
\rowstyle{\color{gray}}
& (MG5\_aMC)  & (NLO) & (\textsc{Pythia} 8) & (\NNPDF[3.0nlo]) & (A14)   \\
$t\bar t t\bar t$      & MG5\_aMC & NLO & \textsc{Pythia} 8 & \NNPDF[3.1nlo]~\cite{NNPDF:2017mvq} & A14 \\
\rowstyle{\color{gray}}
& (\textsc{Sherpa 2.2.10}) & (\textsc{MePs@Nlo}) & (\textsc{Sherpa}) &  (\NNPDF[3.0nnlo]) & (\textsc{Sherpa} default) \\
$\ttbar \to W^+bW^-\bar{b}\ell^{+}\ell^{-}$ & MG5\_aMC & LO & \textsc{Pythia} 8 & \NNPDF[3.0lo] & A14   \\
$t (Z/\gamma^*)$       & MG5\_aMC & NLO & \textsc{Pythia} 8 & \NNPDF[2.3lo]  & A14  \\
$t W (Z/\gamma^*)$     & MG5\_aMC & NLO & \textsc{Pythia} 8 & \NNPDF[2.3lo]  & A14  \\
Single top             & \textsc{Powheg-Box} & NLO & \textsc{Pythia} 8 & \NNPDF[3.0nlo] & A14   \\
($t$-, $Wt$-, $s$-channel) &  &  &                                     &  &  \\
$W+$jets               & \textsc{Sherpa} 2.2.1 & \textsc{MePs@Nlo} & \textsc{Sherpa} & \NNPDF[3.0nlo] & \textsc{Sherpa} default \\ %
$VH$                   & \textsc{Powheg-BOX} & NLO  & \textsc{Pythia} 8 & \NNPDF[3.0nlo] & A14 \\
$t\bar t t$            & MG5\_aMC & LO & \textsc{Pythia} 8 & \NNPDF[2.3lo] & A14 \\
$t\bar t W^+ W^-$      & MG5\_aMC & LO  & \textsc{Pythia} 8 & \NNPDF[2.3lo]  & A14  \\
$t \bar{t} ZZ$         & MG5\_aMC & LO  & \textsc{Pythia} 8 & \NNPDF[2.3lo] & A14 \\
$t \bar{t} HH$         & MG5\_aMC & LO  & \textsc{Pythia} 8 & \NNPDF[2.3lo] & A14 \\
$t \bar{t} WH$         & MG5\_aMC & LO  & \textsc{Pythia} 8 & \NNPDF[2.3lo] & A14 \\
\hline\hline
\end{tabular}
\end{center}
\end{table}


%
%
%
\section{Event reconstruction and object identification}
\label{sec:objects}

Candidate events are required to have at least one $pp$ interaction vertex. Interaction vertices are reconstructed from at least two tracks with transverse momentum $\pt$ larger than $500~\mev$ that are consistent with originating from the beam collision region in the $x$--$y$ plane. If more than one primary vertex is found in the event, the candidate with the highest scalar sum of the squared transverse momenta of the associated tracks is selected as the hard-scatter primary vertex~\cite{ATL-PHYS-PUB-2015-026}.

Electron candidates are reconstructed from energy clusters in the electromagnetic calorimeter matched to a track in the ID~\cite{EGAM-2018-01}. They are required to satisfy $\pt>10$~GeV and $|\eta_{\mathrm{cluster}}|<2.47$, excluding the transition region between the endcap and barrel calorimeters ($1.37<|\eta_{\mathrm{cluster}}|<1.52$).
\enquote{Loose} and \enquote{Tight} electron identification working points are used, based on a likelihood discriminant employing calorimeter, tracking and combined information that provide separation between electrons and jets.

The reconstruction of muon candidates is based on tracking information from the MS and the ID, as well as energy deposits in the calorimeter system~\cite{MUON-2018-03}. Muons are required to satisfy $\pt > 10~\gev$ and $|\eta| < 2.5$. \enquote{Loose} and \enquote{Medium} muon identification working points are used.

Electron (muon) candidates are matched to the primary vertex by requiring that the significance of their transverse impact parameter, $d_0$,\footnote{The transverse impact parameter, $d_0$, is
defined in the $x$--$y$~plane as the distance of closest approach of the track to the beamline.} satisfies $|d_0|/\sigma(d_0)<5\,(3)$, where $\sigma(d_0)$ is the measured uncertainty in $d_0$, and by requiring that their longitudinal impact parameter, $z_0$,\footnote{The longitudinal impact parameter, $z_0$, is defined as the distance in $z$ between the primary vertex and the point on the track used to evaluate $d_0$.} satisfies $|z_0 \sin\theta|<0.5$~mm.

Lepton candidates are also required to be isolated in the tracker and in the calorimeter to further suppress leptons from heavy-flavour (HF) hadron decays, misidentified jets, or photon conversions (collectively referred to as \enquote{non-prompt leptons}). The track-based lepton isolation criterion is based on the quantity $I_R = \sum \pt^{\textrm{trk}}$, where the scalar sum includes all tracks (excluding the lepton candidate itself) within a cone of size $\Delta R<R_{\textrm{cut}}$ around the direction of the lepton. The value of $R_{\textrm{cut}}$ is the smaller of $r_{\textrm{min}}$ and $10~\gev/\pt^\ell$, where $r_{\textrm{min}}$ is set to 0.2 (0.3) for electron (muon) candidates and $\pt^\ell$ is the lepton's $\pt$. All lepton candidates must satisfy $I_R/\pt^\ell < 0.15$.
They are also required to satisfy a calorimeter-based isolation criterion: the sum of the transverse energy within a cone of size $\Delta R=0.2$ around the lepton, after subtracting contributions from \pileup and the energy deposit of the lepton itself, is required to be less than 20\% (30\%) of the electron's (muon's) $\pt^\ell$.

These selection criteria largely suppress the contribution from non-prompt leptons. However, several channels considered in this search have additional suppression requirements targeting the main types of non-prompt leptons. Non-prompt leptons from hadron decays that contain bottom- or charm-quarks (referred to as \enquote{HF non-prompt leptons}) are further rejected using a boosted decision tree (BDT) discriminant (referred to as the non-prompt-lepton BDT~\cite{HIGG-2017-02}), based on isolation and lifetime information about a track-jet that matches the selected electron or muon (referred to as a \enquote{light lepton}).
Three working points (WPs) based on the non-prompt-lepton BDT are used: \textit{Tight}, \textit{VeryTight}, and \textit{Tight--not-VeryTight}.
The \textit{Tight} WP allows prompt muons (barrel/endcap electrons) satisfying the calorimeter- and track-based isolation criteria to be selected with an efficiency that is about 60\% ($60\%/70\%$) for $\pt \sim 20~\gev$ and reaches a plateau of 95\% ($95\%/90\%$) for $\pt \sim 40\ (40/65)~\gev$.
The prompt-lepton efficiency of the \textit{VeryTight} WP for muons (barrel/endcap electrons) that satisfy the calorimeter- and track-based isolation criteria is approximately 55\% ($55\%/60\%$) for $\pt \sim 20~\gev$ and reaches a plateau of 90\% ($85\%/83\%$) for $\pt \sim 40\ (40/65)~\gev$. The corresponding rejection factor\footnote{The rejection factor is defined as the reciprocal of the efficiency.} for muons (electrons) from the decay of $b$-hadrons ranges from 33 to 50 (20 to 50) for the \textit{Tight} WP, and from 50 to 100 (33 to 66) for the \textit{VeryTight} WP, depending on \pt and $\eta$, after resolving ambiguities between overlapping reconstructed objects.
The \textit{Tight--not-VeryTight} WP allows the selection of non-prompt leptons and is part of the event selection for control regions enriched in HF non-prompt-lepton background, as described in Section~\ref{sec:background}.

To suppress electrons with an incorrect charge assignment, a BDT discriminant based on calorimeter and tracking quantities~\cite{EGAM-2018-01} is used. An efficiency of approximately 96\% in the barrel region and 81\% in the endcaps is obtained, with rejection factors of 19 in the barrel region and 40 in the endcaps.
The electron candidates are separated into three classes: \enquote{material conversion}, \enquote{internal conversion}, and \enquote{non-conversion}.
Most electrons arising from material conversions, i.e.\ from photon conversions in the detector material, are rejected by the standard electron identification selection, but additional requirements are imposed to remove residual material-conversion candidates.
These candidates have a reconstructed displaced vertex with radius $r > 20$~mm that includes the track associated with the electron.\footnote{The beampipe and insertable B-layer inner radii are 23.5~mm and 33~mm, respectively.} The invariant mass of the associated track and the closest (in $\Delta\eta$) opposite-sign charge track reconstructed in the silicon detector, calculated at the conversion vertex, is required to be lower than $100~\mev$. Internal conversion candidates, which correspond to the internal photon conversions ($\gamma^* \to \ell^{+}\ell^{-}$), must fail to satisfy the requirements for material conversions, and the di-track invariant mass, calculated here at the primary vertex, is also required to be lower than $100~\mev$.

The various requirements applied to the different lepton categories used are summarised in Table~\ref{tbl:tightleps}. After the initial categorisation based on \enquote{loose} leptons (corresponding to $L$), the best lepton working point to further optimise the event selection is chosen depending on the main background processes and available number of data events in each category. The various choices for the signal and control regions are described in Section~\ref{sec:events}.

\begin{table}
\begin{center}
\caption{\label{tbl:tightleps} Description of the loose inclusive ($L$), loose with tighter identification ($L*$), medium inclusive ($M$), medium exclusive ($M_{\mathrm{ex}}$), and tight ($T$)  lepton definitions. The electron $e^{*}$ is required to fulfil, in addition to the corresponding lepton definition requirements, those corresponding to an internal or material conversion candidate.}
\resizebox{\textwidth}{!}{
\begin{tabular}{l||c|c|c|c|c||c|c|c|c|c}
\hline\hline
& \multicolumn{5}{c||}{Electron} & \multicolumn{5}{c}{Muon} \\
\hline
Lepton definition                  		& $L$ & $L*$ & $M$ & $M_{\mathrm{ex}}$ & $T$ &  $L$ & $L*$ & $M$ & $M_{\mathrm{ex}}$ & $T$ \\
\hline
Identification                     		& Loose & \multicolumn{4}{c||}{Tight} & Loose & \multicolumn{4}{c}{Medium} \\
\hline
$|d_0|/\sigma_{d_0}$		        		&  \multicolumn{5}{c||}{$<5$} & \multicolumn{5}{c}{$<3$ } \\
\hline
$|z_0 \sin \theta|$ [mm]  				&  \multicolumn{10}{c}{$< 0.5$}  \\
\hline
Isolation                          		&  \multicolumn{5}{c||}{Yes} & \multicolumn{5}{c}{Yes} \\
\hline
Non-prompt lepton WP               		&  \multicolumn{2}{c|}{--} & \textit{Tight}  & \textit{Tight--not-}  & \textit{VeryTight} &  \multicolumn{2}{c|}{--} & \textit{Tight}  & \textit{Tight--not-}  & \textit{VeryTight}  \\
&  \multicolumn{2}{c|}{}   &               	 & \textit{VeryTight}    &                    &  \multicolumn{2}{c|}{}   &                 &
\textit{VeryTight}    &  \\
\hline
Charge-misassignment veto 				&  \multicolumn{2}{c|}{--} & \multicolumn{3}{c||}{Yes}               & \multicolumn{5}{c}{--} \\
\hline
Conversion candidate veto 				&  -- & \multicolumn{4}{c||}{Yes (except $e^{*}$)} 					 & \multicolumn{5}{c}{--} \\
\hline\hline
\end{tabular}
}
\end{center}
\end{table}

The constituents for jet reconstruction are identified by combining measurements from both the ID and the calorimeter using a particle flow (PFlow) algorithm~\cite{PERF-2015-09}. Jet candidates are reconstructed from these PFlow objects using the \antikt~algorithm~\cite{Cacciari:2008gp,Cacciari:2011ma} with a radius parameter of $R=0.4$. They are corrected to particle level by the application of jet energy scale (JES) and resolution (JER) calibrations, derived from 13 TeV data and simulation~\cite{JETM-2018-05}. Only jet candidates with $\pT> 25$~\GeV\ and within $|\eta|<2.5$ are selected. To reduce the effect of \pileup, each jet with $\pT<60$~\GeV\ and $|\eta|<2.4$ is required to have an origin compatible with the primary vertex, as defined by the jet vertex tagger (JVT)~\cite{PERF-2014-03} criteria. A set of quality criteria is also applied to reject events containing at least one jet arising from non-collision sources or detector noise~\cite{ATLAS-CONF-2015-029}.

Jets containing $b$-hadrons are identified ($b$-tagged) via an algorithm~\cite{FTAG-2019-07} that uses a deep-learning neural network based on the distinctive features of $b$-hadron decays, primarily the impact parameters of tracks and the displaced vertices reconstructed in the ID. Additional input to this network is provided by discriminating variables constructed by a recurrent neutral network, which exploits the spatial and kinematic correlations between tracks originating from the same $b$-hadron. A multivariate $b$-tagging discriminant value is calculated for each jet.
In this search, a jet is considered $b$-tagged if it passes the working point corresponding to 85\%, 77\%, 70\%, or 60\% average expected efficiency to tag a $b$-quark jet,
with a light-jet\footnote{`Light jet' refers to a jet originating from the hadronisation of a light quark ($u$, $d$, $s$) or a gluon.} rejection factor of about 40 to 2500, and a charm-jet ($c$-jet) rejection factor of about 3 to 40, as determined for jets with $\pt >20~\gev$ and $|\eta|<2.5$ in simulated $\ttbar$ events.
The $b$-tagging distribution obtained by ordering the resulting five exclusive bins from the four working points from higher to lower $b$-jet efficiency is referred to as `pseudo-continuous' $b$-tagging score, and it is used as input to the multivariate analysis discriminant described in Section~\ref{sec:events}.
The notation \bveryloose, \bloose, \bmedium, and \btight is used to denote a $b$-tagged jet (\bjet) that satisfies the corresponding working point.
Correction factors derived from dedicated calibration samples enriched in \bjet{}s, $c$-tagged jets, or light-tagged jets, are applied to the simulated event samples~\cite{FTAG-2018-01,FTAG-2020-08,FTAG-2019-02}.

To uniquely identify objects, a sequential \enquote{overlap removal} procedure is performed. Electrons and muons that satisfy the $L$ criteria are considered in this procedure, as well as jets that satisfy the JVT requirement.

If two electrons are separated by $\Delta R<0.1$, only the one with the higher \pt\ is kept. If an electron and a muon overlap within $\Delta R<0.1$, the muon is removed if it is reconstructed only from an ID track and calorimeter energy deposits consistent with a minimum-ionising particle (i.e.\ if it is \enquote{calo-tagged}), otherwise the electron is removed. If an electron and a selected jet are found within $\Delta R < 0.2$, the jet is removed. For each electron in the event a \pt-dependent variable-size cone of maximum size $\Delta R=0.4$ is defined. If a selected jet, surviving all previous overlap criteria, is found in this cone, the lepton is rejected. The same procedure is also applied between jets and muons, with the exception that, if a muon and a jet overlap with $\Delta R <$ 0.2, the jet is removed, unless the number of tracks in the jet is more than two.

The missing transverse momentum $\vec{p}_{\mathrm{T}}^{\mathrm{miss}}$ (with magnitude $\met$) is defined as the negative vector sum of the $\pt$ of all selected and calibrated objects in the event that fulfilled the overlap removal procedure, and an additional term to account for the momenta of soft particles that are not associated with any of the selected objects~\cite{JETM-2020-03}.
This soft term is calculated from inner-detector tracks matched to the primary vertex, which makes it more resilient to contamination from \pileup interactions.


%

%
\section{Search strategy}
\label{sec:events}

%
Events are firstly required to satisfy a minimal preselection and then are categorised into orthogonal SRs based on different criteria such as number of leptons and a multi-class DNN classifier.
This categorisation provides a set of regions that are sensitive to all the signal production and decay modes considered in this search.
Orthogonal CRs are defined to constrain the normalisation of the main backgrounds.
Dedicated kinematic selections are applied to the CRs to improve the purity of the targeted backgrounds.
A maximum-likelihood fit is performed across the electron (muon) SRs to test for a possible \vlle (\vllmu) signal, together with the CRs to constrain in situ the leading backgrounds simultaneously.

Candidate events are selected by a combination of single-lepton and dilepton triggers, requiring the electrons or muons to satisfy identification criteria similar to those used in the offline reconstruction and isolation requirements~\cite{TRIG-2018-01,TRIG-2018-05}.
Single-electron triggers require a minimum $\pt$ threshold of 24 (26)~$\gev$ in the 2015 (2016, 2017 and 2018) data-taking period(s), while single-muon triggers have a lowest $\pt$ threshold of 20 (26)~$\gev$ in 2015 (2016--2018).
The dielectron triggers require two electrons with minimum $\pt$ thresholds ranging from 12~$\gev$ in 2015 to 24~$\gev$ in 2017--2018, whereas the dimuon triggers use asymmetric $\pt$ thresholds for leading (subleading) muons: 18 (8)~$\gev$ in 2015 and 22 (8)~$\gev$ in 2016--2018. Finally, an electron+muon trigger requires events to have an electron candidate with a 17~$\gev$ threshold and a muon candidate with a 14~$\gev$ threshold for all periods.

In the offline selection at least two leptons in the event are required to be matched, within $\Delta R < 0.15$, to the corresponding leptons reconstructed by the trigger and to have a $\pt$ exceeding the trigger $\pt$ threshold by at least 1~$\gev$.

Three orthogonal event categories are defined according to the number of $L$ leptons in the event: opposite-sign charge dilepton (\llOS), three-lepton (\lll), and four-lepton (\llll) categories. The four-lepton category is inclusive and contains events with higher lepton multiplicity, while the other two are exclusive. %
VLL signals from both doublet and singlet models can be classified into distinct topologies in the \llOS and \lll channels. Signals from the VLL doublet model are characterised mostly by low \met, while those from the singlet model can have larger \met due to a neutrino from the VLL in the final state.
To minimise the migration of signal events in each decay mode across categories while maximising rejection against the SM background, the analysis uses a multi-class DNN, trained separately in the \llOS and \lll channels to classify events into the background or one of the signal categories.

The training of the DNN is done using the \textsc{Keras} library~\cite{chollet2015keras} with \textsc{Tensorflow} as a backend~\cite{tensorflow2015-whitepaper} and Adam optimiser~\cite{kingma2017adam}.
The networks consist of five input features, two dense fully connected layers of 22 (30) nodes with rectified linear units as activation functions, interleaved with a drop-out layer with 20\% rate, and six (four) output nodes with a soft-max activation function for the categorisation of \llOS (\lll) events. The network is trained with a batch size of 2000 and up to 100 epochs, using all the available signal mass points. To avoid discarding signal events in the evaluation, a two-fold cross-validation is used with the events divided by even/odd event number.

The five input features are the number of jets, the sum of the pseudo-continuous $b$-tagging scores of all jets, the number of hadronic $W/Z$ bosons, the number of hadronic $H$ bosons, and the event \met, as shown in Table~\ref{table:varNN}.
The hadronic $H$ boson candidates are reconstructed by requiring one (for boosted scenarios) or two \bveryloose jets to have an invariant mass within 90--140 GeV, compatible with the Higgs boson mass.
The hadronic $W/Z$ boson candidates are similarly identified by requiring one or two jets to have an invariant mass within 60--110 GeV, compatible with the $W$ or $Z$ boson mass.
The order in which the matching of jet(s) as originating from $H$ or $W/Z$ bosons is tested is the following: 1 jet matched to $H$ boson candidate, 2 jets matched to $H$ boson candidate, 1 jet matched to $W/Z$ boson candidate, 2 jets matched to $W/Z$ boson candidate.
These variables are independent of lepton flavour and independent of the mass of the VLL, ensuring the same training can be used for both \vlle and \vllmu for all mass points.
Good modelling of the input variables is observed in the \ttbar and \Zjets correction regions as defined in Section~\ref{sec:background}.

\begin{table}[tb]
\small
\begin{center}
\caption{Input variables to the DNN trainings in the \llOS and \lll channels.}
\begin{tabular}{l}
\hline
Variable \\
\hline
Number of jets ($N_\text{jets}$) \\
Sum of pseudo-continuous $b$-tagging scores of jets \\ %
Number of hadronic $W$/$Z$ bosons \\ %
Number of hadronic $H$ bosons \\ %
Missing transverse energy (\met) \\
\hline
\end{tabular}
\label{table:varNN}
\end{center}
\end{table}

The output categories correspond to the signal topologies, with an additional category in each channel corresponding to the `SM-like' events, defined to capture events that fall into none of the other signal categories. Each event is categorised according to the highest class probability.

In the \llOS channel, three low-\met and two high-\met signal topologies are defined, sensitive to both VLL models:

\begin{itemize}
\item $\ell\ell \mathrm{HH}$: low-\met topology targeting two hadronic Higgs boson candidates. Characterised by a high number of $b$-tagged jets.
\item $\ell\ell \mathrm{HV}$: low-\met topology targeting a hadronic Higgs boson candidate and a hadronic $W/Z$ boson candidate. Characterised by a high number of light- and heavy-flavour jets.
\item $\ell\ell \mathrm{VV}$: low-\met topology targeting two hadronic $W/Z$ boson candidates. Characterised by a high number of light-flavour jets.
\item $\ell\nu \mathrm{HW}$: high-\met topology targeting a hadronic Higgs boson candidate and a leptonic $W$ boson candidate. Characterised by heavy-flavour jets.
\item $\ell\nu \mathrm{WZ}$: high-\met topology targeting a hadronic $Z$ boson candidate and a leptonic $W$ boson candidate. Characterised by light-flavour jets.
\end{itemize}

Similarly in the \lll channel, two low-\met and one high-\met signal topologies are defined:

\begin{itemize}
\item $\ell\ell \mathrm{HH+HW}$: low-\met topology targeting a hadronic Higgs boson candidate and a leptonic $W$ boson candidate (prompt or from the other Higgs boson decay). Characterised by a high number of $b$-tagged jets.
\item $\ell\ell \mathrm{HV+VV}$: low-\met topology targeting a Higgs boson $H \to WW$ candidate, with one of the $W$ bosons decaying into a lepton and a neutrino, and a hadronic $W/Z$ boson candidate. Characterised by a high number of light-flavour jets.
\item $\ell\nu \mathrm{HV}$: high-\met topology targeting a Higgs boson $H \to WW$ candidate, with both $W$ bosons decaying into a lepton and a neutrino, and a hadronic $W/Z$ boson candidate. Characterised by a low number of jets.
\end{itemize}

The advantage of this classification is that the analysis is sensitive to not only the VLL doublet and singlet models, but also to any similar non-VLL topology that fulfils one of the above descriptions. In the training all background processes are included, normalised to their respective cross-sections and including the corrections described in Section~\ref{sec:background}, as well as the \vlle or \vllmu signal samples in the doublet and singlet scenarios for all masses, sub-divided using truth information into the aforementioned signal templates.

The `SM-like' categories in \llOS and \lll are expected to be low in signal yield but capture a significant contribution of the SM background. These categories are not included in the final fit to the data, rejecting a large fraction of the SM background contamination.

The \llll channel is subdivided into four signal categories based on the presence of one or two opposite-sign charge same-flavour (OSSF) lepton pairs, and the multiplicity of electrons or muons. Table~\ref{tab:selectionSR} shows a summary of the signal region categories.

\begin{table}[tb]
\begin{center}
\caption{Event selection summary in the signal regions.
Leptons are ordered by decreasing \pt in the \llOS and \llll regions. In the \lll regions the lepton with opposite-sign charge is taken first, followed by the two same-sign leptons in decreasing \pt order. In $2e2\mu$ ($2\mu2e$) events the sum of the \pt of the two electrons (muons) is larger than the sum of the \pt of the two muons (electrons).
The splitting of regions are based on DNN classifications in the \llOS and \lll regions, where `H' (`V') stands for the Higgs boson ($W$ and $Z$ bosons).
}
\resizebox{\textwidth}{!}{
\def\arraystretch{1.4}
\begin{tabular}{l|c|c|c}
\hline\hline
Signal regions                     & \llOS                                 & \lll                                         & \llll \\
\hline\hline
Lepton flavour					   & $e$: ($ee$) 						   & $e$: $(2e1\mu+3e)$       					  & $e$: $(2e2\mu+3e1\mu+4e)$ \\
& $\mu$: ($\mu\mu$)					   & $\mu$: $(2\mu1e+3\mu)$						  & $\mu$: $(2\mu2e+3\mu1e+4\mu)$ \\
Lepton definition				   & $(L*,L*)$                             & $(L*,M,M)$                                   & $(L*,L*,L*,L*)$ \\
Minimum lepton \pt[GeV]           & (20, 20)                              & (10, 20, 20)                                 & (10, 10, 10, 10) \\
\mllOSSF[GeV]                     & $>15$                                 & \multicolumn{2}{c}{--} \\
$|\mllOSSF - m_Z|$ [GeV]           & $>10$                                 & $>10$                                        & $>10$ for at least 1 OSSF pair \\
\njets                             & $\ge$ 2                               & $\ge$ 1                                      & $\ge$ 0 (1OSSF)  \\
&                                       &                                              & $\ge$ 1 (2OSSF)  \\
Other                              & $\Delta R (\ell,\ell) > 1$            & Total lepton charge $= \pm 1$                & Total lepton charge $= 0$  \\

\hline\hline
Region split                       & {\small($\ell\ell \mathrm{HH}$, $\ell\ell \mathrm{HV}$, $\ell\ell \mathrm{VV}$, $\ell\nu \mathrm{HV}$, $\ell\nu \mathrm{VV}$) $\times$ ($e, \mu$)} & {\small($\ell\ell \mathrm{HH+HV}$, $\ell\ell \mathrm{HV+VV}$, $\ell\nu \mathrm{HV}$) $\times$ ($e, \mu$)} & {\small(1OSSF, 2OSSF) $\times$ ($e, \mu$)} \\
\hline
Region naming                      & \twoeehh, \twommhh       			   &   \threeehhhw, \threemhhhw      	 & \eeem, \mmme \\
& \twoeehv, \twommhv     			   &   \threeevvhz, \threemvvhz     	 & \eeee, \mmmm  \\
& \twoeevv, \twommvv     			   &   \threeenuhw, \threemnuhw    	 	 &               \\
& \twoenuhv, \twomnuhv                  &                                 	 &               \\
& \twoenuvv, \twomnuvv                  &                                 	 &               \\
\hline\hline
\end{tabular}
}
\label{tab:selectionSR}
\end{center}
\end{table}

Multiple CRs are defined to fit the normalisation of the leading backgrounds. These regions are orthogonal to the signal regions and among each other based on different requirements on the lepton working points, dilepton invariant mass, and jet and $b$-jet multiplicities. First, a region enriched in \ttW is defined by selecting two same-sign leptons with the tight definition and at least two \btight jets. A region enriched in $\WZ$ and \ttZ is defined by selecting events with three leptons (a same-sign pair of $M$ leptons and an opposite-sign charge $L*$ lepton), from which one OSSF lepton pair is required to be compatible with a $Z$ boson, $|\mllOSSF - m_Z| < 10~\GeV$. A dedicated region dominated by $ZZ$ is defined by requiring four $L*$ leptons where both OSSF pairs fulfil $|\mllOSSF - m_Z| < 10~\GeV$. Two CRs enriched in photon conversions from $Z \rightarrow \mu \mu \gamma^*(\rightarrow ee)$ are defined, according to the identification of the electron as a material conversion or internal conversion candidate. Finally, four CRs enriched in HF non-prompt leptons are defined, requiring exactly one \btight jet to be orthogonal to the \ttW CR. Events with two same-sign leptons are categorised according to the criteria $(T,M_{\mathrm{ex}})$ and $(T,L*)$ for the leading and subleading leptons in $\pt$, and further split according to the fake-lepton-candidate flavour, which is assumed to be the subleading lepton. These two CRs allow to derive constraints on the background from HF non-prompt leptons for the \lll channel and for the \llOS and \llll channels, respectively.

The full description of the kinematic selections applied to each CR is given in Table~\ref{tab:selectionCR}. As described in Section~\ref{sec:background}, background corrections are derived in orthogonal regions and applied to the corresponding simulated processes before the simultaneous fit to data.

\begin{table}[tb]
\begin{center}
\caption{Event selection summary in the CRs. The notation $e^*$ is used to denote material conversion or internal conversion candidates, as described in Section~\ref{sec:objects}.
Leptons are ordered by decreasing \pt in the \llSS and \llll regions. In the \lll regions the lepton with opposite-sign charge is taken first, followed by the two same-sign leptons in decreasing \pt order.
In the HF non-prompt lepton region naming, `$2\ell$tt(e)' (`$2\ell$tt($\mu$)') is the CR enriched in non-prompt electrons (muons) from semileptonic $b$-decays originating mostly from \ttbar and with the lepton flavours for the leading and subleading leptons corresponding to `$ee, \mu e$' (`$\mu\mu, e \mu$'). The additional $(T,M_{\mathrm{ex}})$ and $(T,L*)$ subscripts refer to the lepton definitions required for the leading and subleading leptons in each region.
}
\resizebox{\textwidth}{!}{
\def\arraystretch{1.4}
\begin{tabular}{l|c|c|c|c|c}
\hline\hline
Control regions                    & \ttW     					& \WZ and \ttZ      &  \ZZ     				& Conversions 			& HF non-prompt \\
\hline\hline
\njets                             & $\geq 2$ 					& \multicolumn{3}{c|}{$\geq 0$}              						& $\geq 2$ \\
\nbjets                            & $\geq$ 2 \btight			& \multicolumn{2}{c|}{--}     				& 0 \btight 			& 1 \btight \\
Lepton requirement                 & \llSS                      & \lll         		& 2e2$\mu$, 4e, 4$\mu$  & $\mu\mu e^*$ 			& \llSS \\
Lepton definition                  & $(T,T)$ 					& $(L*,M,M)$ 		& $(L*,L*,L*,L*)$		& $(L*,M,M)$ 			& $(T,M_{\mathrm{ex}})$, $(T,L*)$  \\
Minimum lepton \pt[GeV]           & (20, 20)    				& (10, 20, 20)      & (10, 10, 10, 10)      & (10, 20, 20)          & (20, 20) \\
\mllOSSF[GeV]                     & $>15$						& \multicolumn{4}{c}{--} \\
$|\mllOSSF - m_Z|$ [GeV]           & --							& \multicolumn{2}{c|}{$<10$} 				& $>10$ 				& -- \\
$|\mlll - m_Z|$ [GeV]              & \multicolumn{3}{c|}{--}  												& $<10$ 				& -- \\
\hline\hline
Region split                       & \multicolumn{3}{c|}{--}					& internal / material 	& subleading $e/\mu$ $\times$ ($(T,M_{\mathrm{ex}})$, $(T,L*)$) \\
\hline
Region naming                      & \llSSttW  					& \lllVVttZ 		& \llllZZ 				& \lllintc 				& \llttetm, \llttetl \\
&         					&         			&						& \lllmatc 				& \llttmutm, \llttmutl \\
\hline\hline
\end{tabular}
}
\label{tab:selectionCR}
\end{center}
\end{table}

Figure~\ref{fig:sketch_strategy} illustrates the categorisation and definition of the SRs and CRs that are fitted simultaneously.
A total of 19 analysis regions are defined for each VLL search (electron or muon), with 10 SRs (5 for \llOS, 3 for \lll, and 2 for \llll) and 9 CRs. In each region, a given kinematic variable is fitted to improve the sensitivity to the targeted signal process (in the case of the SRs) or to improve the modelling of a particular background process (in the case of the CRs).
The sum of the \pt of the leptons plus the event \met (\htlepmet) is fitted in the signal regions and is connected to the VLL mass; the $b$-jet multiplicity (\nbjets) is fitted in the diboson and \ttZ CR and provides discrimination between these two background processes; the total event yield is fitted in the other CRs.

\begin{figure}[tb]
\centering
\includegraphics[width=1.1\textwidth]{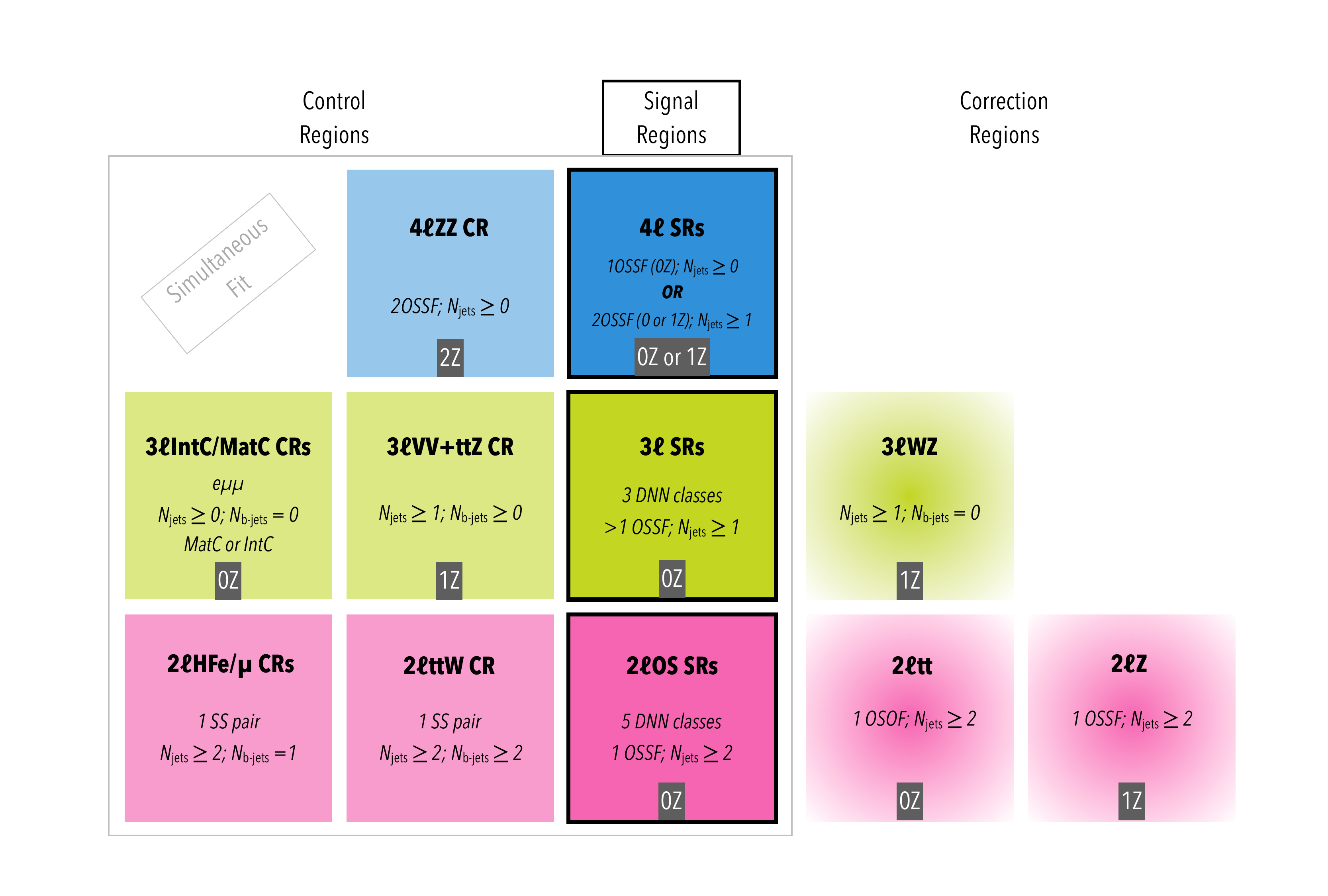}\\
\caption{Illustrative sketch of the definition of the SRs and CRs. The $2\ell \mathrm{HF}_{\mathrm{e}/\mu}$ CRs include the \llttetm, \llttetl, \llttmutm, and \llttmutl regions. The \llOS, \lll and \llll SRs include the various regions shown in Table~\ref{tab:selectionSR}. The control regions correspond to those described in Table~\ref{tab:selectionCR} and the correction regions are presented in Table~\ref{tab:selectionCorrectionRegions}. The terms `OSSF' and `OSOF' refer to opposite-sign charge lepton pairs with same and opposite lepton flavours, respectively, whereas `SS' refers to a same-sign charge lepton pair.}
\label{fig:sketch_strategy}
\end{figure}


%

%
\section{Background estimation}
\label{sec:background}

Section~\ref{sec:bkg_irreducible} describes the irreducible backgrounds, where prompt leptons are produced from $W/Z$ boson decays, leptonic $\tau$-lepton decays, or internal conversions. Section~\ref{sec:bkg_reducible} introduces the reducible backgrounds, containing prompt leptons with misassigned charge or at least one non-prompt lepton in the event.

All background processes are estimated by using the simulation samples described in Section~\ref{sec:datamc}. Before the simultaneous fit to data, the event kinematics of the simulated $\ttbar$, $\Zjets$ and $\VV$ backgrounds require dedicated corrections derived from data control samples (described in Table~\ref{tab:selectionCorrectionRegions}) to better describe the data. During the simultaneous fit to data discussed in Section~\ref{sec:results}, the yields of $\ttW$, $\ttZ$, $\WZ$, $\ZZ$ and non-prompt-lepton backgrounds, are adjusted via normalisation factors.

\begin{table}[htb]
\begin{center}
\caption{Event selection summary in the regions used for deriving data-driven corrections and for cross-checks of the \ttbar, \Zjets and \WZ background processes.
Leptons are ordered by decreasing \pt in the \llOS regions. In the \lll region the lepton with opposite-sign charge is taken first, followed by the two same-sign leptons in decreasing \pt order.
}
\def\arraystretch{1.4}
\begin{tabular}{l|c|c|c}
\hline\hline
Correction regions    				& \ttbar                                 & \Zjets                                     & \WZ \\
\hline\hline
Lepton flavour					   	& \llOS $e\mu$						   	 & \llOS $ee$, $\mu\mu$						  & \lll \\
Lepton definition				   	& $(L*,L*)$                              & $(L*,L*)$                                  & $(L*,T,T)$ \\
Minimum lepton \pt[GeV]           	& \multicolumn{2}{c|}{(20, 20)}							                              & (10, 20, 20)   \\
\mll[GeV]	                       	& $>15$ (OS)                             & $>15$ (OSSF)								  & -- \\
$|\mll - m_Z|$ [GeV]          	   	& $>10$ (OS)						   	 & \multicolumn{2}{c}{$<10$ (OSSF)} \\
\njets                             	& \multicolumn{2}{c|}{$\ge$ 2}					                                      & $\ge$ 1 \\
\nbjets							   	& \multicolumn{2}{c|}{--}					                              & 0 \bloose \\
\hline\hline
\end{tabular}
\label{tab:selectionCorrectionRegions}
\end{center}
\end{table}

\subsection{Irreducible backgrounds}
\label{sec:bkg_irreducible}

Background contributions with prompt leptons originate from a wide range of physics processes with the relative importance of individual processes varying by channel. The main irreducible backgrounds originate from $\ttbar$, $\Zjets$, $\ttW$, $\ttdy$, $\WZ$ and $\ZZ$ production, and have final states and kinematic properties similar to the VLL signal. Smaller contributions originate from the following rare processes: $\ttH$, $\tttt$, $\tZ$, $\tW$, $\tWZ$, $\ttWW$, $\VVV$, and $\ttt$ production.

\subsubsection{\ttbar background}

The \ttbar process is one of the main prompt backgrounds in the \llOS signal regions. Three distinct corrections are applied: first a theory-based correction to correct the distributions of the \pT and of the invariant mass of the $t\bar{t}$ system ($m(t\bar{t})$) at parton level, then data-driven corrections to improve the modelling of the number of extra HF jets and the jet multiplicity in different corners of the \met phase space.

Previous studies~\cite{Czakon_2017} have shown that the latest theoretical predictions at NNLO QCD and NLO EW for the top quark \pT are significantly softer than the spectrum from the nominal \textsc{Powheg+Pythia8} and alternate Monte Carlo samples considered in this analysis, and significantly different in $d\sigma(t\bar{t})/dm(t\bar{t})$. An iterative, recursive reweighting procedure is used to correct the parton level distributions of $p_{\mathrm{T}}(t)$, $p_{\mathrm{T}}(\bar{t})$, $m(t\bar{t})$ and $p_{\mathrm{T}}(t\bar{t})$ in each of the \ttbar MC samples.
Figure~\ref{fig:tt_thcorrection} shows the sum of the \pT of the jets (\HTjets) distributions before and after the corrections are applied.

\begin{figure}[!h]
\centering
\subfloat[]{\includegraphics[width=0.35\textwidth]{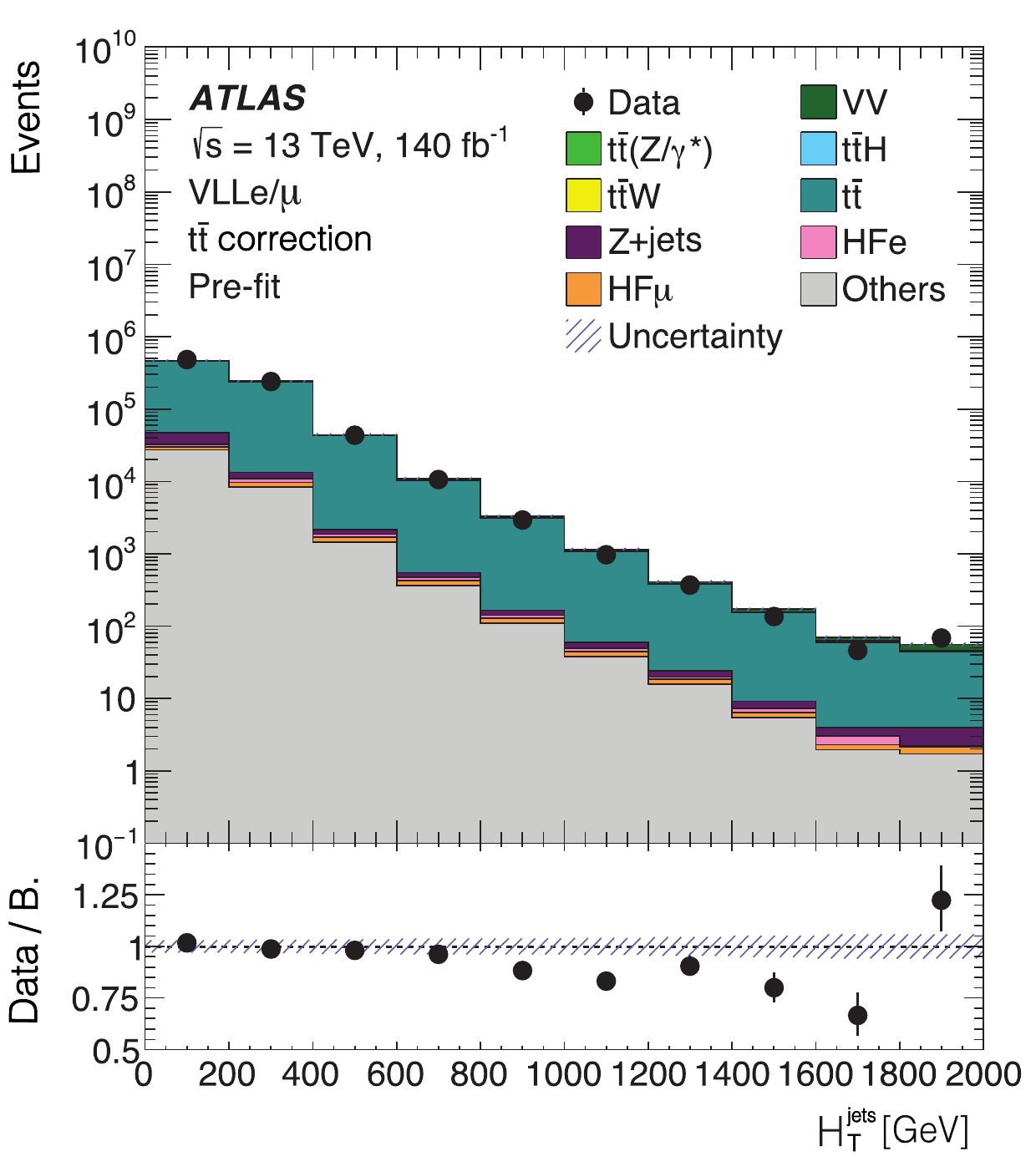}}
\subfloat[]{\includegraphics[width=0.35\textwidth]{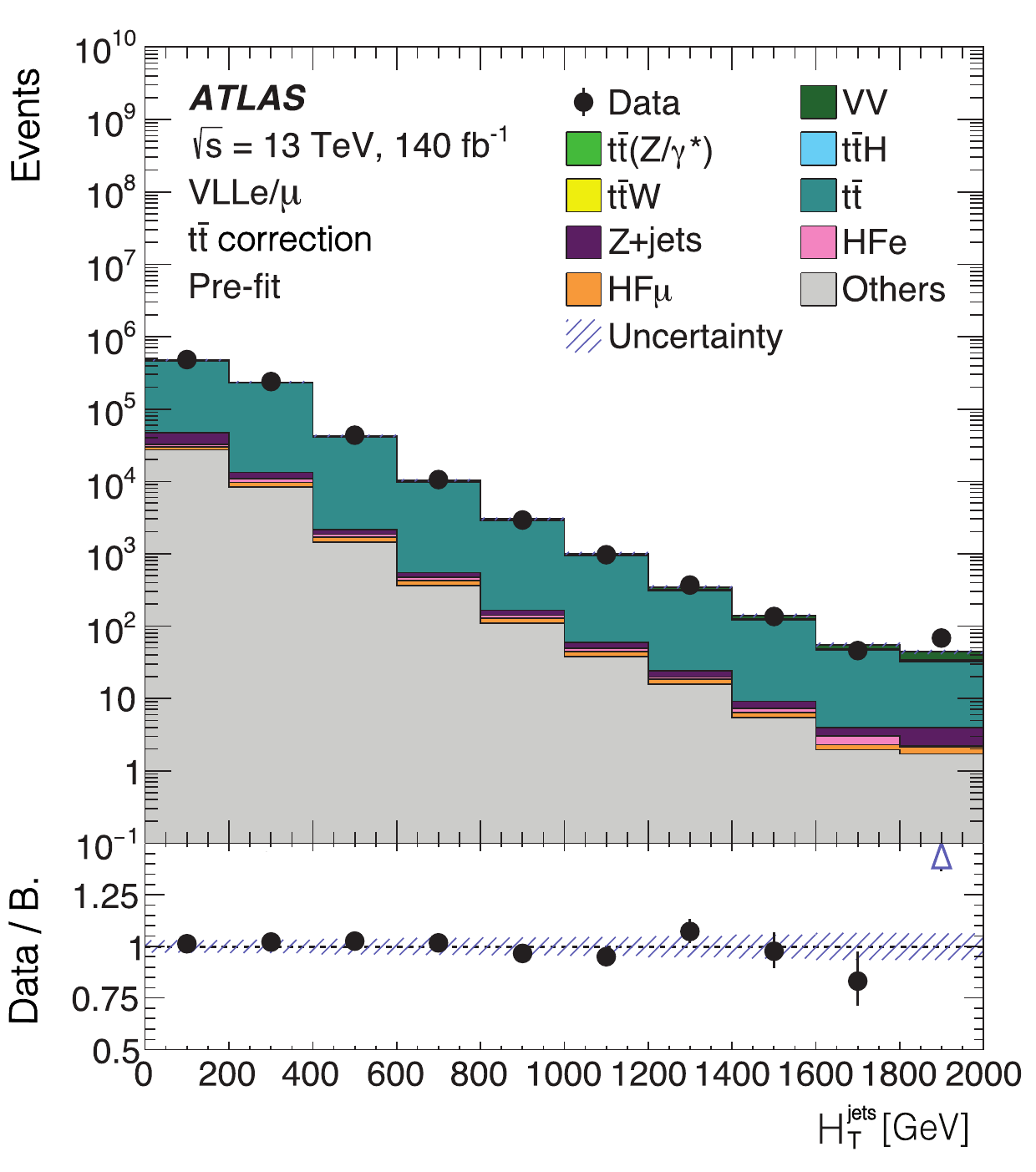}}
\caption{Distributions of \HTjets before the fit to data in the \ttbar correction region (a) before any corrections are applied and (b) after the theory-based \ttbar corrections are implemented.
The ratio of the data to the background prediction (`B.') is shown in the lower panel.
The `Others' contribution is dominated by the $\tW$ production. Only statistical uncertainties are shown. The last bin in each distribution contains the overflow.
\label{fig:tt_thcorrection}}
\end{figure}

Even though significant improvement in the agreement of MC to data is observed after the reweighting in many kinematic variables related to the top/anti-top \pT, the agreement in jet multiplicity and multiplicity of b-tagged jets remains suboptimal. Additional data-driven corrections are derived for these distributions in the \ttbar $e \mu$ correction region (see Table~\ref{tab:selectionCorrectionRegions}) in an iterative procedure. First, corrections to the fractions of $t\bar{t}$ + light flavour jets ($t\bar{t}$ + LF), $t\bar{t}$ + $b$-tagged jets ($t\bar{t} + b$) and $t\bar{t}$ + $c$-tagged jets ($t\bar{t} + c$) are estimated with dedicated normalisation factors, by fitting the sum of pseudo-continuous b-tagging scores to data. The resulting values are $0.98 \pm 0.01$ for \ttbar + LF, $1.30 \pm 0.04$ for \ttbar + $b$ and $1.40 \pm 0.05$ for \ttbar + $c$.
Second, a bin-to-bin rescaling to data is performed for the jet multiplicity in four \met bins: 0--100~\gev, 100--200~\gev, 200--300~\gev, and $>$300~\gev. The derived correction is less than 1\% for 2 or 3 jets and up to 10\% for 6 jets in the regions with \met $< 300$~\gev. For higher jet multiplicities and higher \met, the correction can be up to 48\%. After correcting for the number of jets, a fit to the sum of pseudo-continuous b-tagging scores is repeated to check that the best-fit values of \ttbar + LF, \ttbar + $b$ and \ttbar + $c$ are not affected.

The distributions of the fitting variable, \htlepmet, in the different \llOS DNN classes (see Table~\ref{tab:selectionSR}) in the \ttbar correction region (see Table~\ref{tab:selectionCorrectionRegions}) are shown in Figure~\ref{fig:ttbar_VRs}.

\begin{figure}[!h]
\centering
\subfloat[]{\includegraphics[width=0.33\textwidth]{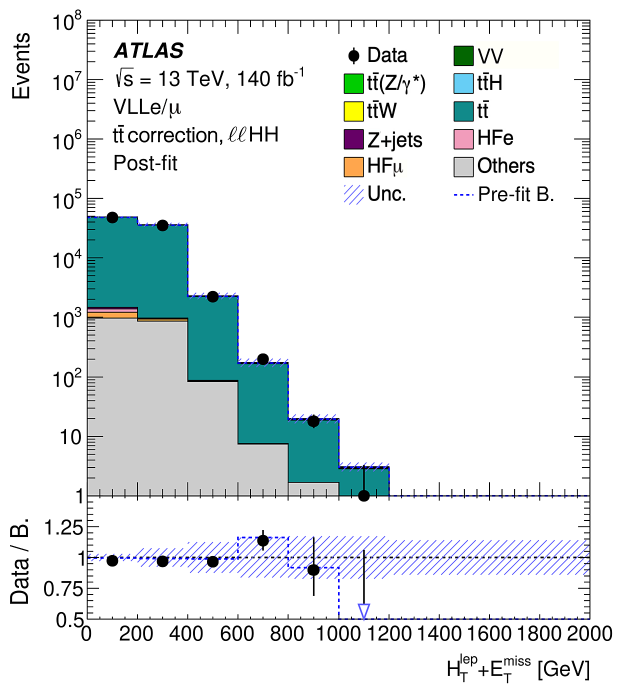}}
\subfloat[]{\includegraphics[width=0.33\textwidth]{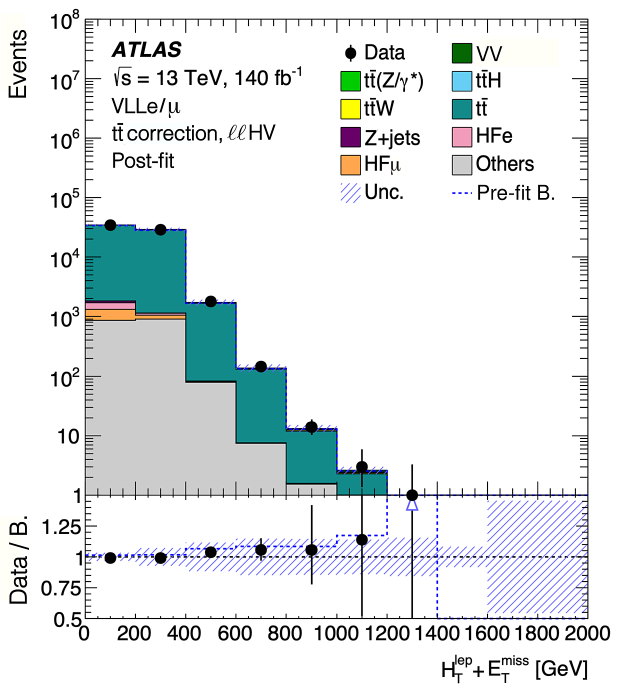}}
\subfloat[]{\includegraphics[width=0.33\textwidth]{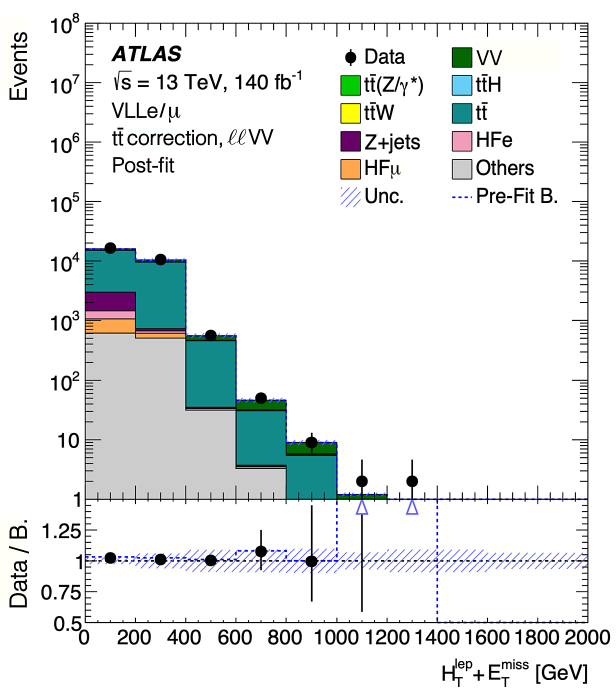}} \\
\subfloat[]{\includegraphics[width=0.33\textwidth]{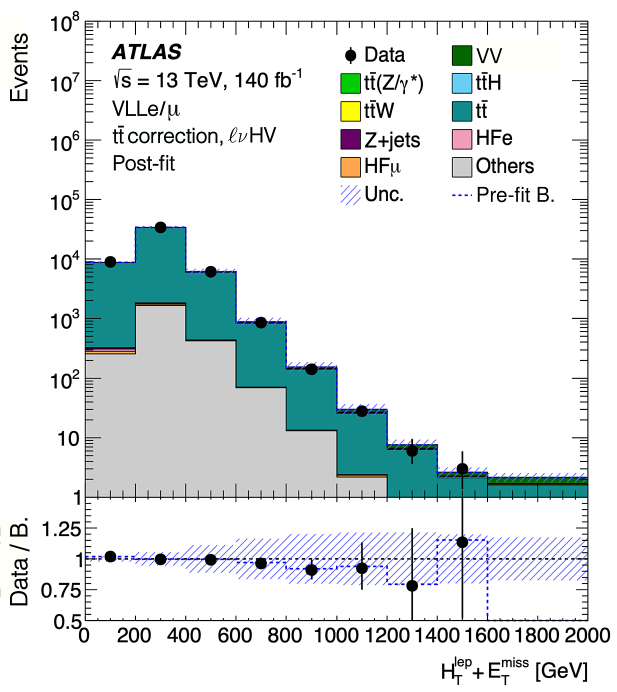}}
\subfloat[]{\includegraphics[width=0.33\textwidth]{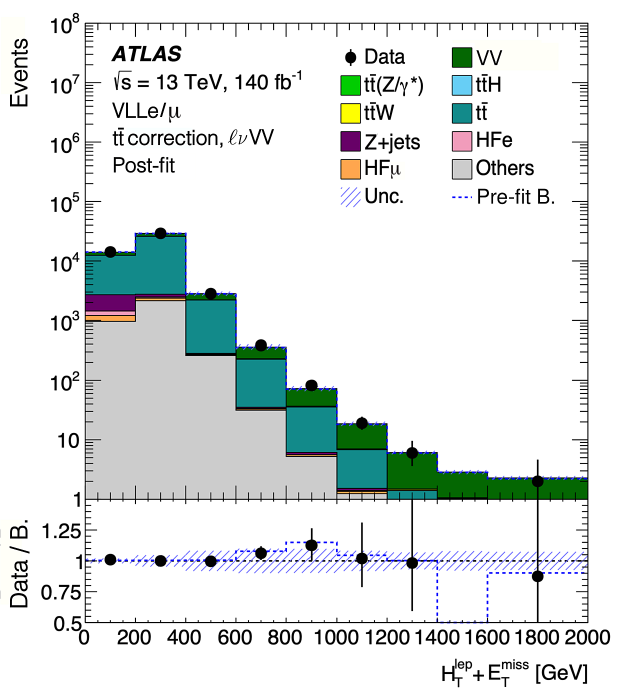}}
\caption{Distributions of the fitting variable, \htlepmet, before the fit to data in the different DNN classes within the \ttbar \llOS correction region ((a) $\ell\ell HH$, (b) $\ell\ell HV$, (c) $\ell\ell VV$, (d) $\ell\nu HV$, (e) $\ell\nu VV$) after all \ttbar corrections are applied.
The background contributions after the likelihood fit to data (`post-fit') for the background-only hypothesis are shown as filled histograms.
The ratio of the data to the background prediction (`B.') is shown in the lower panel, separately for post-fit background (black points) and pre-fit background (dashed blue line).
The `Others' contribution is dominated by the $\tW$ production. The blue hashed band shows both statistical and systematic uncertainties.
\label{fig:ttbar_VRs}}
\end{figure}

Four normalisation factors affecting the \ttbar contributions in the corresponding \llOS categories are free-floated in the simultaneous fit to data using the CRs and the \vlle (\vllmu) SRs and measured to be for the background-only hypothesis:
$\hat{\lambda}_{\ttbar_{\ell\ell \mathrm{HH}}} = 0.98 \pm 0.07$ ($0.92 \pm 0.06$),
$\hat{\lambda}_{\ttbar_{\ell\ell \mathrm{HV}}} = 0.97 \pm 0.08$ ($0.91 \pm 0.06$),
$\hat{\lambda}_{\ttbar_{\ell\nu \mathrm{VV}, \ell\ell \mathrm{VV}}} = 1.01 \pm 0.07$ ($1.06 \pm 0.07$) and
$\hat{\lambda}_{\ttbar_{\ell\nu \mathrm{HV}}} = 0.98 \pm 0.04$ ($1.00 \pm 0.04$),
where the uncertainty includes both statistical and systematic contributions.
The \ttbar events satisfying the \lll and \llll selections contain at least one non-prompt lepton. The normalisation of these events is treated separately and discussed in Section~\ref{sec:bkg_reducible_nonprompt}.

\subsubsection{\Zjets background}

Two types of data-driven corrections are derived for \Zjets in the corresponding correction region from Table~\ref{tab:selectionCorrectionRegions}: first, the jet multiplicity is corrected to data, followed by a correction to the angular separation between the leptons coming from the $Z$ boson ($\Delta R_{\ell\ell}$).

The number of jets distribution exhibits good MC agreement with data at low jet multiplicities but requires a correction of up to $25 \%$ at high jet multiplicities. A data-driven correction is then derived per bin in the jet multiplicity distribution.

Next, the $\Delta R_{\ell\ell}$ distribution is examined in different bins of jet multiplicity. The $\Delta R_{\ell\ell} < 1$ region shows poorly modelled MC background for all jet multiplicities, most likely originating from a suboptimal modelling of the boosted $Z/\gamma^*$+jets process, and therefore events in this region are vetoed from the \llOS SRs. For the remaining bins, a bin-by-bin correction is derived as a function of jet multiplicity, with corrections as large as $13 \% (25 \%)$ at low (high) $\Delta R_{\ell\ell}$.

Two normalisation factors are assigned to the \Zjets contributions in the \llOS categories with the largest contamination from this background process (\twoeevv, \twommvv, \twoenuvv, \twomnuvv) and measured to be for the background-only hypothesis using the CRs and the \vlle (\vllmu) SRs:
${\hat{\lambda}_{\Zjets_{\ell\ell \mathrm{VV}}} = 0.90 \pm 0.14}$ ($0.81 \pm 0.11$) and
${\hat{\lambda}_{\Zjets_{\ell\nu \mathrm{VV}}} = 0.77 \pm 0.19}$ ($0.70 \pm 0.14$),
where the uncertainty includes both statistical and systematic contributions.

\subsubsection{\ttW background}

Despite the use of state-of-the-art simulations, accurate modelling of additional QCD and QED radiation in $\ttW$ production remains challenging. Disagreement between the data and the pre-fit prediction from the simulation is observed in the \ttW CR. Therefore, an overall normalisation factor is assigned to the \ttW background, and is determined during the likelihood fit. The measured normalisation factor for the background-only hypothesis using the CRs and the \vlle (\vllmu) SRs is $\hat{\lambda}_{\ttW} = 1.33 \pm 0.22$ ($1.31 \pm 0.22$), where the uncertainty includes both statistical and systematic contributions. This measured normalisation factor is compatible with the one determined by the \ttW production cross-section measurement~\cite{TOPQ-2019-30}.

\subsubsection{\VV and \ttdy backgrounds}

A data-driven correction to the \VV jet multiplicity spectrum is derived from the \WZ correction region described in Table~\ref{tab:selectionCorrectionRegions}. This region is enriched in \WZ + LF jets; however the jet multiplicity mismodelling is assumed to be independent from the flavour of the additional jets in the event.

The \lllVVttZ and \llllZZ CRs are used in the likelihood fit to improve the prediction of the background contribution from the $\WZ$ + LF jets, $\WZ$ + HF jets, $\ZZ$, and $\ttdy$ processes. The number of $b$-jets provides good discrimination in the \lllVVttZ CR between the \WZ and \ttdy processes and is the variable used in this region in the fit. The event yields are fitted in the \llllZZ CR.
The measured normalisation factors for the background-only hypothesis using the CRs and the \vlle (\vllmu) SRs are: $\hat{\lambda}_\textrm{\WZ + LF} = 0.99 \pm 0.06$ ($0.97 \pm 0.06$), $\hat{\lambda}_\textrm{\WZ + HF} = 0.96 \pm 0.29$ ($0.83 \pm 0.26$), $\hat{\lambda}_\textrm{\ZZ} = 1.03 \pm 0.03$ ($1.03 \pm 0.03$), and $\hat{\lambda}_\textrm{\ttZ} = 1.25 \pm 0.17$ ($1.20 \pm 0.16$), where the uncertainty includes both statistical and systematic contributions.

Figure~\ref{fig:VVttZCR_Nbjpost} shows the $b$-jet multiplicity distribution in the \lllVVttZ CR after the likelihood fit to data.

\begin{figure}[t]
\centering
\includegraphics[width=0.40\textwidth]{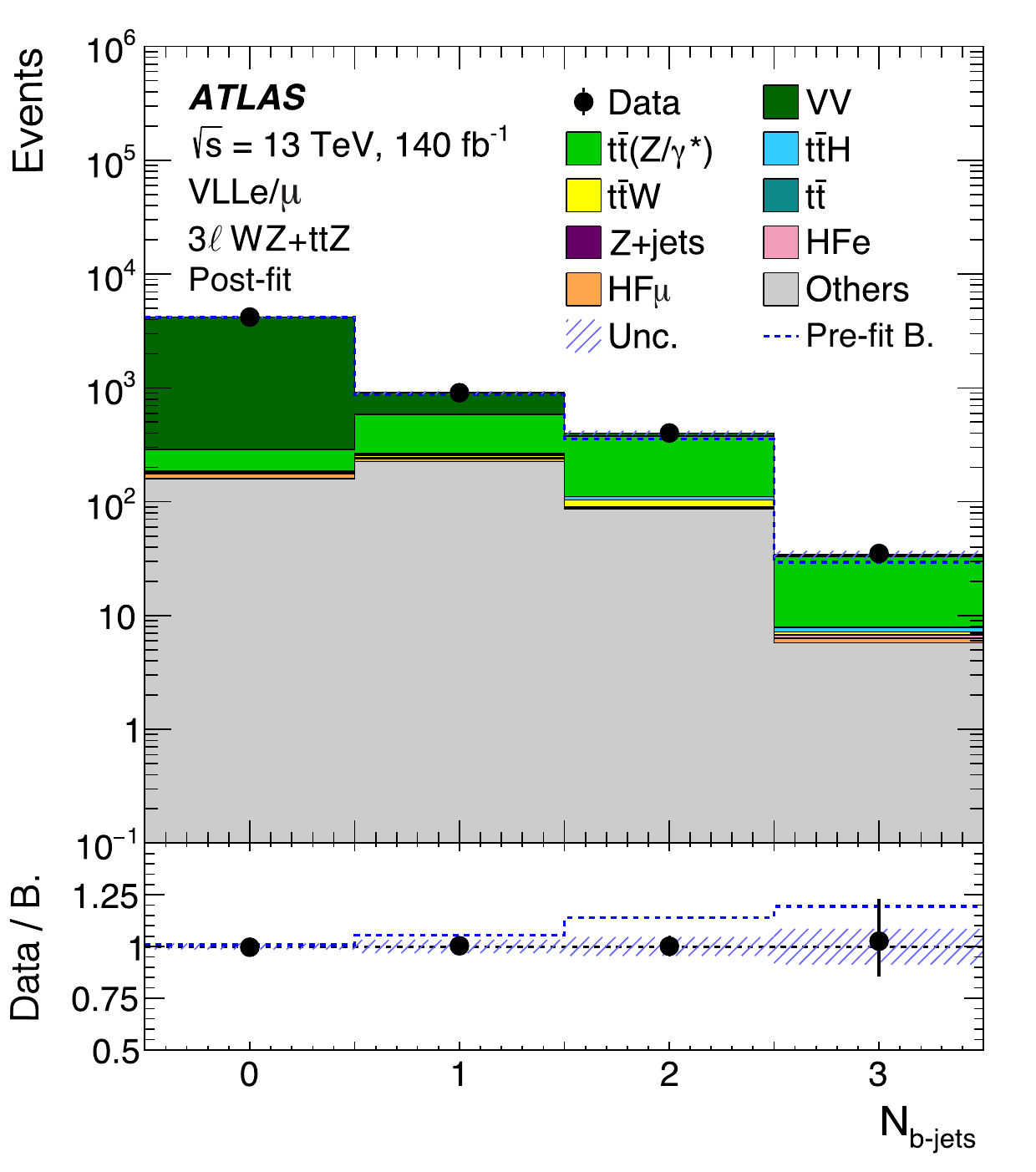}
\caption{
Comparison between data and the background prediction for the distribution of the $b$-jet multiplicity in the \lllVVttZ CR after the \VV jet multiplicity correction.
The background contributions after the likelihood fit to data (`post-fit') for the background-only hypothesis are shown as filled histograms.
The ratio of the data to the background prediction (`B.') is shown in the lower panel, separately for post-fit background (black points) and pre-fit background (dashed blue line).
The `Others' contribution is dominated by the $\tZ$ production.
The size of the combined statistical and systematic uncertainty in the background prediction is indicated by the blue hatched band.
The last bin contains the overflow.
} \label{fig:VVttZCR_Nbjpost}
\end{figure}

\subsubsection{Other irreducible backgrounds}
\label{sec:other_irreducible_backgrounds}

The rate of the background from internal conversions with $m(e^+e^-)<1~\gev$ is estimated by using the two dedicated CRs, \lllintc and \lllmatc, with a purity of 86\% and 14\%, respectively. The total yield in each category is used in the likelihood fit to determine the normalisation factor, which is measured for the background-only hypothesis using the CRs and the \vlle (\vllmu) SRs to be $\hat{\lambda}^\textrm{IntC}_e = 1.04 \pm 0.13$ ($1.04 \pm 0.14$), where the main contribution to the uncertainty comes from the statistics. The contribution from muons from internal conversions for $m(\mu^+\mu^-)<1~\gev$ is negligible in this analysis.

\subsection{Reducible backgrounds}
\label{sec:bkg_reducible}

\subsubsection{Non-prompt leptons}
\label{sec:bkg_reducible_nonprompt}

Non-prompt leptons originate from material conversions, LF and HF hadron decays, or the improper reconstruction of other particles, and their relative composition depends on the lepton quality requirements and event categories. These backgrounds are generally small in all SRs and CRs and thus are estimated from simulation, with the normalisation determined by the likelihood fit.
The main contribution to the non-prompt-lepton background is from $\ttbar$ production, with much smaller contributions from $V$+jets and single-top-quark processes. %
The non-prompt leptons in the simulated samples are labelled according to whether they originate from HF or LF hadron decays, or from a material conversion candidate. The HF category includes leptons from both bottom and charm decays.

Several of the event categories introduced in Section~\ref{sec:events} are designed to be enriched in specific processes and are used to derive normalisation factors to improve their modelling by the simulation.
The \lllmatc CR is enriched in material conversions with a purity of 85\% and only the total event yield is used.

Given the different lepton quality requirements used in the \llOS and \llll SRs compared with the \lll SR, as well as in the CRs, four \ll CRs enriched in contributions from HF non-prompt leptons in \ttbar events are defined requiring a leading $T$ lepton, with two of them using the $M_{\mathrm{ex}}$ lepton definition for the subleading lepton and the other two using the looser lepton definition $L*$. Normalisation factors for five non-prompt-lepton background contributions are estimated from the likelihood fit. The normalisation factor for HF non-prompt leptons is estimated separately for electrons and muons, and for joint $(M, M_{\mathrm{ex}}, T)$ lepton definitions (denoted by `tight') and for $L*$ (denoted by `loose'), i.e. $\lambda^\textrm{had}_{e, \textrm{tight}}$, $\lambda^\textrm{had}_{e, \textrm{loose}}$, $\lambda^\textrm{had}_{\mu, \textrm{tight}}$, and $\lambda^\textrm{had}_{\mu, \textrm{loose}}$. An additional normalisation factor is determined for the material conversions background, $\lambda^\textrm{MatC}_e$.
The measured normalisation factors for the background-only hypothesis using the CRs and the \vlle (\vllmu) SRs are:
$\hat{\lambda}^\textrm{had}_{e, \textrm{tight}} = 0.88 \pm 0.27$ ($0.88 \pm 0.28$),
$\hat{\lambda}^\textrm{had}_{e, \textrm{loose}} = 0.90 \pm 0.03$ ($0.90 \pm 0.04$),
$\hat{\lambda}^\textrm{had}_{\mu, \textrm{tight}} = 0.99 \pm 0.11$ ($1.00 \pm 0.11$),
$\hat{\lambda}^\textrm{had}_{\mu, \textrm{loose}} = 1.00 \pm 0.03$ ($0.99 \pm 0.03$),
$\hat{\lambda}^\textrm{MatC}_e = 1.16 \pm 0.08$ ($1.16 \pm 0.08$),
where the uncertainties include systematic effects but are dominated by the statistical uncertainty.

Backgrounds with leptons with the charge incorrectly assigned affect primarily the same-sign $\ll$ \ttW and HF non-prompt lepton control regions and predominantly arise from $t\bar{t}$ production, where one electron undergoes a hard bremsstrahlung and an asymmetric conversion ($e^\pm \to e^\pm \gamma^* \to e^\pm e^+ e^-$) or a mismeasured track curvature. This background process has negligible contributions in this analysis and is estimated from MC simulation. The muon charge misassignment rate is also negligible in the $\pt$ range relevant to this analysis.


%

%
\section{Systematic uncertainties}
\label{sec:systematics}

Several sources of systematic uncertainty may affect the signal and background yields in each signal and control region and are described in the following subsections.
Given the low background yields and good signal-to-background separation provided by the final discriminating variable used in the signal-enriched event categories, the search sensitivity is limited by the number of data events rather than by the systematic uncertainties in the background estimate.

\subsection{Experimental uncertainties}

Experimental systematic uncertainties related to the trigger efficiency, lepton reconstruction, identification and isolation, jet calibration, $b$-tagging calibration, and \met measurement are considered in the search.

The uncertainty in the measurement of the combined 2015--2018 integrated luminosity is 0.83\% \cite{DAPR-2021-01}, obtained using the LUCID-2 detector \cite{LUCID2} for the primary measurements, complemented by the ones using the inner detector and calorimeters.

Uncertainties associated with the lepton selection arise from the trigger, reconstruction, identification and isolation efficiencies, and the lepton momentum scale and resolution~\cite{PERF-2015-10,EGAM-2018-01,MUON-2018-03,MUON-2022-01}.
Uncertainties in the calibration of the non-prompt lepton BDT are estimated through a $Z\to\ell\ell$ tag-and-probe method and cover uncertainties related to the $Z(\to\ell\ell)$+jets MC modelling, the template cut/shape, the $m_{\ell\ell}$ window, the tag-and-probe lepton selections, the multijet background, the non-prompt lepton background, the luminosity, the \xs{}s of the considered
processes, and the limited number of events in simulation and data.

Uncertainties associated with the jet selection arise from the JES, the JVT requirement and the JER~\cite{JETM-2018-05,PERF-2014-03}. The JES and its uncertainties are derived by combining information from test-beam data, collision data and simulation~\cite{JETM-2018-05}. The JES (JER) have 30 (13) components included in the fit. The uncertainties in the JES, JER and JVT increase at lower jet \pt.

The efficiency of the flavour-tagging algorithm is measured for each jet flavour using control samples in data and in simulation. From these measurements, correction factors are derived to correct the tagging rates in the simulation~\cite{FTAG-2018-01,FTAG-2020-08,FTAG-2019-02}. Experimental uncertainties in these correction factors are taken as uncorrelated between $b$-jets, $c$-jets, and
light-flavour jets. An additional uncertainty is assigned to account for the extrapolation of the $b$-tagging efficiency measurement from the \pt region used to determine the correction factors to regions with higher transverse momentum.

The treatment of the uncertainties associated with reconstructed objects is common to all analysis channels and applies to all signal and background samples and thus these are considered as fully correlated among different analysis regions and samples.

\subsection{Theoretical uncertainties}

The modelling uncertainties in the main irreducible backgrounds are assessed through comparisons with alternative MC samples, as listed in Table~\ref{tab:mcconfig}. Additional uncertainties are evaluated from renormalisation and factorisation scale variations by a factor of 0.5 and 2, relative to the nominal scales, for the \ttbar, \Zjets, \ttW, \ttZ, and diboson samples. An additional 20\% uncertainty is assigned to the \ttW electroweak contribution~\cite{PhysRevD.103.094014}.

For the \ttbar process, four additional uncertainties are considered related to the reweighting method itself, derived by comparing the nominal reweighted SM \ttbar\ sample to a sample obtained through the alternative reweighting obtained by varying the renormalisation and the factorisation scales separately by a factor of 0.5 and 2, applied on the (anti-)top quark \pT or on the $m(t\bar{t})$ distribution independently.
Related to the reweighting of the \ttbar + LF, \ttbar + $b$ and \ttbar + $c$ contributions, uncertainties of 3\% and 4\% are assigned to events originating from \ttbar + $\geq 1 b$ and \ttbar + $\geq 1 c$, respectively.
The statistical uncertainty related to reweighting the jet multiplicity distribution is expected to be very small due to the large statistics of \ttbar in the VR and is not considered.
All alternative \ttbar MC samples are reweighted to the same higher-order predictions as the nominal \POWHEG[v2]+\PYTHIA[8.230] MC sample.
In addition to the comparison to the alternative MC sample shown in Table~\ref{tab:mcconfig}, the nominal predictions are also compared with those obtained from an alternative sample generated as the nominal sample but setting the $p_{\mathrm{T}}^{\mathrm{hard}}$ parameter in \PYTHIA to 1 instead of 0~\cite{ATL-PHYS-PUB-2023-029}. This parameter regulates the definition of the vetoed region of the showering to avoid holes or overlaps in the phase space filled by \POWHEG and \PYTHIA.
An uncertainty related to the choice of the \hdamp parameter is estimated by comparing the predictions of the nominal sample to those obtained with an alternative sample with the \hdamp parameter increased by a factor of 1.5 compared with its nominal value. %
Variations in the initial state radiation (ISR) are estimated by varying the factorisation and renormalisation scales independently up and down by a factor of two.
Similarly, the uncertainty related to final-state radiation (FSR) is assessed by varying the renormalisation scale for final-state parton-shower emissions up and down by a factor of two.
Finally, the uncertainty associated with the A14 tune is derived by varying the A14 tune (Var3c), which affects the renormalisation scale variations in the ISR PS. No theory reweighting is applied to this systematic.

For the \Zjets process, the uncertainty related to the upper cut-off of perturbative calculations for PS evolution is known as the re-summation scale (QSF). This uncertainty is evaluated at truth level by varying the nominal value of 2 GeV by a factor of 4 up and $1/4$ down. Fiducial cuts are applied at truth level to define a phase space close to that used at reconstruction level in the signal regions where \Zjets is a dominant background. Additionally, the uncertainty related to the choice of the CKKW merging scale, i.e. the scale for calculating the overlap between jets from the ME and the PS, is derived similarly; the nominal value of 20 GeV is varied down to 15 GeV and up to 30 GeV and differences relative to the nominal distribution are evaluated at truth level.

The statistical uncertainty in the fitted parameters for the $VV$ jet-multiplicity correction is propagated as an uncertainty in the diboson background. Finally, additional normalisation uncertainties are included for all processes whose normalisation is not obtained from the fit. In particular, for the \tttt, \ttH, and $tZ$ processes, cross-section uncertainties of 20\%~\cite{Frederix:2017wme}, 11\%~\cite{deFlorian:2016spz}, and 5\%~\cite{TOPQ-2018-01} are assigned, respectively, while for \ttt, $tWZ$, \ttWW, and triboson backgrounds a 50\% cross-section uncertainty is assigned as a conservative estimate, since they are small backgrounds and have low impact on the search.

Uncertainties in the modelling of the signal samples are evaluated through independent variations of the factorisation and renormalisation scales by a factor of two. Additional uncertainties due to PDF effects are estimated through an ensemble of eigenvariations of the NNPDF set, and by taking the differences relative to alternative PDF sets~\cite{Butterworth:2015oua}.

\subsection{Reducible background uncertainties}

The normalisation of HF non-prompt leptons for processes where the non-prompt lepton is $M$ or $T$ is obtained from regions including one sub-leading $M_{\mathrm{ex}}$ lepton. An uncertainty of 20\% in the extrapolation from $M_{\mathrm{ex}}$ to $M$ and $T$ leptons is applied from the comparison of the relative efficiency between nominal and alternative \ttbar\ MC samples.
Validation regions with looser lepton requirements and further enriched in non-prompt leptons are defined. A good agreement between data and background prediction is observed in all kinematic variables except for the number of $b$-jets. Based on this disagreement, an \nbjets-dependent uncertainty is added to the HF non-prompt background, ranging from 6\%--40\% for 1--3 additional $b$-jets in the non-prompt muon regions, and 10\%--80\% in the non-prompt electron regions.

The modelling of internal and material conversions is tested in dedicated validation regions with two tight same-sign leptons, requiring one of them to be a conversion candidate. Additional uncertainties of 10\% and 50\% are assigned to the material and internal conversion backgrounds, respectively, evaluated from the data to background agreement in the validation regions.

A systematic uncertainty of 20\% is assigned to the background from electrons with a misidentified charge and accounts for the disagreement observed between the MC simulation for this background and data in the same-sign charge $ee$ inclusive region.


%

%
\section{Results}
\label{sec:results}

A maximum-likelihood fit is performed on all bins in the 19 signal and control regions considered in this search to simultaneously determine the signal and background yields that are most consistent with the data. The \htlepmet is used as the discriminating variable in the signal regions, while the \nbjets and the total event yield are fitted in control regions. Two separate fits are performed for the \vlle and \vllmu signal hypotheses, and for each of the two fits the 10 SRs for electrons or the 10 SRs for muons introduced in Table~\ref{tab:selectionSR} are simultaneously fitted with the 9 CRs to data.

The likelihood function ${\cal L}(\mu,\vec{\lambda}, \vec{\theta})$ is constructed as a product of Poisson probability terms over all bins considered in the search, and depends on: the signal-strength parameter, $\mu$, a multiplicative factor applied to the predicted yield for the VLL signal; $\vec{\lambda}$, the normalisation factors for several backgrounds; $\vec{\theta}$, a set of nuisance parameters (NPs), encoding systematic uncertainties in the signal and background expectations~\cite{Cranmer:2012sba}. Systematic uncertainties can impact the estimated signal and background rates, the migration of events between categories, and the shape of the fitted distributions.
Both $\mu$ and $\vec{\lambda}$ are treated as free parameters in the likelihood fit. The NPs $\vec{\theta}$ allow variations of the expectations for signal and background according to the systematic uncertainties, subject to Gaussian or Poisson constraints in the likelihood fit.
Their fitted values represent the deviations from the nominal expectations that globally provide the best fit to the data.
Statistical uncertainties in each bin due to the limited size of the simulated samples are taken into account by dedicated parameters using the Beeston--Barlow `lite' technique~\cite{Conway:2011}.

The test statistic $q_\mu$ is defined as the profile likelihood ratio:
$q_\mu = -2\ln({\cal L}(\mu,{\hat{\vec{\lambda}}}_\mu, {\hat{\vec{\theta}}}_\mu)/{\cal L}(\hat{\mu},\hat{\vec{\lambda}}_{\hat{\mu}}, \hat{\vec{\theta}}_{\hat{\mu}}))$,
where $\hat{\mu}$, $\hat{\vec{\lambda}}_{\hat{\mu}}$, and $\hat{\vec{\theta}}_{\hat{\mu}}$ are the values of the parameters that maximise the likelihood function,
and ${\hat{\vec{\lambda}}}_\mu$ and ${\hat{\vec{\theta}}}_\mu$ are the values of the parameters that maximise the likelihood function for a given value of $\mu$.
The test statistic $q_\mu$ is evaluated with the {\textsc RooFit} package~\cite{Verkerke:2003ir}.
A related statistic is used to determine the probability that the observed data are incompatible with the background-only hypothesis (i.e.~the discovery test) by setting $\mu=0$ in the profile likelihood ratio ($q_0$).
The $p$-value (referred to as $p_0$) representing the probability of the data being compatible with the background-only hypothesis is estimated by integrating the distribution of $q_0$ from background-only pseudo-experiments, approximated using the asymptotic formulae given in Ref.~\cite{Cowan:2010js}, above the observed value of $q_0$. Some model dependence exists in the estimation of the $p_0$, as a given signal scenario must be assumed in the calculation of the denominator of $q_0$, even if the overall signal normalisation is allowed to float and is fitted to data. The observed $p_0$ is checked for each explored signal scenario.
Upper limits on the signal production cross-section for each of the signal scenarios considered are derived by using $q_\mu$ in the CL$_{\textrm{s}}$ method~\cite{Junk:1999kv,Read:2002hq}.
For a given signal scenario, values of the production cross-section (parameterised by $\mu$) yielding CL$_{\textrm{s}} < 0.05$, where CL$_{\textrm{s}}$ is computed using the asymptotic approximation~\cite{Cowan:2010js}, are excluded at $\geq 95\%$ CL.

A comparison of the predicted numbers of background events, obtained from the combined likelihood fit in the background-only hypothesis, and the observed data is shown in Figures~\ref{fig:summarySR_e} and~\ref{fig:summarySR_mu} for the 10 \vlle and 10 \vllmu SRs, respectively, and in Figure~\ref{fig:summaryCR} for the 9 CRs. The corresponding post-fit yields for the \vlle and \vllmu SRs are reported in Tables~\ref{tab:SRspostfityields_VLLe} and~\ref{tab:SRspostfityields_VLLmu}.

Comparisons between data and the background prediction for the \htlepmet distributions used in the different SRs are shown in Figures~\ref{fig:htlepmet_vlle} and~\ref{fig:htlepmet_vllmu} for the \vlle and \vllmu searches, respectively.
The binning used for the \htlepmet distributions in the different SRs represents a compromise between preserving enough discrimination in the fit between the background and the signal and keeping the MC statistical uncertainty in the background prediction per bin below 30\%.
As shown in Section~\ref{sec:background}, the fitted normalisation factors for each background are compatible between the \vlle and \vllmu searches.

No significant deviations from the SM expectations are observed in any of the SRs considered.
The smallest p-value for each of the signal benchmarks considered is 0.34 (0.40) for \vlles 800~\gev (\vlled 1.3~\tev) and 0.14 (0.18) for \vllmus 150~\gev (\vllmud 150~\gev), which corresponds to a local significance of 0.42$\sigma$ (0.25$\sigma$) and 1.07$\sigma$ (0.91$\sigma$), respectively.
Limits at 95\% CL on the cross-section of a \vlle or a \vllmu signal as a function of the mass of the VLL are set. Figures~\ref{fig:Limits_VLLesinglet},~\ref{fig:Limits_VLLedoublet},~\ref{fig:Limits_VLLmusinglet} and~\ref{fig:Limits_VLLmudoublet} show the limits on the cross-section for the models: \vlles, \vlled, \vllmus, and \vllmud.
The \vlles is observed (expected) to be excluded at $95\%$ CL for masses up to 320 (300) GeV while the \vlled is observed (expected) to be excluded at $95\%$ CL for masses up to 1220 (1230) GeV. The \vllmus is observed (expected) to be excluded at $95\%$ CL for masses up to 400 (330) GeV while the \vllmud is observed (expected) to be excluded at $95\%$ CL for masses up to 1270 (1250) GeV.
The search is dominated by statistical uncertainties: the expected limits obtained when including only statistical uncertainties represent approximately 90\% of the total limits throughout the probed mass range. Within the systematic uncertainties, the leading ones for the search of a \vlled or \vllmud with a 600 GeV mass are the \Zjets QSF and signal PDF and scale variation uncertainties.
The \llOS channel is the most sensitive signal region at high VLL masses, whereas the \lll (\llll) channel sets stronger expected exclusion limits at lower VLL masses in the VLL doublet (singlet) scenario.

\begin{figure*}[!h]
\centering
\subfloat[]{\includegraphics[width=0.7\textwidth]{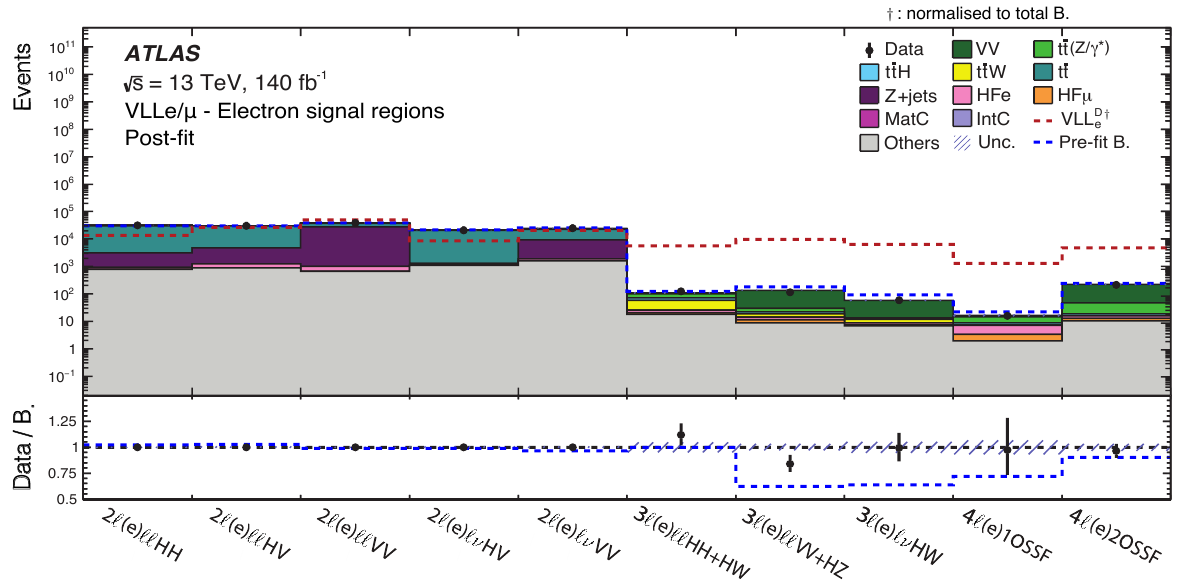}\label{fig:summarySR_e}} \\
\subfloat[]{\includegraphics[width=0.7\textwidth]{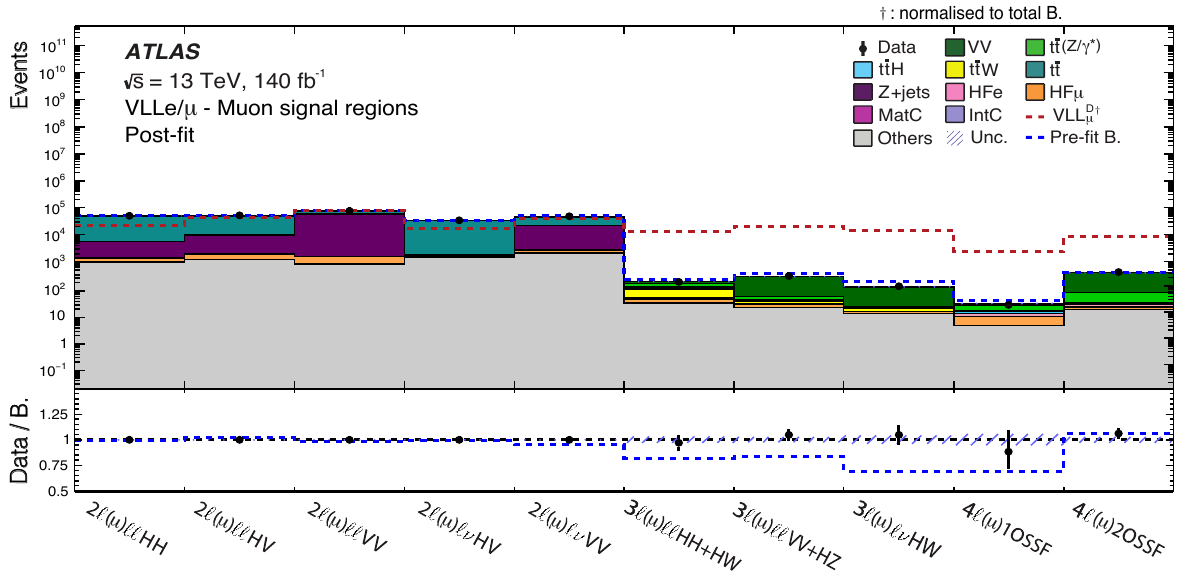}\label{fig:summarySR_mu}} \\
\subfloat[]{\includegraphics[width=0.7\textwidth]{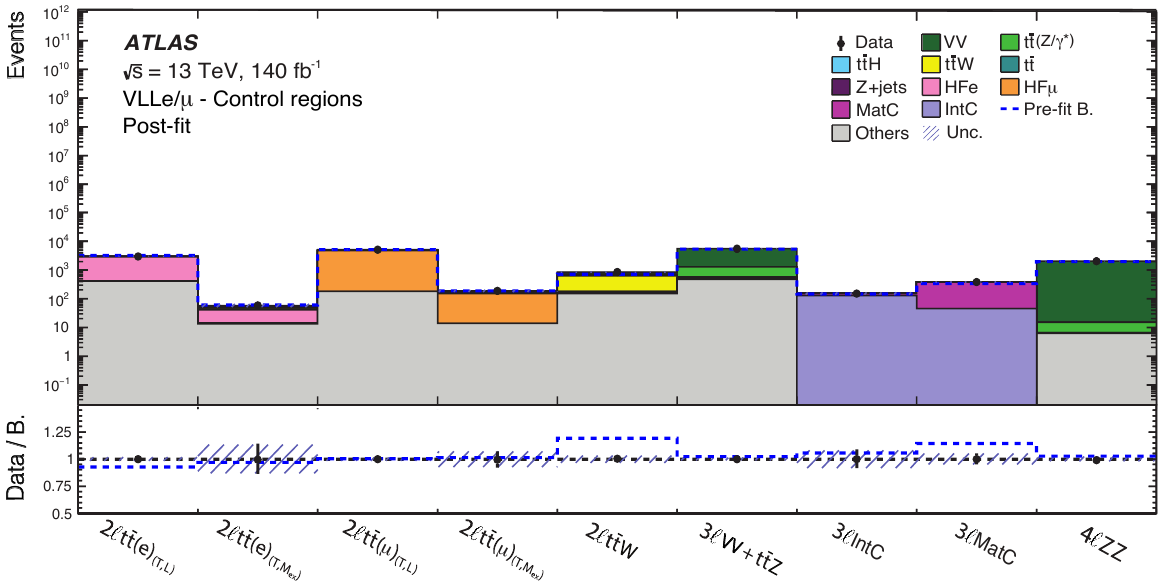}\label{fig:summaryCR}}
\caption{Comparison between data and the background prediction for the event yields in the (a) 10 \vlle SRs, (b) 10 \vllmu SRs, and (c) 9 CRs, after a background-only fit to data (`post-fit') in the (a) \vlle SRs and CRs and (b, c) \vllmu SRs and CRs. The background contributions in the CRs after the likelihood fit to data in the \vlle SRs and CRs are comparable to those in (c).
Distributions for the \vlled and \vllmud signal points for a VLL mass of 600 GeV are overlaid for comparison in (a) and (b), respectively.
The lower panels show the ratio of data to the background estimate (`B.'), separately for post-fit background (black points) and pre-fit background (dashed blue line).
The size of the combined statistical and systematic uncertainty in the background prediction is indicated by the blue hatched band.
}
\end{figure*}

\clearpage

\begin{table}[tb]
\begin{center}
\caption{Summary of observed and predicted yields in the ten \vlle signal region categories. The background prediction is shown after the combined likelihood fit to data under the background-only hypothesis across all control region and signal region categories.
The expected signal yields for \vlles and \vlled for a VLL mass of 200 GeV and 600 GeV, respectively, that are obtained by using their theoretical cross-sections are also shown with their pre-fit uncertainties, assuming $\mu$=1.
The uncertainties correspond to the combined statistical and systematic uncertainties in the predicted yields.
The `Others' contribution is dominated by $\tW$, $VVV$, and $\tWZ$ in the \ll, \lll, and \llll SRs, respectively.
Dashes refer to components that are negligible or not applicable. \label{tab:SRspostfityields_VLLe}}
\footnotesize
\begin{tabular}{l|D{,}{\,\pm\,}{6.6}|D{,}{\,\pm\,}{6.6}|D{,}{\,\pm\,}{6.6}|D{,}{\,\pm\,}{6.6}|D{,}{\,\pm\,}{6.6}}
\hline
Channel 	& \multicolumn{1}{c|}{\twoeehh} & \multicolumn{1}{c|}{\twoeehv} & \multicolumn{1}{c|}{\twoeevv} & \multicolumn{1}{c|}{\twoenuhv} & \multicolumn{1}{c}{\twoenuvv} \\
\hline
VV     	& 53.5,8.4 	& 98,18 		& 1124,72 	& 148,11 		& 2336,54 \\
\ttdy   	& 165,93 		& 182,92 		& 54,26 		& 59,42 		& 30,21 \\
\ttH   	& 106,15 		& 56.5,8.4 	& 12.4,2.1 	& 14.9,1.9 	& 5.49,0.85 \\
\ttW   	& 119,16 		& 140,17 		& 39.6,6.3 	& 57.5,5.9 	& 29.8,4.3 \\
\ttbar   	& 27630,680 	& 25530,720 	& 8970,510 	& 19460,340 	& 13360,380 \\
Z+jets   	& 2160,640 	& 3410,720 	& 27070,560 	& 101,34 		& 7260,360 \\
HFe   	& 156.6,6.0 	& 377,14 		& 356,14 		& 77.7,3.0 	& 269,10 \\
HF$\mu$   & \multicolumn{1}{c|}{--} & \multicolumn{1}{c|}{--} & \multicolumn{1}{c|}{--} & \multicolumn{1}{c|}{--} & \multicolumn{1}{c}{--} \\
Others   	& 770,170 	& 890,170 	& 654,69 		& 1110,320 	& 1620,180 \\
\hline
Total  	& 31160,180 	& 30670,190 	& 38280,210 	& 21020,150 	& 24900,160 \\
\hline
Data   	& 31162 			& 30677 			& 38279 			& 21022 			& 24890 \\
\hline\hline
\vlles 200 GeV & 63.8,2.9 & 88.8,5.0  & 137.2,7.7 & 34.9,1.7 & 119.5,2.8 \\
\vlled 600 GeV & 67.2,3.2 & 132.1,6.9  & 244,12 & 42.4,1.9 & 100.7,4.3 \\
\hline
\end{tabular}
\\[10pt]
\begin{tabular}{l|D{,}{\,\pm\,}{6.6}|D{,}{\,\pm\,}{6.6}|D{,}{\,\pm\,}{6.6}|D{,}{\,\pm\,}{6.6}|D{,}{\,\pm\,}{6.6}}
\hline
Channel 	& \multicolumn{1}{c|}{\threeehhhw} & \multicolumn{1}{c|}{\threeevvhz} & \multicolumn{1}{c|}{\threeenuhw} & \multicolumn{1}{c|}{\eeem} & \multicolumn{1}{c}{\eeee} \\
\hline
VV     	& 10.7,1.1 	& 106.8,3.3 	& 45.2,2.5 	& 2.10,0.19 	& 174.8,6.9 \\
\ttdy   	& 26.8,3.2 	& 7.4,1.5 	& 1.28,0.19 	& 5.47,0.60 	& 27.7,2.8 \\
\ttH   	& 13.8,1.9 	& 2.70,0.49 	& 0.58,0.10 & 1.75,0.24 	& 2.95,0.42 \\
\ttW   	& 32.9,3.1 	& 4.99,0.71 	& 3.10,0.37 	& \multicolumn{1}{c|}{--} & \multicolumn{1}{c}{--} \\
\ttbar  	& \multicolumn{1}{c|}{--} & \multicolumn{1}{c|}{--} & \multicolumn{1}{c|}{--} & \multicolumn{1}{c|}{--} & \multicolumn{1}{c}{--} \\
Z+jets   	& \multicolumn{1}{c|}{--} & \multicolumn{1}{c|}{--} & \multicolumn{1}{c|}{--} & \multicolumn{1}{c|}{--} & \multicolumn{1}{c}{--} \\
HFe   	& 5.4,1.7 	& 2.87,0.90 	& 1.18,0.37 	& 3.89,0.29 	& 3.11,0.13 \\
HF$\mu$   & 2.33,0.27 	& 2.25,0.26 	& 0.73,0.10 	& 1.33,0.10  	& 2.64,0.10 \\
Others   	& 17.6,4.9 	& 8.9,1.9 	& 6.7,1.4 	& 1.94,0.35 	& 10.3,2.3 \\
\hline
Total  	& 109.2,5.3 	& 135.4,4.2 	& 58.5,3.1 	& 16.4,1.1 	& 221.2,7.3 \\
\hline
Data   	& 122 				& 114 				& 58 				& 16 				& 213 \\
\hline\hline
\vlles 200 GeV & 3.35,0.27 	& 10.51,0.57 	& 3.66,0.21 	& 1.94,0.16 	& 35.0,1.0 \\
\vlled 600 GeV & 27.8,1.4 	& 47.7,2.4 	& 31.7,1.6 	& 6.35,0.34 	& 23.7,1.2 \\
\hline
\end{tabular}
\end{center}
\end{table}

\begin{table}[h!]
\begin{center}
\caption{Summary of observed and predicted yields in the ten \vllmu signal region categories. The background prediction is shown after the combined likelihood fit to data under the background-only hypothesis across all control region and signal region categories.
The expected signal yields for \vllmus and \vllmud for a VLL mass of 200 GeV and 600 GeV, respectively, that are obtained by using their theoretical cross-sections are also shown with their pre-fit uncertainties, assuming $\mu$=1.
The uncertainties correspond to the combined statistical and systematic uncertainties in the predicted yields.
The `Others' contribution is dominated by $\tW$, $VVV$, and $\tWZ$ in the \ll, \lll, and \llll SRs, respectively.
Dashes refer to components that are negligible or not applicable. \label{tab:SRspostfityields_VLLmu}}
\footnotesize
\begin{tabular}{l|D{,}{\,\pm\,}{6.6}|D{,}{\,\pm\,}{6.6}|D{,}{\,\pm\,}{6.6}|D{,}{\,\pm\,}{6.6}|D{,}{\,\pm\,}{6.6}}
\hline
Channel 	& \multicolumn{1}{c|}{\twommhh} & \multicolumn{1}{c|}{\twommhv} & \multicolumn{1}{c|}{\twommvv} & \multicolumn{1}{c|}{\twomnuhv} & \multicolumn{1}{c}{\twomnuvv} \\
\hline
VV   		& 69,10 		& 124,22 		& 1580,110 	& 230,17 		& 3940,120 \\
\ttdy   	& 250,160 	& 250,140 	& 75,43 		& 77,60 		& 39,29 \\
\ttH   	& 159,21 		& 86,12 		& 18.5,3.1 	& 19.9,2.5 	& 7.4,1.1 \\
\ttW   	& 178,25 		& 215,28 		& 58,11 		& 77.0,8.1 	& 40.2,6.4 \\
\ttbar   	& 43840,920 	& 41000,1000 	& 15740,740 	& 32350,540 	& 22530,510 \\
Z+jets   	& 4340,910 	& 8000,1000 	& 57000,800 	& 282,88 		& 18910,480 \\
HFe   	& \multicolumn{1}{c|}{--} & \multicolumn{1}{c|}{--} & \multicolumn{1}{c|}{--} & \multicolumn{1}{c|}{--} & \multicolumn{1}{c}{--} \\
HF$\mu$   & 377,11 		& 765,23 		& 725,21 		& 181.4,5.4 	& 637,19 \\
Others   	& 970,230 	& 1180,250 	& 800,110 	& 1390,460 	& 2160,260 \\
\hline
Total  	& 50175,240 	& 50970,260 	& 75990,330 	& 34610,190 	& 48260,230 \\
\hline
Data   	& 50178 			& 50968 			& 75976 			& 34605 			& 48276 \\
\hline\hline
\vllmus 200 GeV & 95.6,4.7 	& 135.1,6.9  	& 199,12 	& 51.1,1.9 	& 178.3,4.7 \\
\vllmud 600 GeV & 83.9,3.8 	& 167.9,8.1  	& 305,14 	& 66.5,2.9 	& 150.8,6.2 \\
\hline
\end{tabular}
\\[10pt]
\begin{tabular}{l|D{,}{\,\pm\,}{6.6}|D{,}{\,\pm\,}{6.6}|D{,}{\,\pm\,}{6.6}|D{,}{\,\pm\,}{6.6}|D{,}{\,\pm\,}{6.6}}
\hline
Channel	& \multicolumn{1}{c|}{\threemhhhw} & \multicolumn{1}{c|}{\threemvvhz} & \multicolumn{1}{c|}{\threemnuhw} & \multicolumn{1}{c|}{\mmme} & \multicolumn{1}{c}{\mmmm} \\
\hline
VV   		& 23.0,1.9 	& 235.0,6.8 	& 97.9,4.2 	& 3.75,0.27 	& 326,11 		 \\
\ttdy   	& 45.2,4.6 	& 12.5,2.2 	& 1.88,0.27 	& 9.5,1.0   	& 40.7,4.0 	 \\
\ttH   	& 21.4,2.8 	& 4.18,0.89 	& 0.77,0.11 	& 2.90,0.44   & 3.73,0.49 	 \\
\ttW   	& 50.5,4.7 	& 7.4,1.3 	& 4.22,0.69 	& \multicolumn{1}{c|}{--} & \multicolumn{1}{c}{--} \\
\ttbar   	& \multicolumn{1}{c|}{--} & \multicolumn{1}{c|}{--} & \multicolumn{1}{c|}{--} & \multicolumn{1}{c|}{--} & \multicolumn{1}{c}{--} \\
Z+jets   	& \multicolumn{1}{c|}{--} & \multicolumn{1}{c|}{--} & \multicolumn{1}{c|}{--} & \multicolumn{1}{c|}{--} & \multicolumn{1}{c}{--} \\
HFe   	& 2.76,0.88 	& 0.94,0.30 	& 0.39,0.13 	& 2.87,0.20   & 4.39,0.17 	 \\
HF$\mu$   & 12.6,1.4 	& 6.61,0.75 	& 1.97,0.23 	& 5.24,0.27   & 3.98,0.12 	 \\
Others   	& 29.2,3.3 	& 19.9,3.2 	& 12.3,2.3 	& 4.35,0.51   & 17.4,3.8 	 \\
\hline
Total  	& 183.0,6.0 	& 283.9,7.2 	& 117.5,5.0 	& 28,1.6      & 395,11 		 \\
\hline
Data   	& 177 				& 297 				& 123 				& 25				& 420 				 \\
\hline\hline
\vllmus 200 GeV & 6.34,0.44 	& 18.72,0.78 	& 5.85,0.28  	& 2.96,0.17 	& 55.8,1.2 \\
\vllmud 600 GeV & 51.2,2.3 	& 81.4,3.4 	& 58.0,2.6  	& 8.88,0.40 	& 32.9,1.4 \\
\hline
\end{tabular}
\end{center}
\end{table}

\begin{figure*}[h!]
\centering
\subfloat[]{\includegraphics[width=0.24\textwidth]{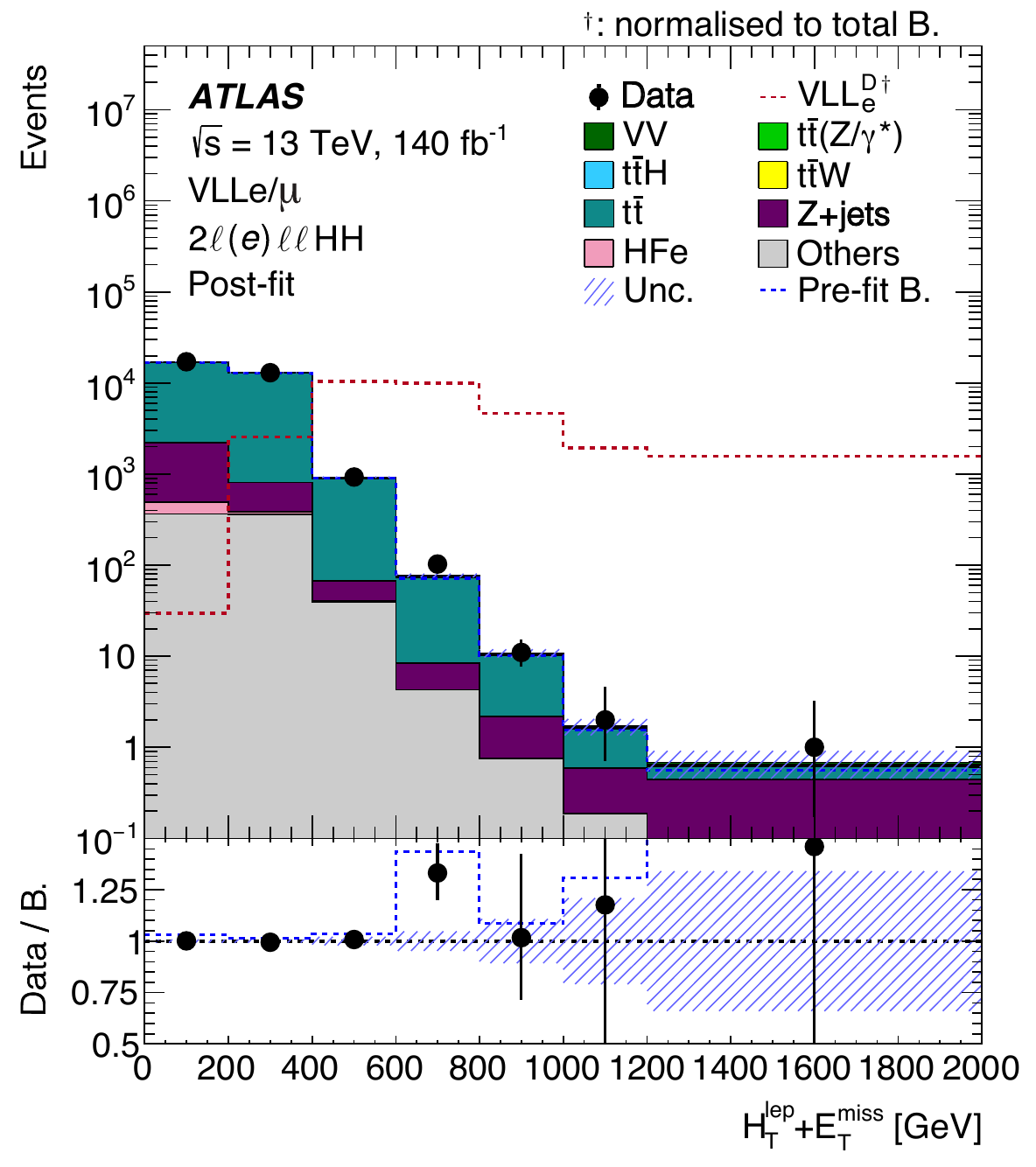}}
\subfloat[]{\includegraphics[width=0.24\textwidth]{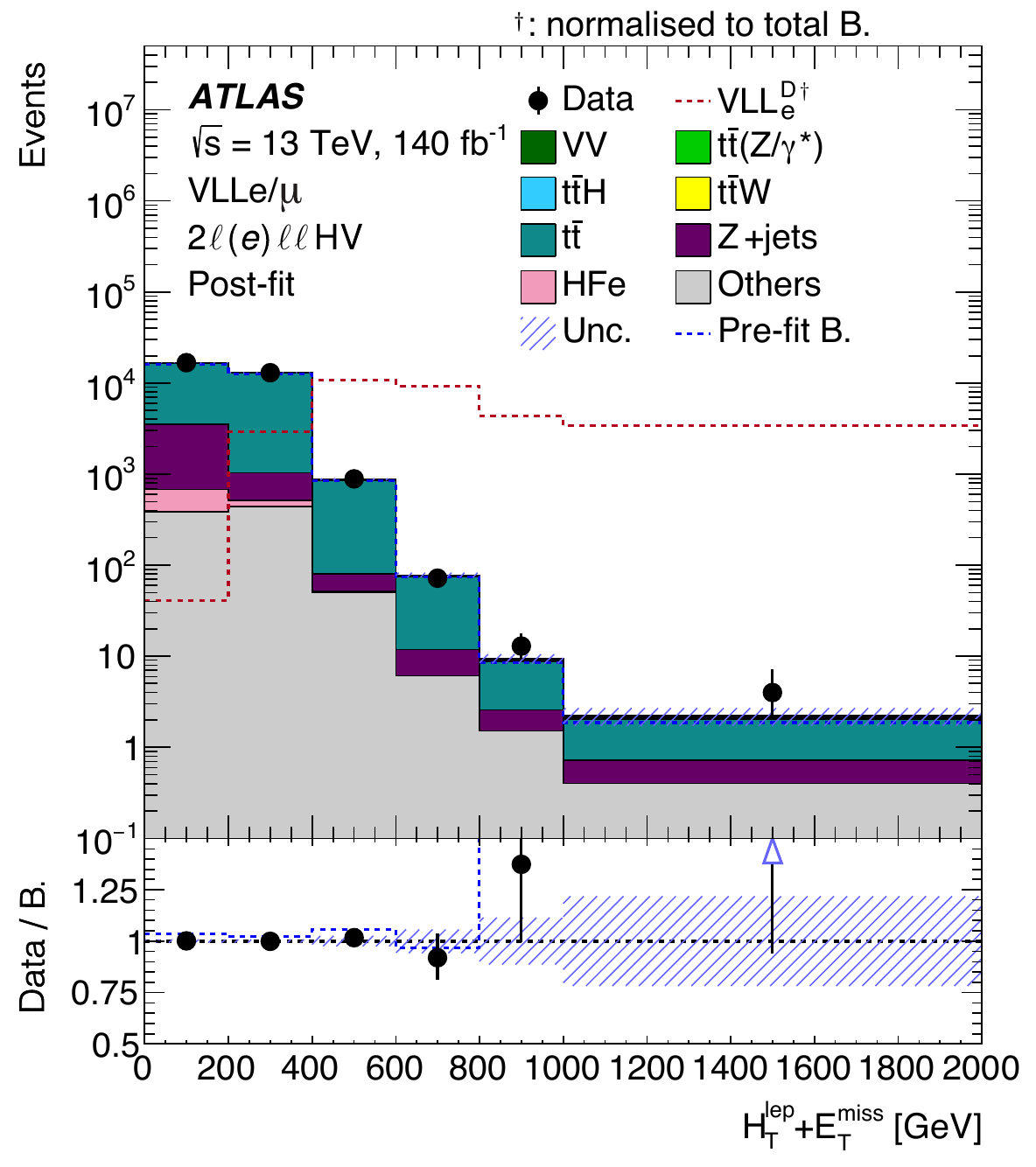}}
\subfloat[]{\includegraphics[width=0.24\textwidth]{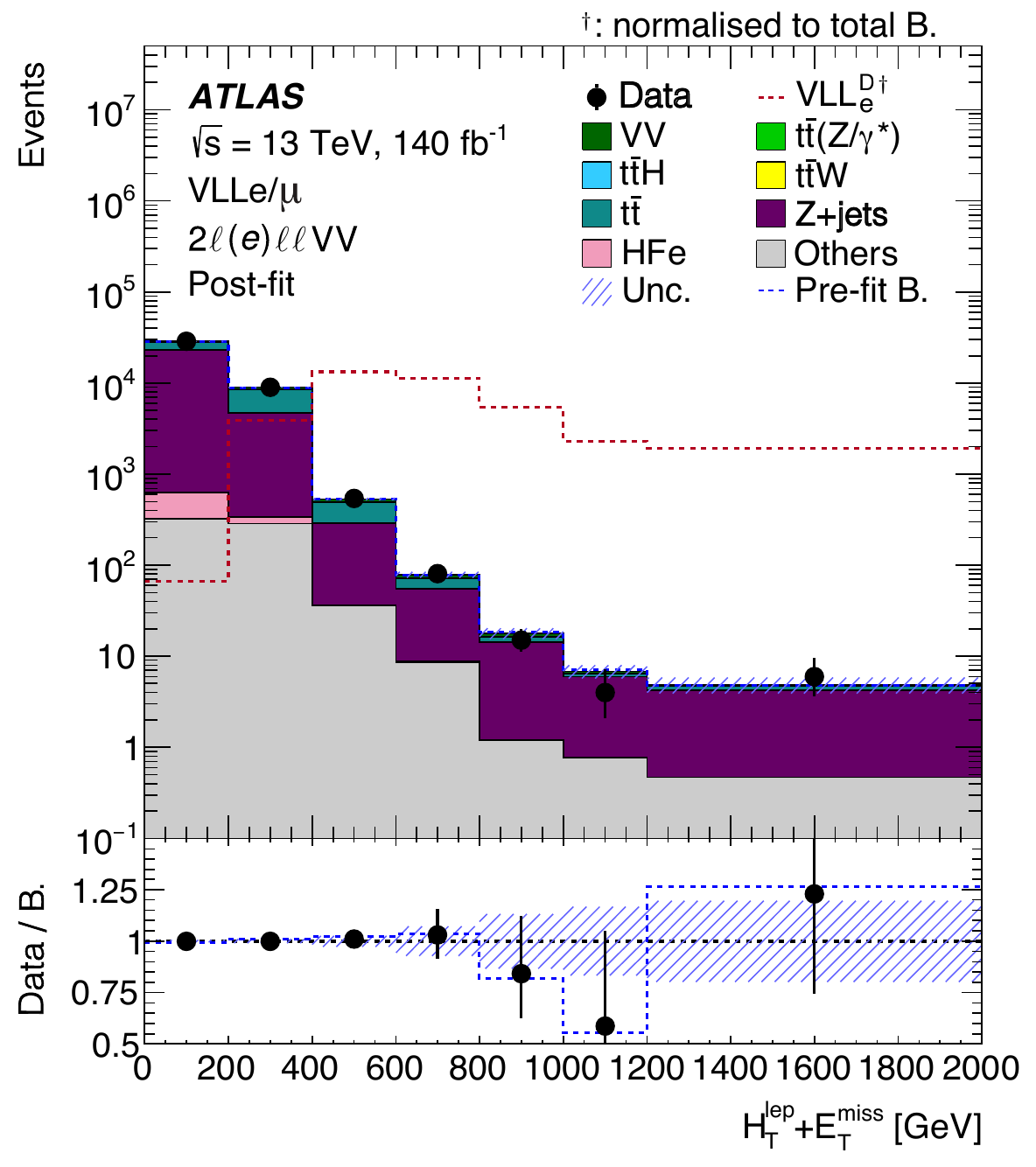}}
\subfloat[]{\includegraphics[width=0.24\textwidth]{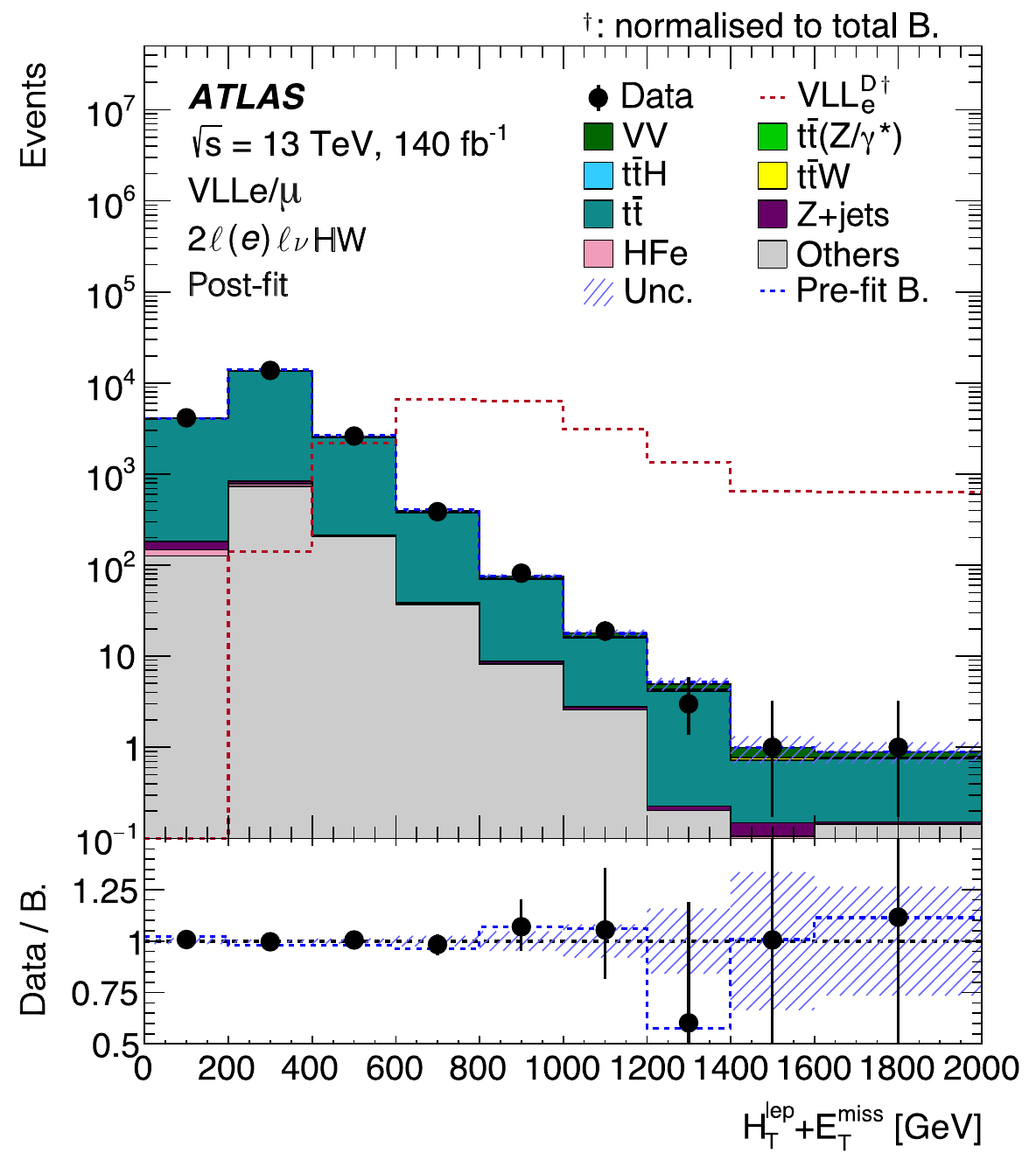}}\\
\subfloat[]{\includegraphics[width=0.24\textwidth]{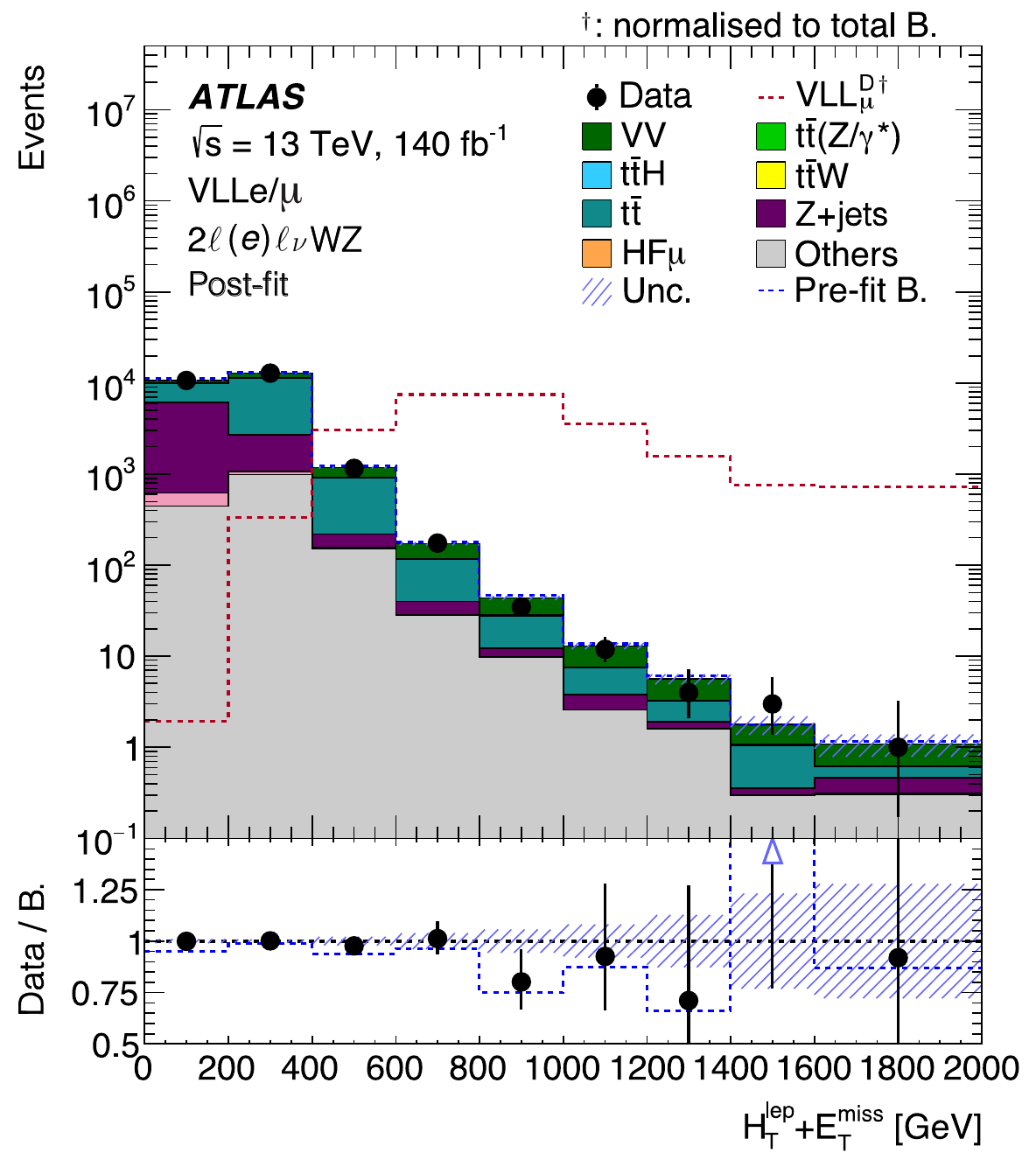}}
\subfloat[]{\includegraphics[width=0.24\textwidth]{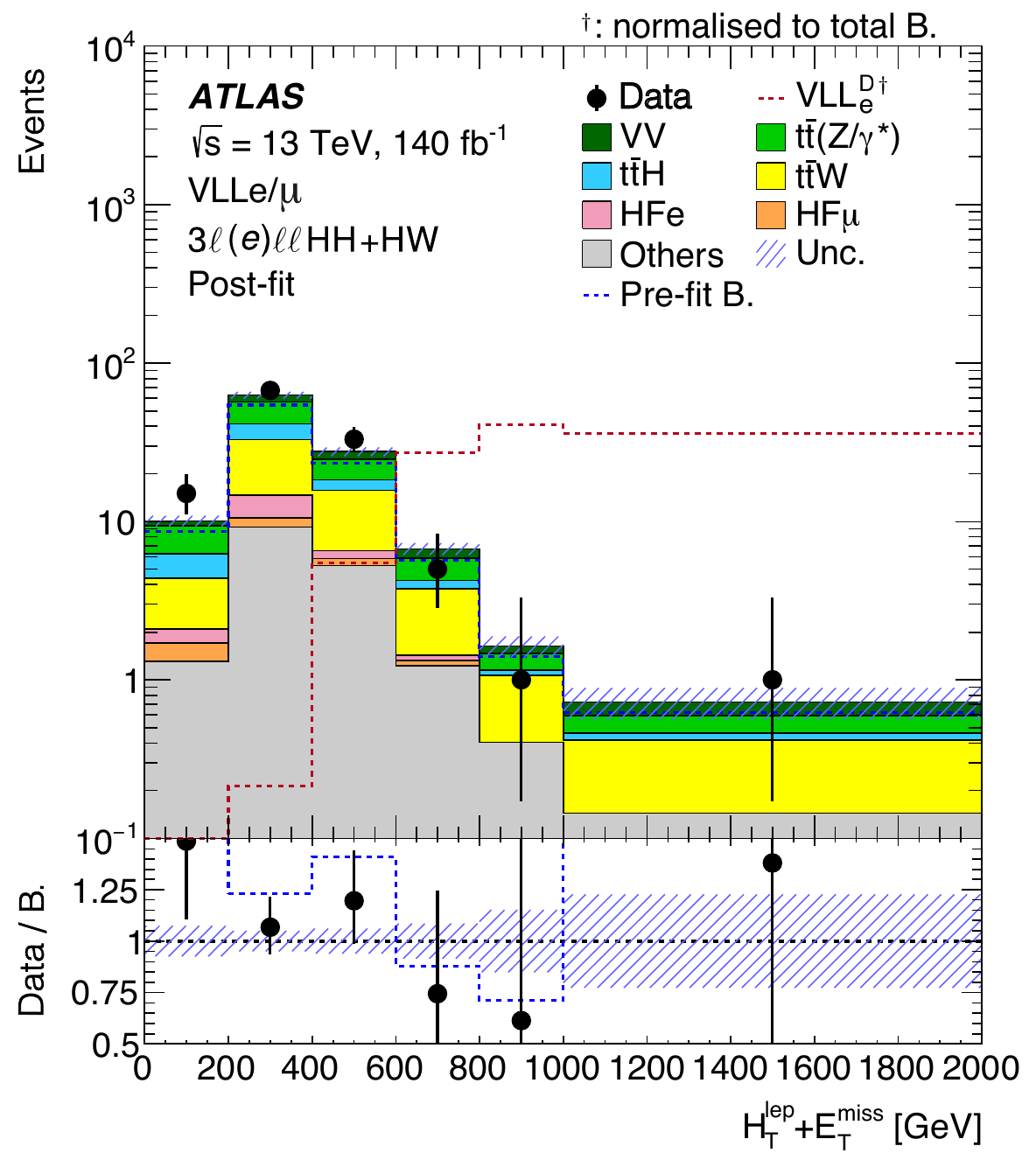}}
\subfloat[]{\includegraphics[width=0.24\textwidth]{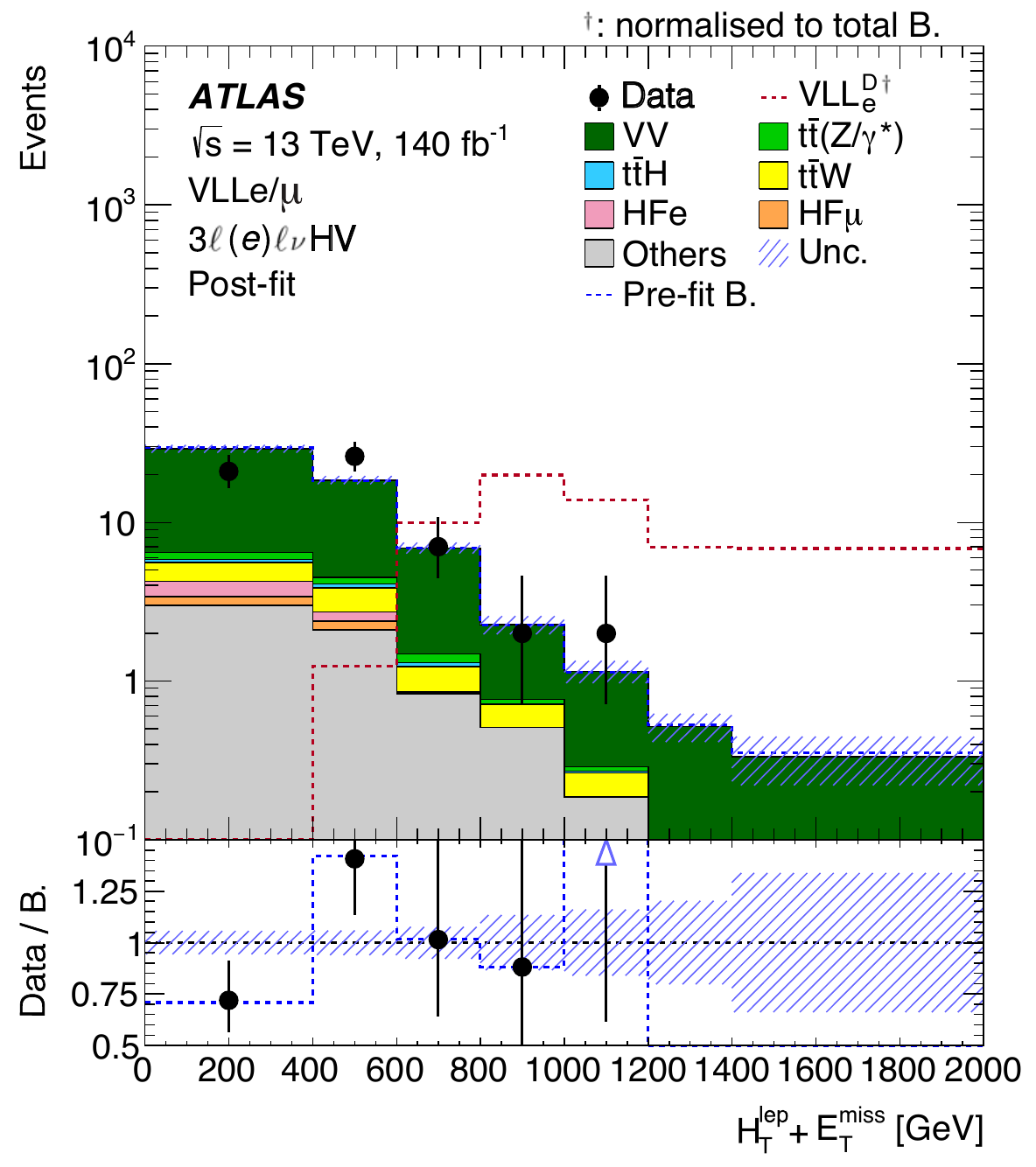}}
\subfloat[]{\includegraphics[width=0.24\textwidth]{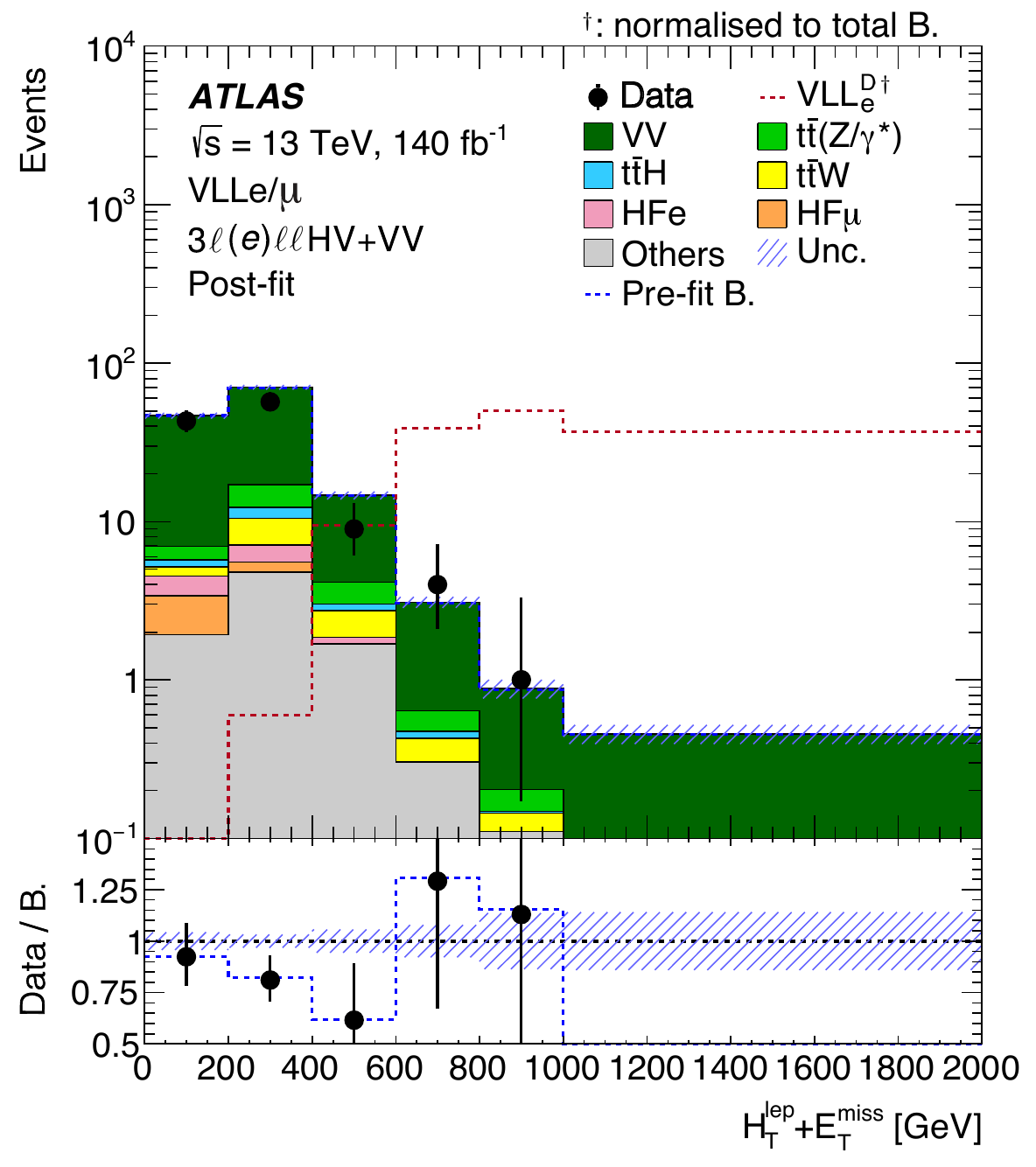}}\\
\subfloat[]{\includegraphics[width=0.24\textwidth]{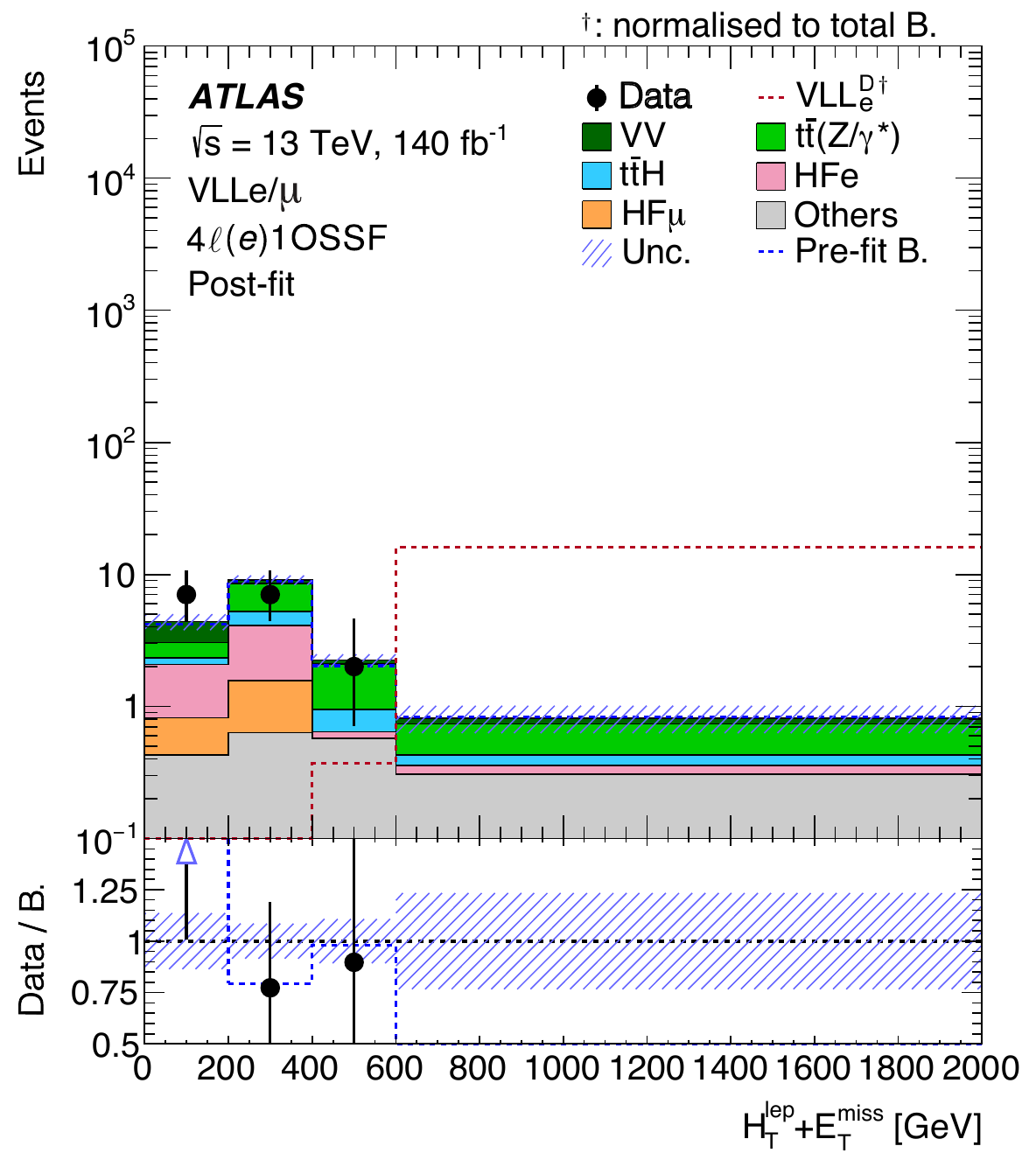}}
\subfloat[]{\includegraphics[width=0.24\textwidth]{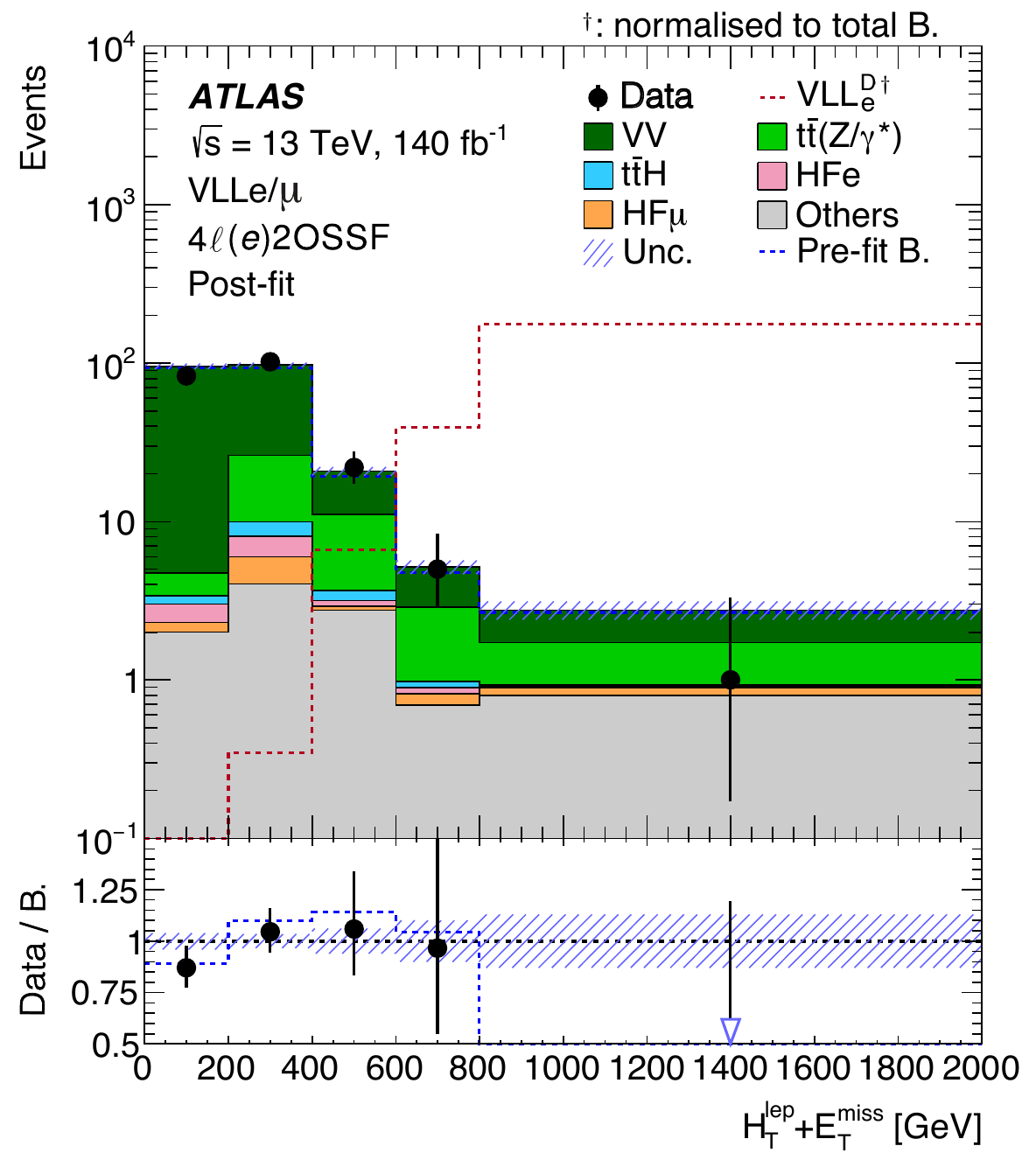}}
\caption{
Comparison between data and the background estimate for the \htlepmet distribution used in different \vlle signal region categories:
(a) \twoeehh, (b) \twoeehv, (c) \twoeevv, (d) \twoenuhv, (e) \twoenuvv, (f) \threeehhhw, (g) \threeenuhw, (h) \threeevvhz, (i) \eeem, (j) \eeee.
Distributions for one \vlled signal point for a VLL mass of 600 GeV normalised to the total background yields are overlaid for comparison.
The lower panels show the ratio of data to the background estimate (`B.'), separately for post-fit background (black points) and pre-fit background (dashed blue line).
The size of the combined statistical and systematic uncertainty in the signal-plus-background prediction is indicated by the blue hatched band.
}
\label{fig:htlepmet_vlle}
\end{figure*}
\begin{figure*}[h!]
\centering
\subfloat[]{\includegraphics[width=0.24\textwidth]{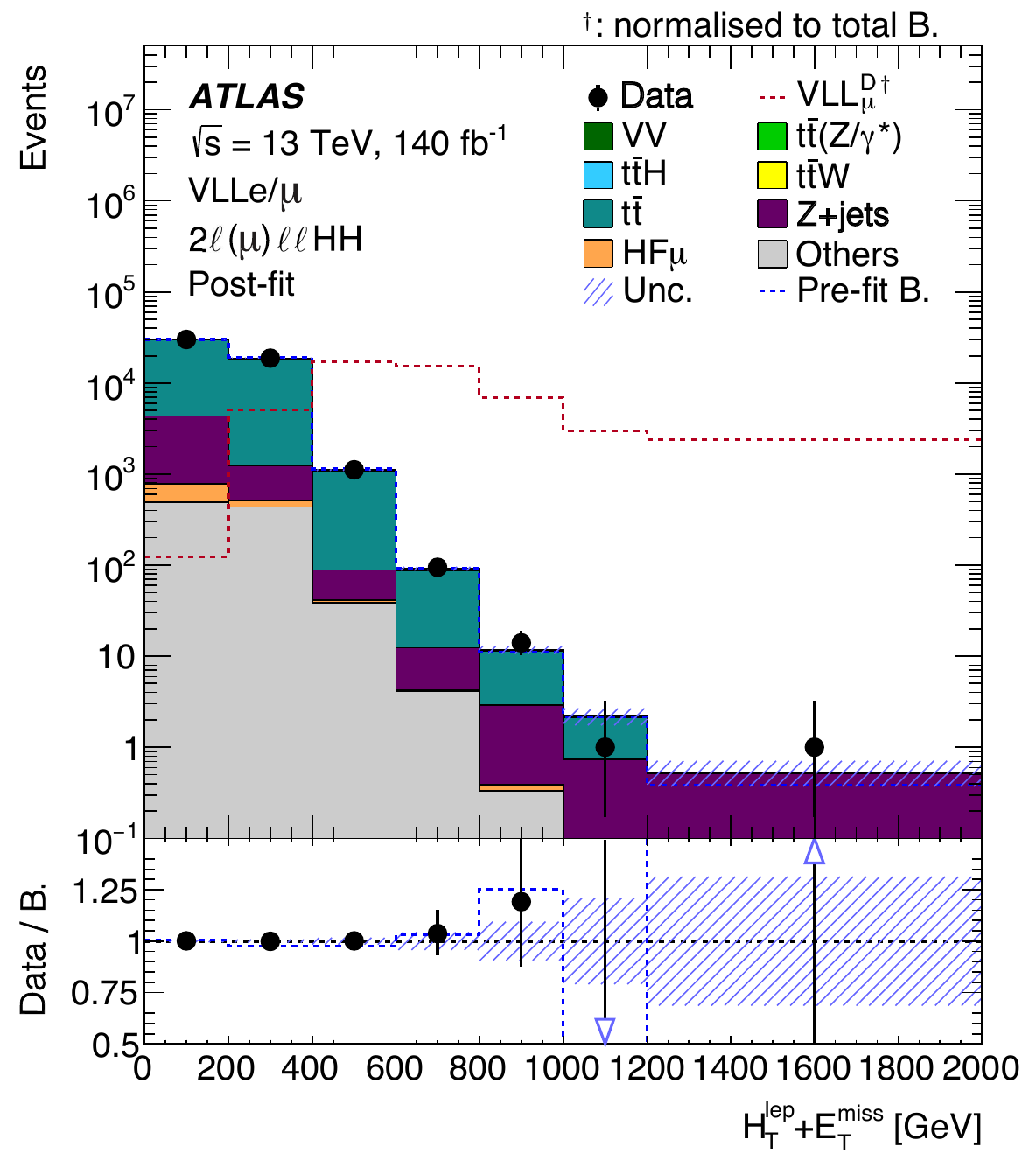}}
\subfloat[]{\includegraphics[width=0.24\textwidth]{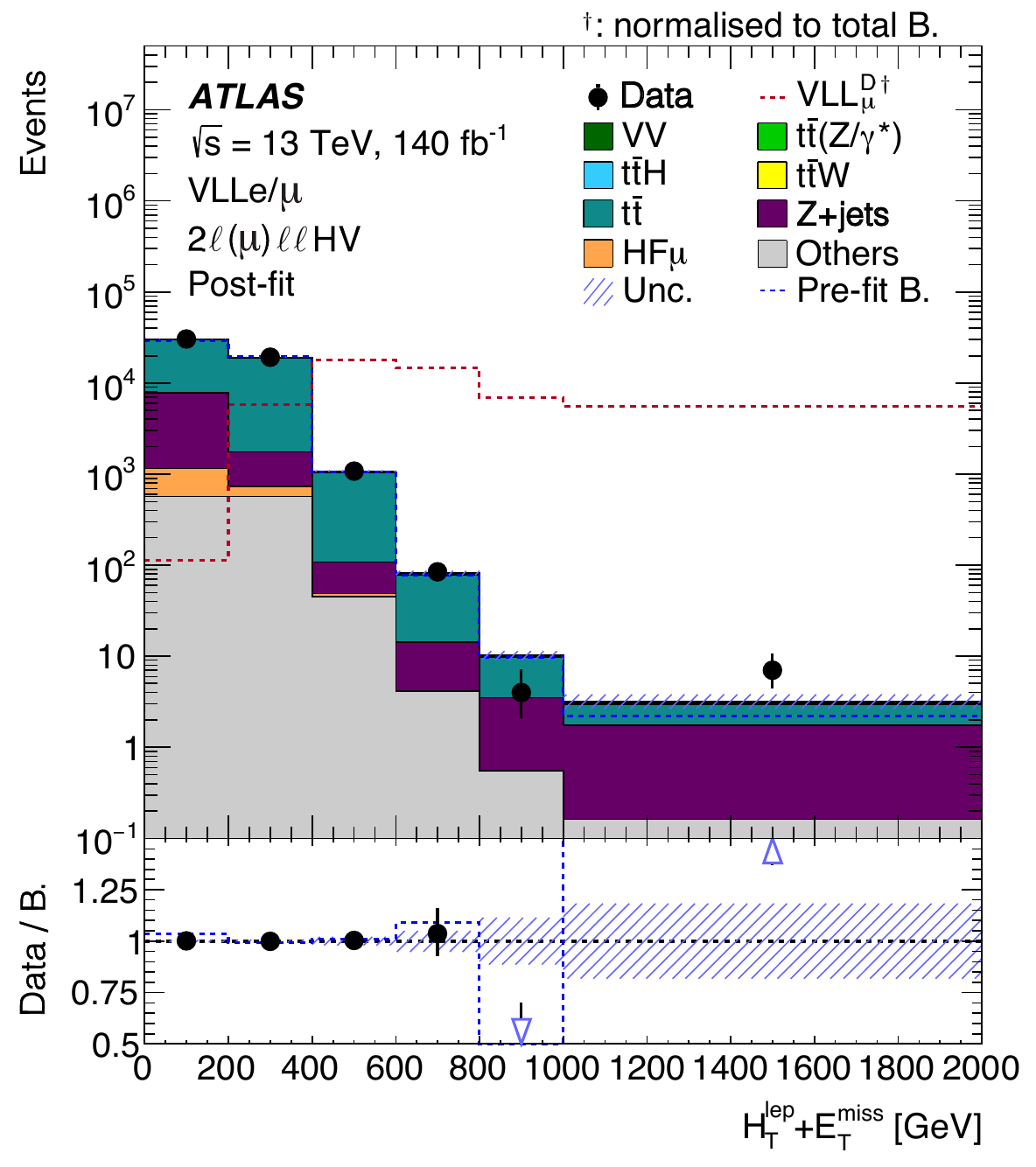}}
\subfloat[]{\includegraphics[width=0.24\textwidth]{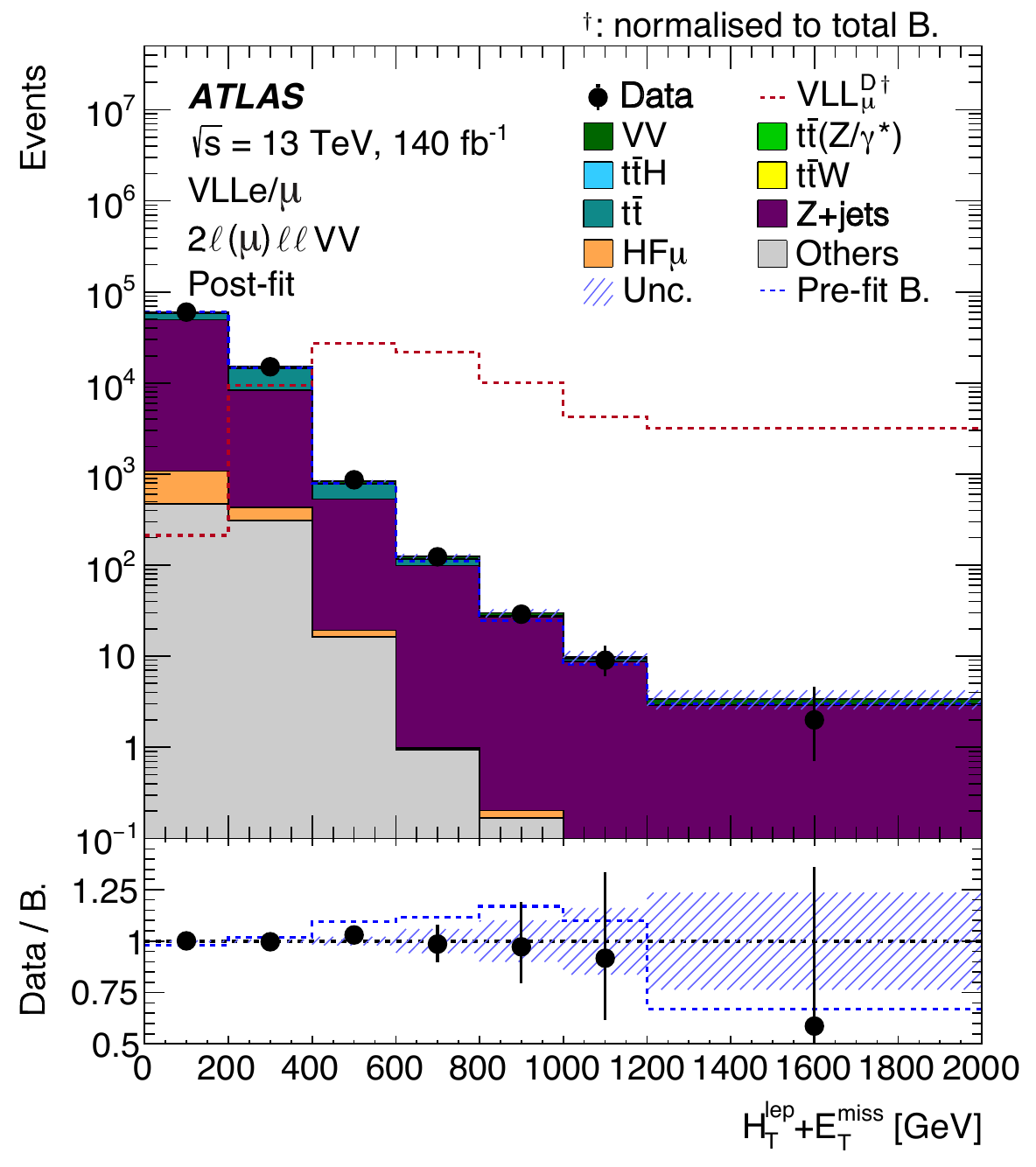}}
\subfloat[]{\includegraphics[width=0.24\textwidth]{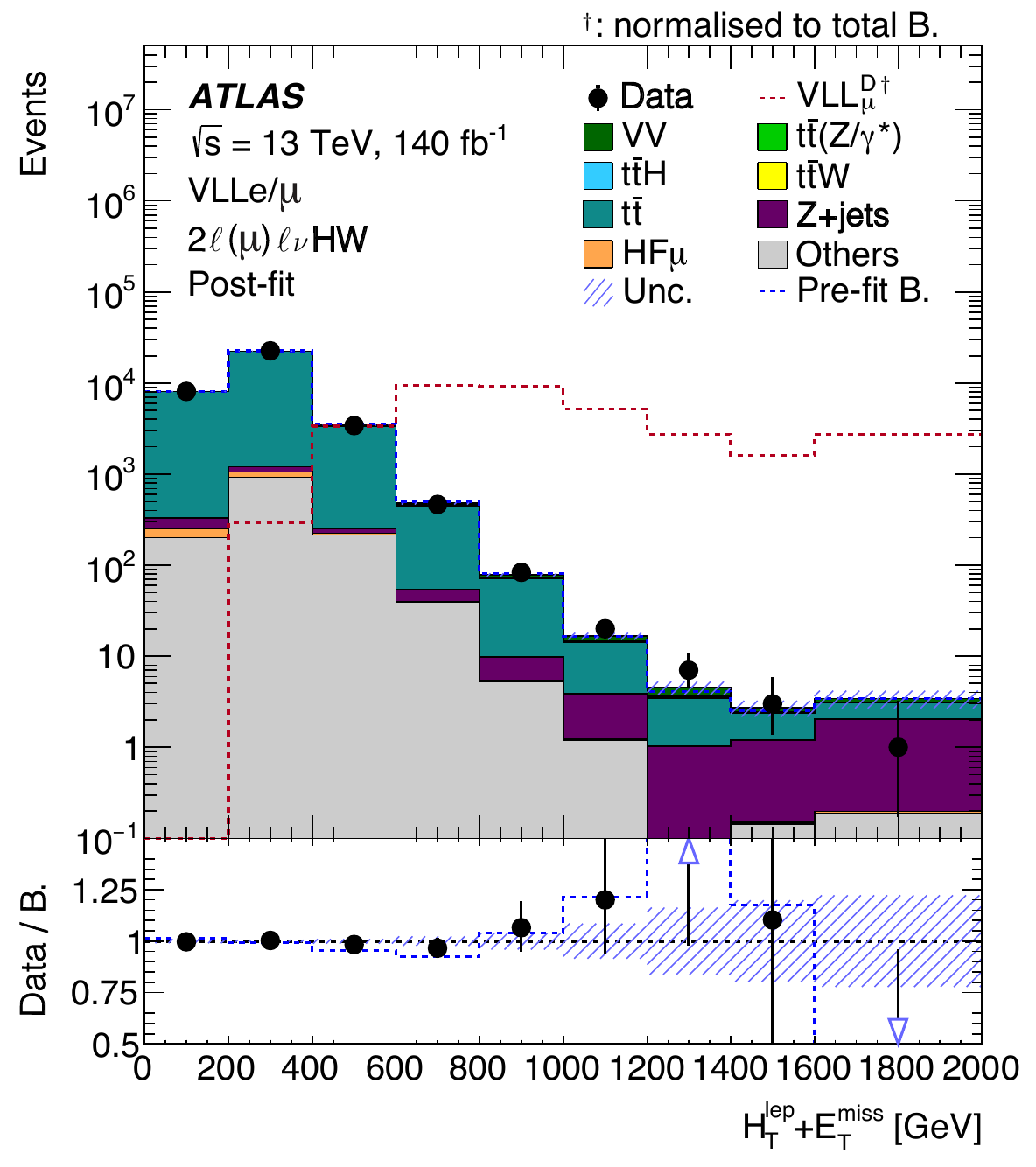}}\\
\subfloat[]{\includegraphics[width=0.24\textwidth]{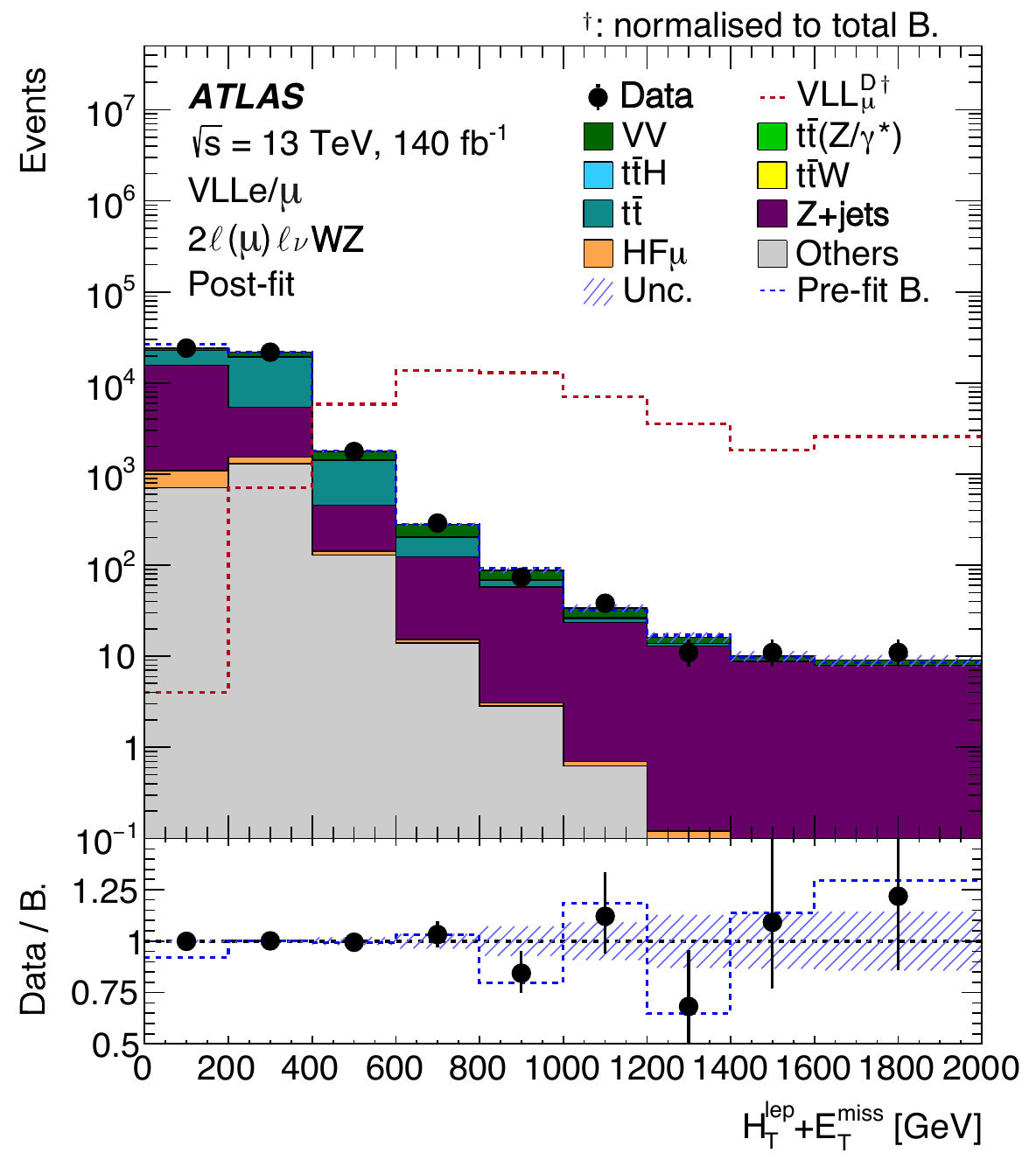}}
\subfloat[]{\includegraphics[width=0.24\textwidth]{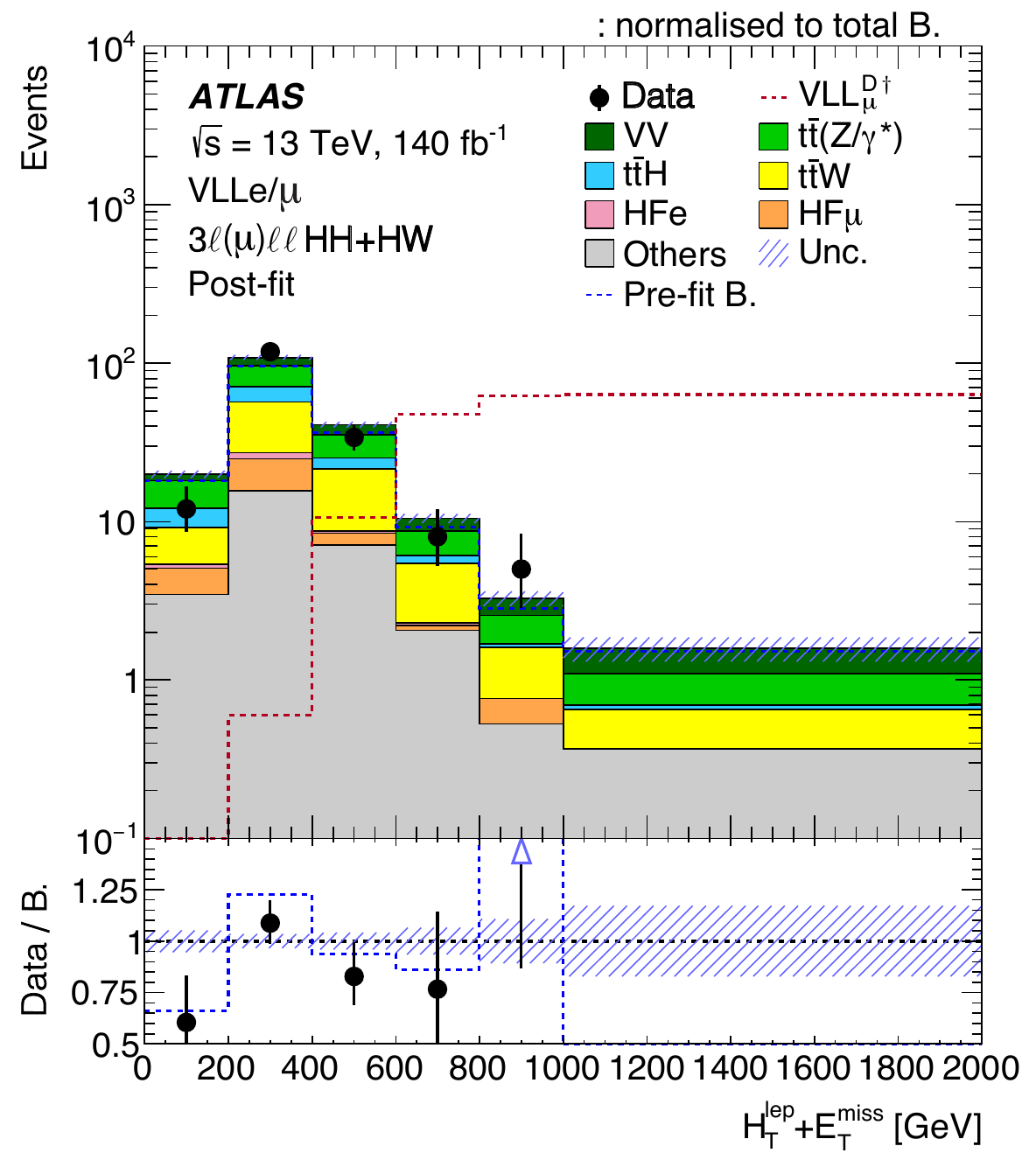}}
\subfloat[]{\includegraphics[width=0.24\textwidth]{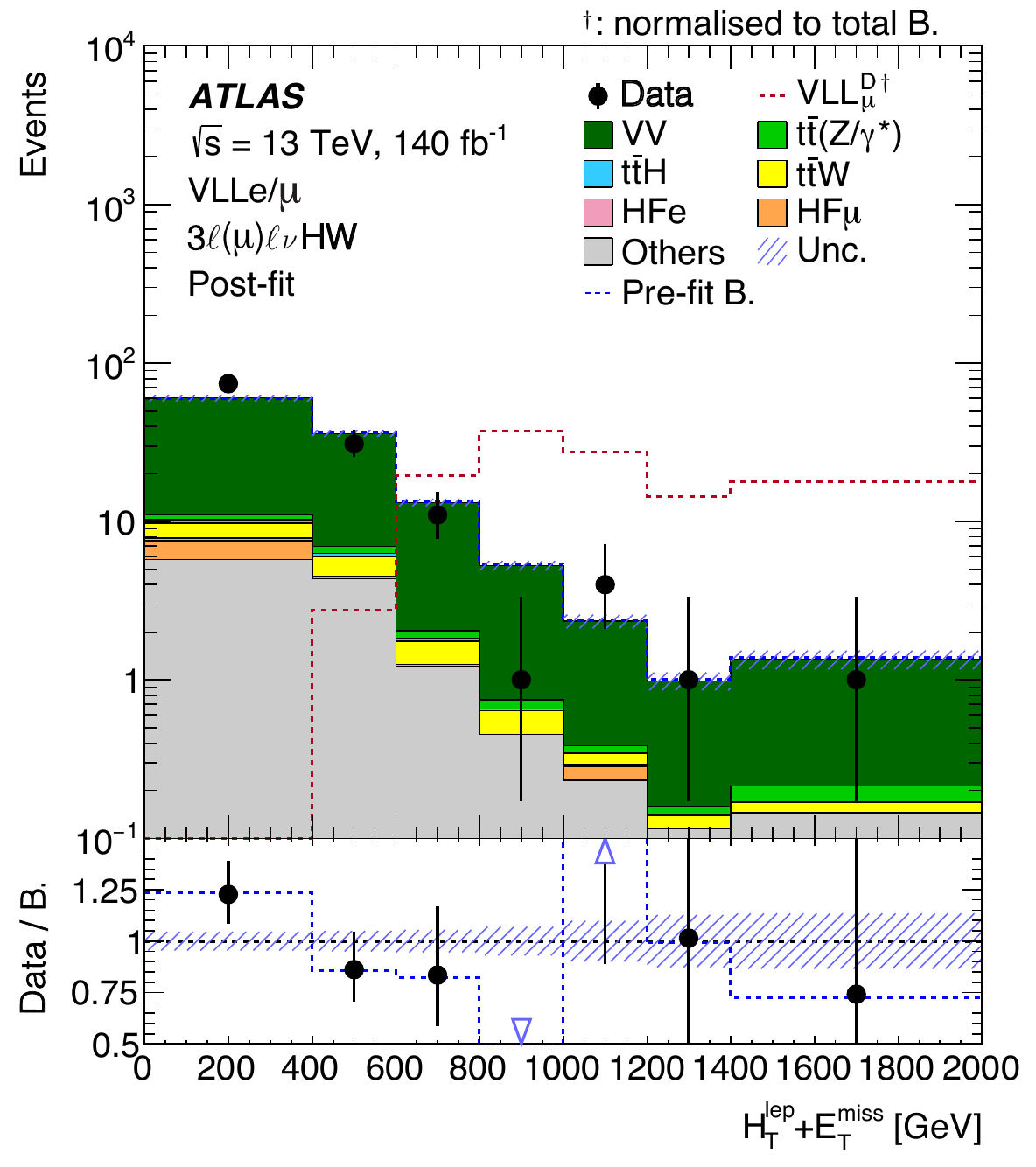}}
\subfloat[]{\includegraphics[width=0.24\textwidth]{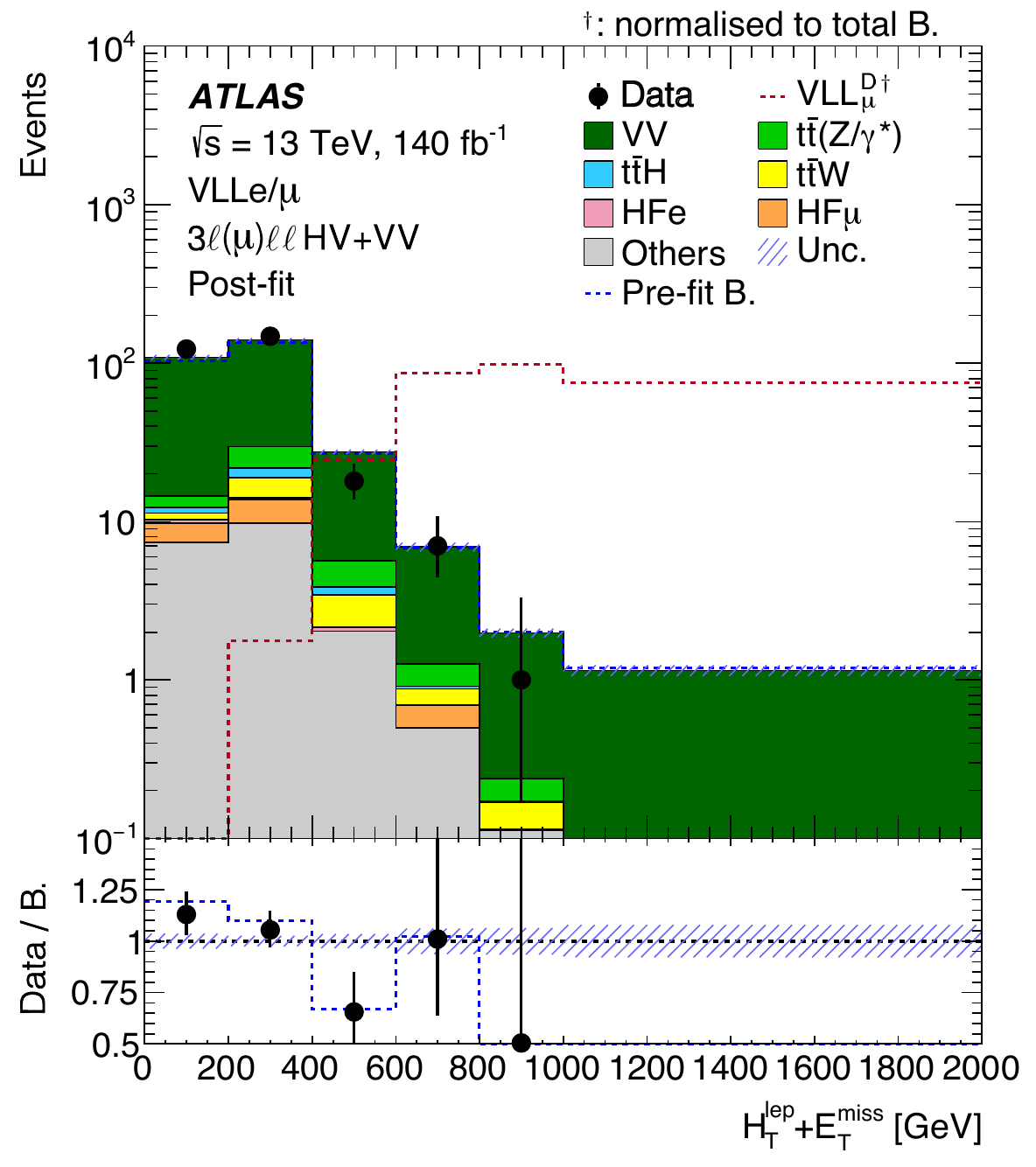}}\\
\subfloat[]{\includegraphics[width=0.24\textwidth]{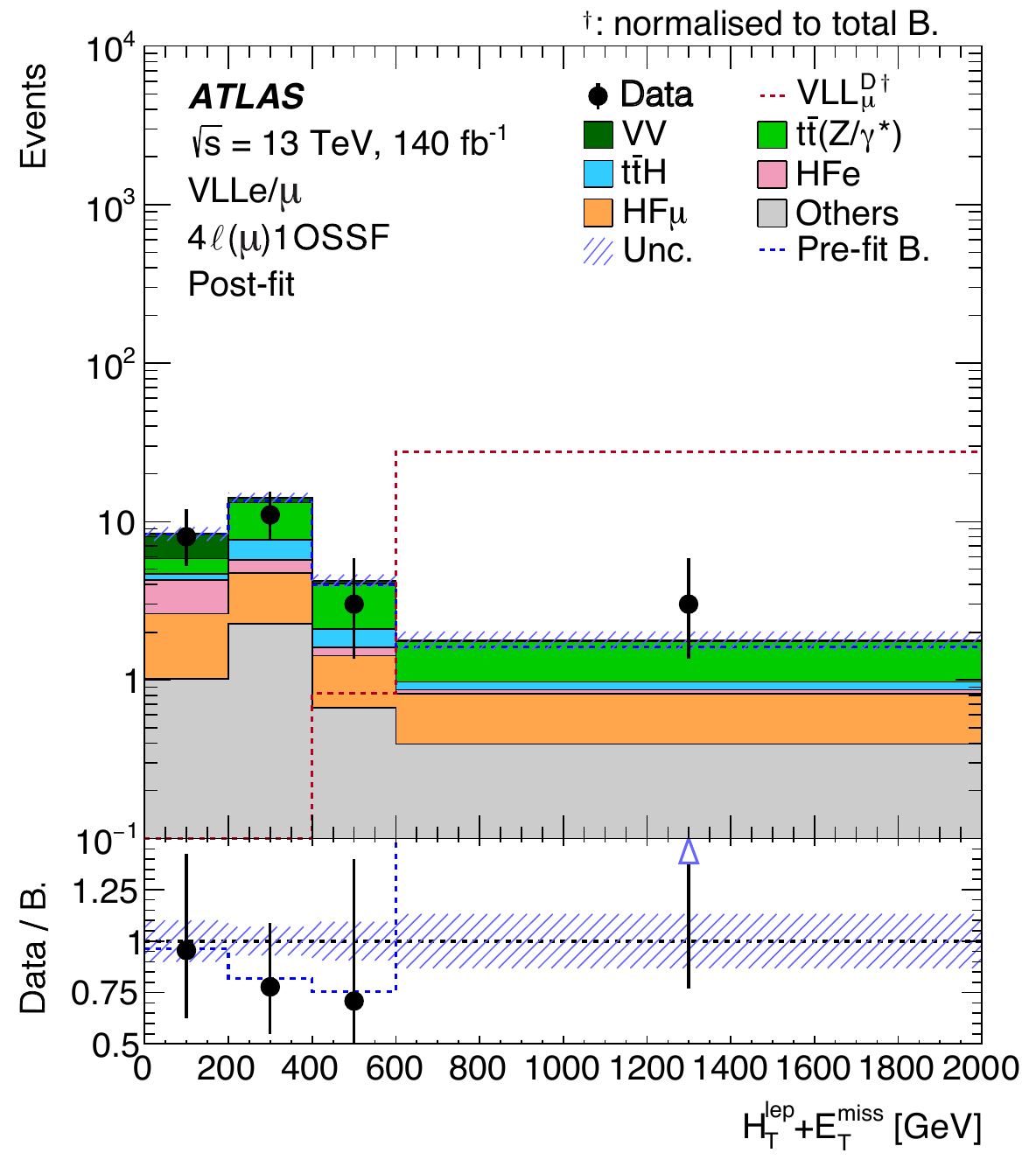}}
\subfloat[]{\includegraphics[width=0.24\textwidth]{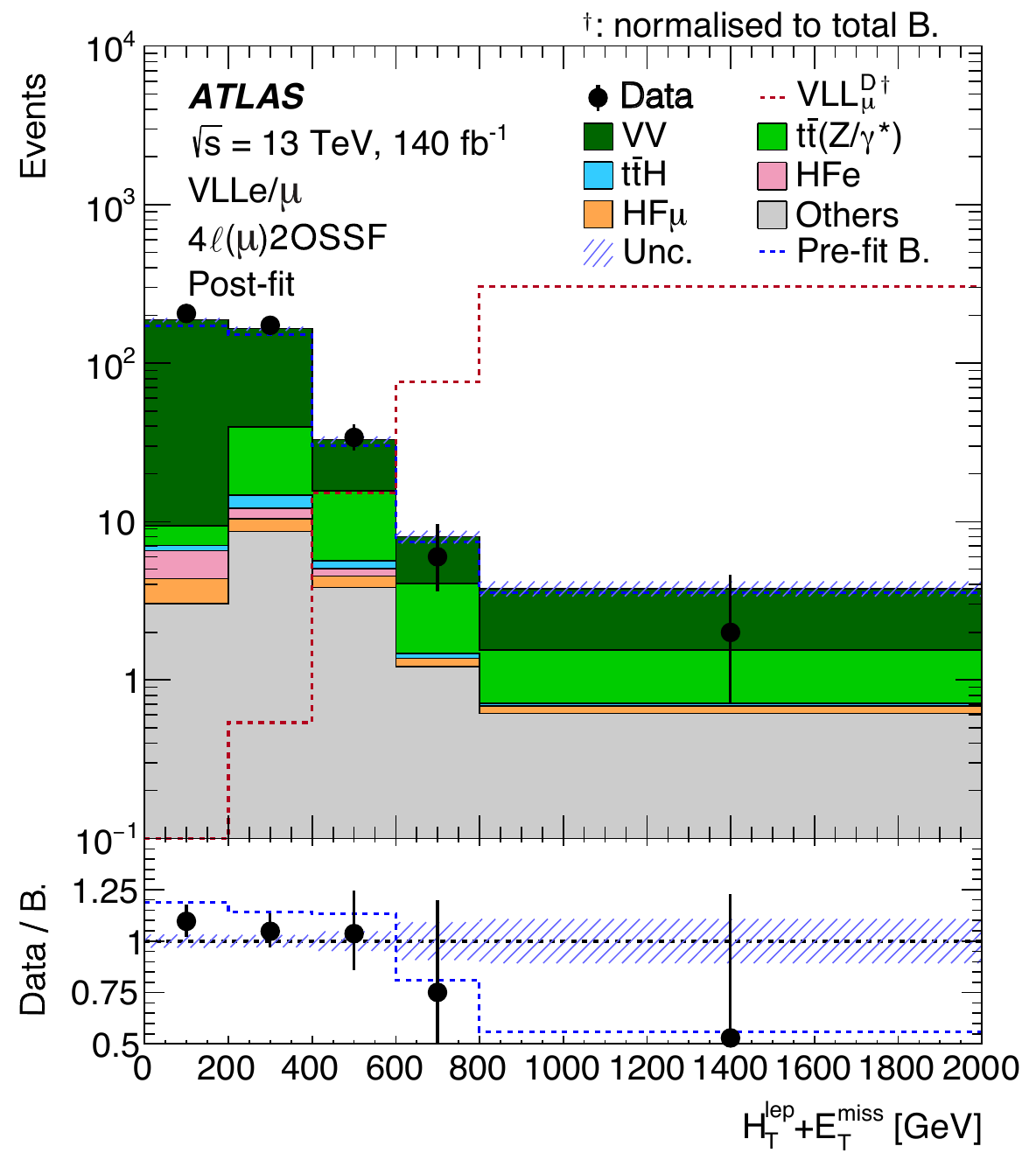}}
\caption{
Comparison between data and the background estimate for the \htlepmet distribution used in different \vllmu signal region categories:
(a) \twommhh, (b) \twommhv, (c) \twommvv, (d) \twomnuhv, (e) \twomnuvv, (f) \threemhhhw, (g) \threemnuhw, (h) \threemvvhz, (i) \mmme, (j) \mmmm.
Distributions for one \vllmud signal point for a VLL mass of 600 GeV normalised to the total background yields are overlaid for comparison.
The lower panels show the ratio of data to the background estimate (`B.'), separately for post-fit background (black points) and pre-fit background (dashed blue line).
The size of the combined statistical and systematic uncertainty in the signal-plus-background prediction is indicated by the blue hatched band.
}
\label{fig:htlepmet_vllmu}
\end{figure*}

\begin{figure}[!h]
\centering
\subfloat[]{\includegraphics[width=0.5\textwidth]{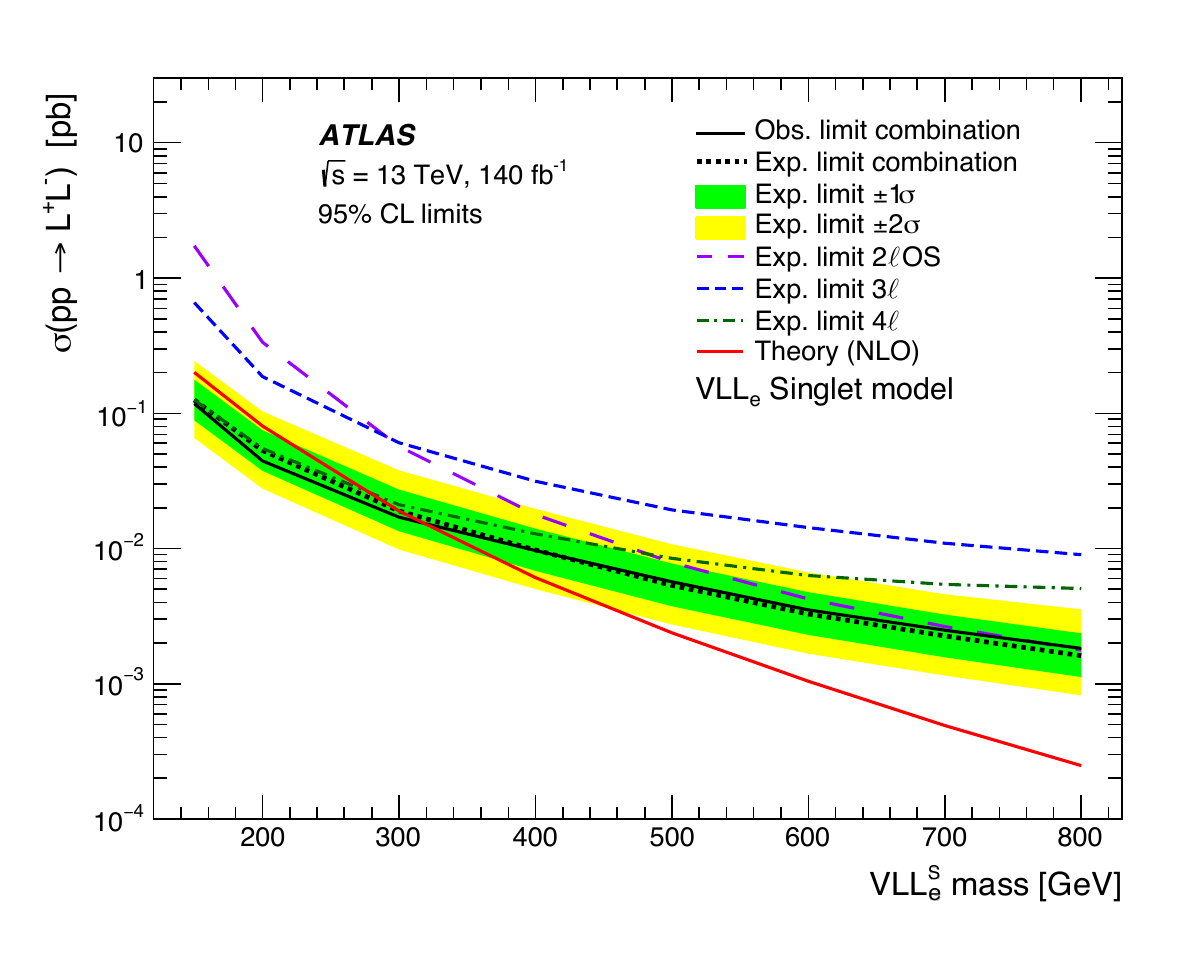}\label{fig:Limits_VLLesinglet}}
\subfloat[]{\includegraphics[width=0.5\textwidth]{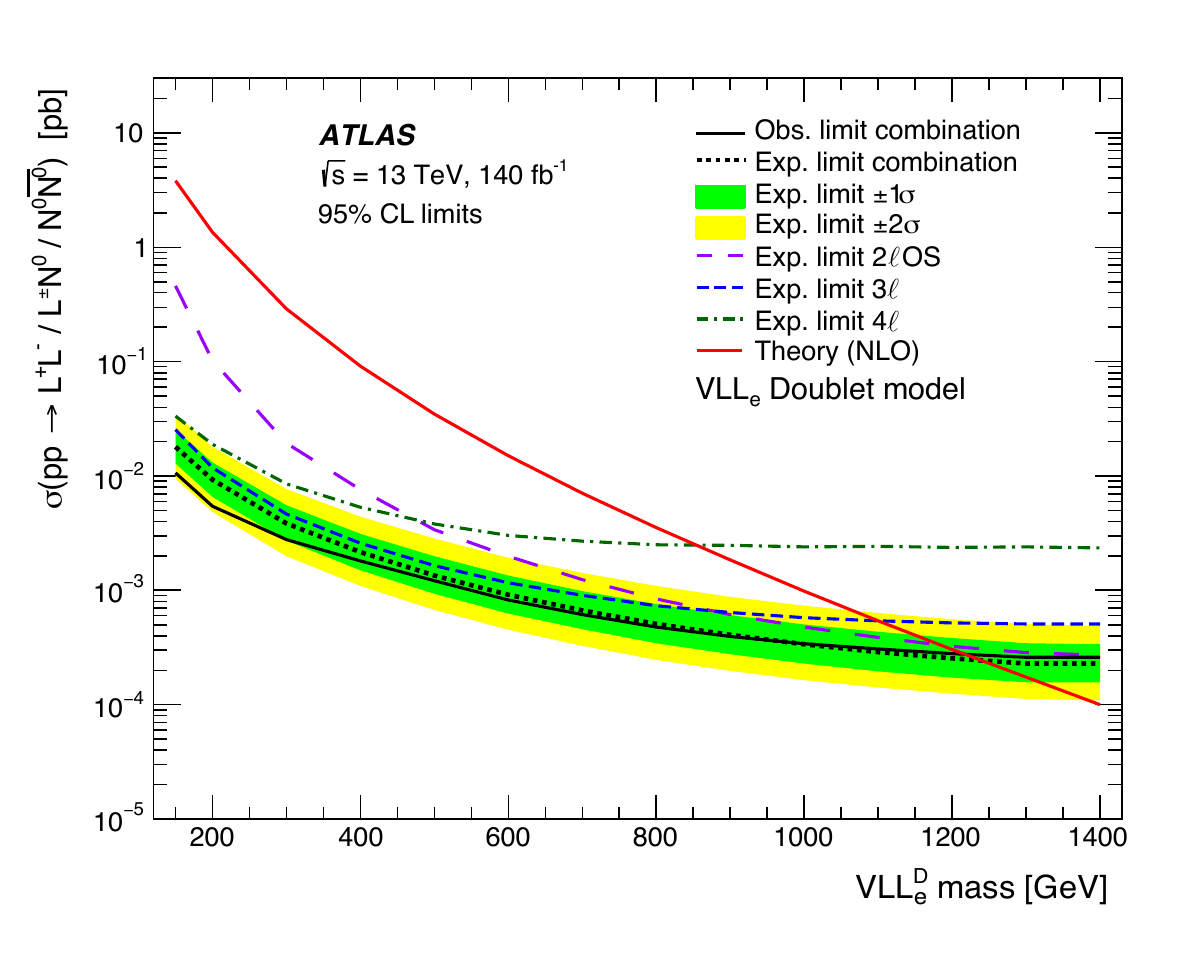}\label{fig:Limits_VLLedoublet}}\\
\subfloat[]{\includegraphics[width=0.5\textwidth]{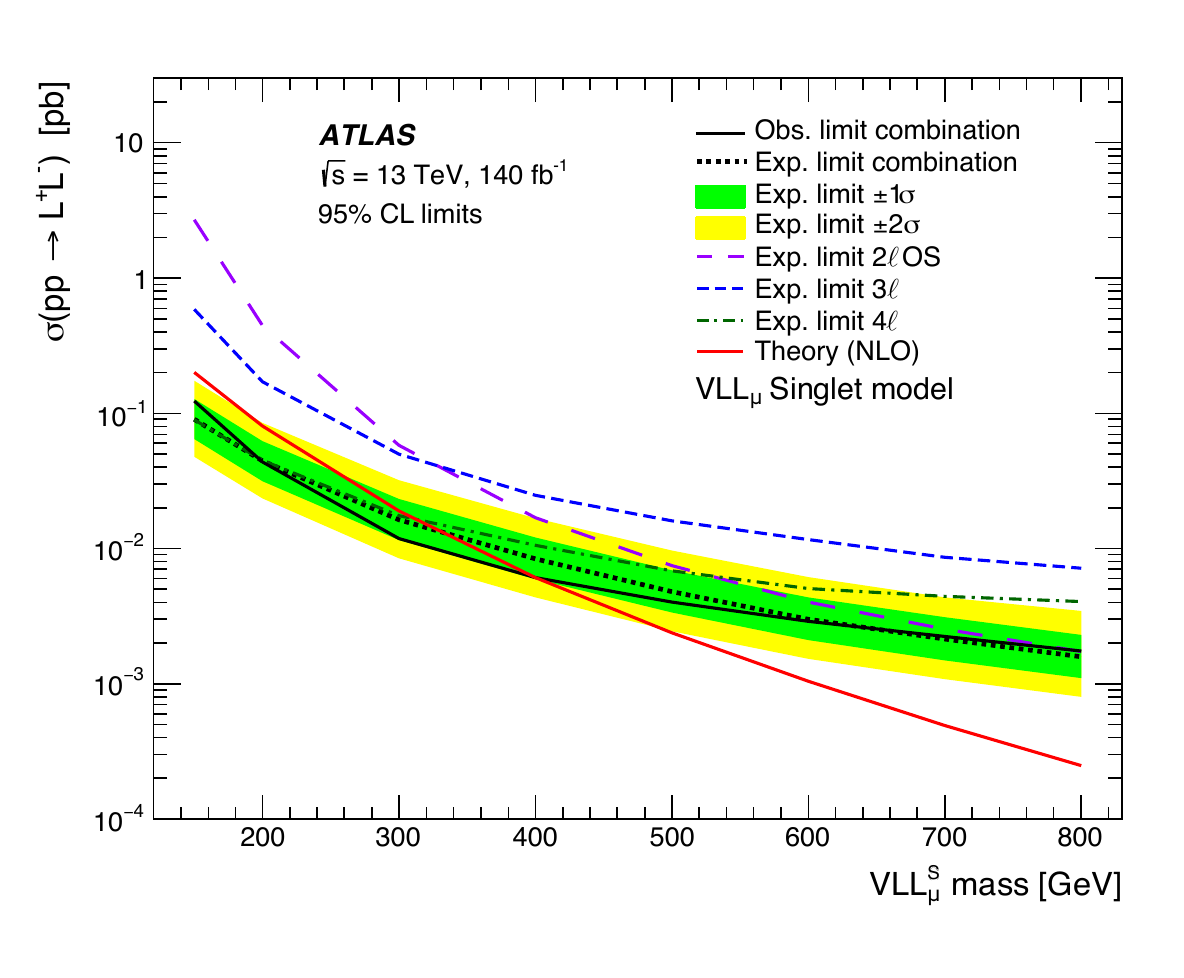}\label{fig:Limits_VLLmusinglet}}
\subfloat[]{\includegraphics[width=0.5\textwidth]{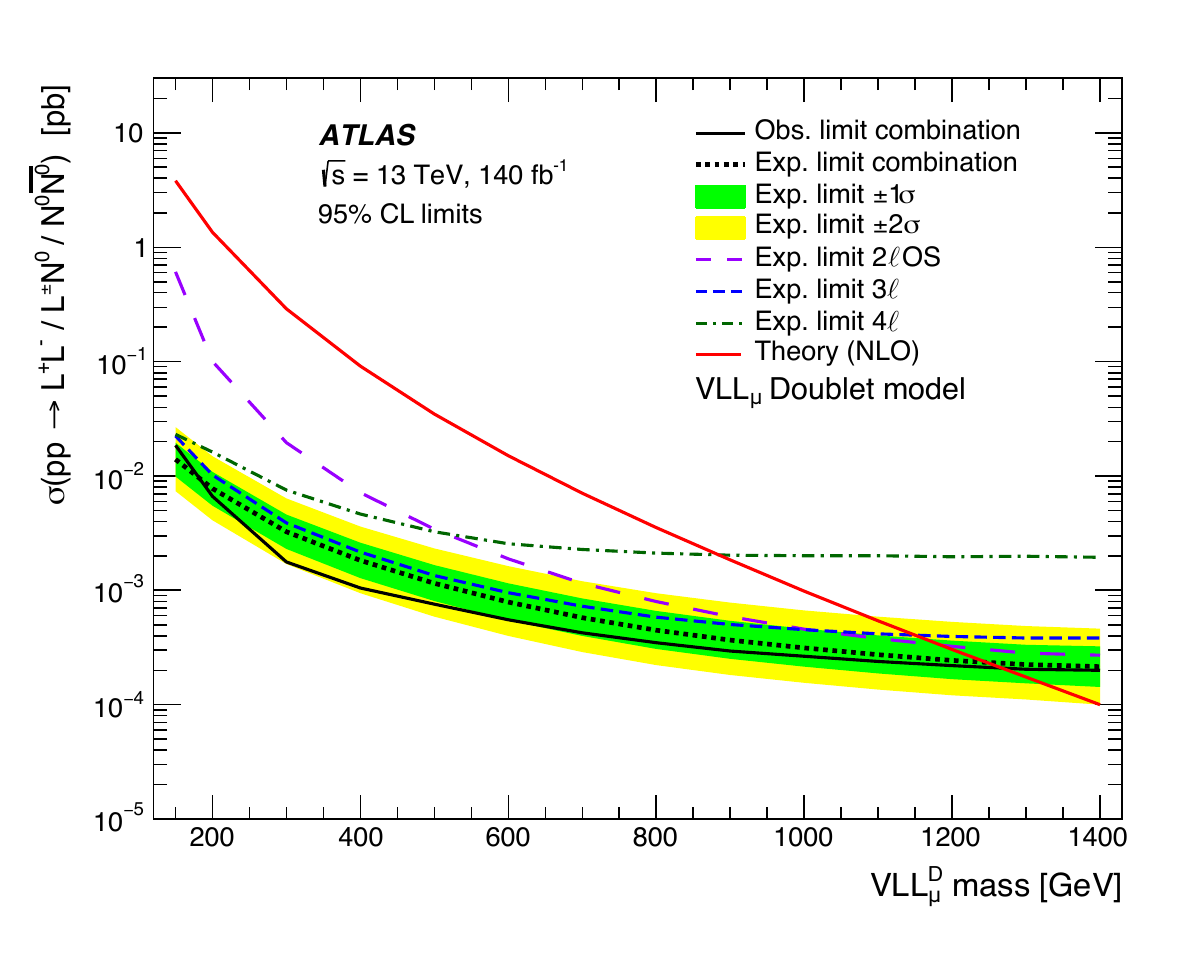}\label{fig:Limits_VLLmudoublet}}
\caption{Combined observed and expected limits at 95\% CL for the (a) \vlles, (b) \vlled, (c) \vllmus, and (d) \vllmud models, as a function of the mass of the VLL.
The inner and outer bands around the expected limit are the $\pm 1 \sigma$ and $\pm 2 \sigma$ variations including all uncertainties, respectively.
The theoretical prediction for the signal production cross-section is also shown as a red solid line.
Expected limits on the signal cross-section from the individual channels are shown in dashed lines for \llOS (long-dashed), \lll (short-dashed), and \llll (dot-dashed).
}
\end{figure}


%

%
%
%

\FloatBarrier

\section{Conclusions}
\label{sec:conclusion}

A search for a doublet or singlet vector-like lepton coupling to the Standard Model first- and second-generation leptons is presented.
The search is based on a data sample of proton--proton collisions recorded at $\sqrt{s}=13~\tev$ by the ATLAS detector during Run 2 of the LHC, corresponding to an integrated luminosity of 140~fb$^{-1}$.
The search is performed in the multilepton (two, three, and four light leptons) final state. A deep-neural-network-based categorisation is performed to enhance the purity of the various signal types and to discriminate signal against the SM background.
The dominant backgrounds originate from \ttbar, \Zjets, \ttW, \ttZ, and \VV, and are estimated from Monte-Carlo simulation and normalised to data in a simultaneous fit of the signal and control regions.
The data are found to be consistent with the Standard Model predictions and exclusion limits are set on the mass of the vector-like electrons and muons, excluding masses below $1220~\gev$ ($1270~\gev$) and $320~\gev$ ($400~\gev$) for vector-like electrons (muons) in the doublet and singlet scenarios at 95\% confidence level, respectively. These are the most stringent limits on vector-like electrons and muons to date, improving the previous mass exclusion limits set with the Run 1 data for the singlet scenario by 140~\gev (230~\gev) for vector-like electrons (muons) and setting limits on the doublet scenario for the first time.


%

%
%
%
%
%
\section*{Acknowledgements}
%

%

%
%

%
%

We thank CERN for the very successful operation of the LHC and its injectors, as well as the support staff at
CERN and at our institutions worldwide without whom ATLAS could not be operated efficiently.

The crucial computing support from all WLCG partners is acknowledged gratefully, in particular from CERN, the ATLAS Tier-1 facilities at TRIUMF/SFU (Canada), NDGF (Denmark, Norway, Sweden), CC-IN2P3 (France), KIT/GridKA (Germany), INFN-CNAF (Italy), NL-T1 (Netherlands), PIC (Spain), RAL (UK) and BNL (USA), the Tier-2 facilities worldwide and large non-WLCG resource providers. Major contributors of computing resources are listed in Ref.~\cite{ATL-SOFT-PUB-2025-001}.

We gratefully acknowledge the support of ANPCyT, Argentina; YerPhI, Armenia; ARC, Australia; BMWFW and FWF, Austria; ANAS, Azerbaijan; CNPq and FAPESP, Brazil; NSERC, NRC and CFI, Canada; CERN; ANID, Chile; CAS, MOST and NSFC, China; Minciencias, Colombia; MEYS CR, Czech Republic; DNRF and DNSRC, Denmark; IN2P3-CNRS and CEA-DRF/IRFU, France; SRNSFG, Georgia; BMBF, HGF and MPG, Germany; GSRI, Greece; RGC and Hong Kong SAR, China; ICHEP and Academy of Sciences and Humanities, Israel; INFN, Italy; MEXT and JSPS, Japan; CNRST, Morocco; NWO, Netherlands; RCN, Norway; MNiSW, Poland; FCT, Portugal; MNE/IFA, Romania; MSTDI, Serbia; MSSR, Slovakia; ARIS and MVZI, Slovenia; DSI/NRF, South Africa; MICIU/AEI, Spain; SRC and Wallenberg Foundation, Sweden; SERI, SNSF and Cantons of Bern and Geneva, Switzerland; NSTC, Taipei; TENMAK, T\"urkiye; STFC/UKRI, United Kingdom; DOE and NSF, United States of America.

Individual groups and members have received support from BCKDF, CANARIE, CRC and DRAC, Canada; CERN-CZ, FORTE and PRIMUS, Czech Republic; COST, ERC, ERDF, Horizon 2020, ICSC-NextGenerationEU and Marie Sk{\l}odowska-Curie Actions, European Union; Investissements d'Avenir Labex, Investissements d'Avenir Idex and ANR, France; DFG and AvH Foundation, Germany; Herakleitos, Thales and Aristeia programmes co-financed by EU-ESF and the Greek NSRF, Greece; BSF-NSF and MINERVA, Israel; NCN and NAWA, Poland; La Caixa Banking Foundation, CERCA Programme Generalitat de Catalunya and PROMETEO and GenT Programmes Generalitat Valenciana, Spain; G\"{o}ran Gustafssons Stiftelse, Sweden; The Royal Society and Leverhulme Trust, United Kingdom.

In addition, individual members wish to acknowledge support from Armenia: Yerevan Physics Institute (FAPERJ); CERN: European Organization for Nuclear Research (CERN DOCT); Chile: Agencia Nacional de Investigaci\'on y Desarrollo (FONDECYT 1230812, FONDECYT 1230987, FONDECYT 1240864); China: Chinese Ministry of Science and Technology (MOST-2023YFA1605700, MOST-2023YFA1609300), National Natural Science Foundation of China (NSFC - 12175119, NSFC 12275265, NSFC-12075060); Czech Republic: Czech Science Foundation (GACR - 24-11373S), Ministry of Education Youth and Sports (FORTE CZ.02.01.01/00/22\_008/0004632), PRIMUS Research Programme (PRIMUS/21/SCI/017); EU: H2020 European Research Council (ERC - 101002463); European Union: European Research Council (ERC - 948254, ERC 101089007, ERC, BARD, 101116429), Horizon 2020 Framework Programme (MUCCA - CHIST-ERA-19-XAI-00), European Union, Future Artificial Intelligence Research (FAIR-NextGenerationEU PE00000013), Italian Center for High Performance Computing, Big Data and Quantum Computing (ICSC, NextGenerationEU); France: Agence Nationale de la Recherche (ANR-20-CE31-0013, ANR-21-CE31-0013, ANR-21-CE31-0022, ANR-22-EDIR-0002); Germany: Baden-Württemberg Stiftung (BW Stiftung-Postdoc Eliteprogramme), Deutsche Forschungsgemeinschaft (DFG - 469666862, DFG - CR 312/5-2); Italy: Istituto Nazionale di Fisica Nucleare (ICSC, NextGenerationEU), Ministero dell'Università e della Ricerca (NextGenEU PRIN20223N7F8K M4C2.1.1); Japan: Japan Society for the Promotion of Science (JSPS KAKENHI JP22H01227, JSPS KAKENHI JP22H04944, JSPS KAKENHI JP22KK0227, JSPS KAKENHI JP23KK0245); Netherlands: Netherlands Organisation for Scientific Research (NWO Veni 2020 - VI.Veni.202.179); Norway: Research Council of Norway (RCN-314472); Poland: Ministry of Science and Higher Education (IDUB AGH, POB8, D4 no 9722), Polish National Agency for Academic Exchange (PPN/PPO/2020/1/00002/U/00001), Polish National Science Centre (NCN 2021/42/E/ST2/00350, NCN OPUS 2023/51/B/ST2/02507, NCN OPUS nr 2022/47/B/ST2/03059, NCN UMO-2019/34/E/ST2/00393, NCN \& H2020 MSCA 945339, UMO-2020/37/B/ST2/01043, UMO-2021/40/C/ST2/00187, UMO-2022/47/O/ST2/00148, UMO-2023/49/B/ST2/04085, UMO-2023/51/B/ST2/00920); Slovenia: Slovenian Research Agency (ARIS grant J1-3010); Spain: Generalitat Valenciana (Artemisa, FEDER, IDIFEDER/2018/048), Ministry of Science and Innovation (MCIN \& NextGenEU PCI2022-135018-2, MICIN \& FEDER PID2021-125273NB, RYC2019-028510-I, RYC2020-030254-I, RYC2021-031273-I, RYC2022-038164-I); Sweden: Carl Trygger Foundation (Carl Trygger Foundation CTS 22:2312), Swedish Research Council (Swedish Research Council 2023-04654, VR 2018-00482, VR 2021-03651, VR 2022-03845, VR 2022-04683, VR 2023-03403), Knut and Alice Wallenberg Foundation (KAW 2018.0458, KAW 2019.0447, KAW 2022.0358); Switzerland: Swiss National Science Foundation (SNSF - PCEFP2\_194658); United Kingdom: Leverhulme Trust (Leverhulme Trust RPG-2020-004), Royal Society (NIF-R1-231091); United States of America: U.S. Department of Energy (ECA DE-AC02-76SF00515), Neubauer Family Foundation.

%
%


%
%
%
%
%
%

%
%
%
\printbibliography
\clearpage
 
\begin{flushleft}
\hypersetup{urlcolor=black}
{\Large The ATLAS Collaboration}

\bigskip

\AtlasOrcid[0000-0002-6665-4934]{G.~Aad}$^\textrm{\scriptsize 104}$,
\AtlasOrcid[0000-0001-7616-1554]{E.~Aakvaag}$^\textrm{\scriptsize 17}$,
\AtlasOrcid[0000-0002-5888-2734]{B.~Abbott}$^\textrm{\scriptsize 123}$,
\AtlasOrcid[0000-0002-0287-5869]{S.~Abdelhameed}$^\textrm{\scriptsize 119a}$,
\AtlasOrcid[0000-0002-1002-1652]{K.~Abeling}$^\textrm{\scriptsize 56}$,
\AtlasOrcid[0000-0001-5763-2760]{N.J.~Abicht}$^\textrm{\scriptsize 50}$,
\AtlasOrcid[0000-0002-8496-9294]{S.H.~Abidi}$^\textrm{\scriptsize 30}$,
\AtlasOrcid[0009-0003-6578-220X]{M.~Aboelela}$^\textrm{\scriptsize 46}$,
\AtlasOrcid[0000-0002-9987-2292]{A.~Aboulhorma}$^\textrm{\scriptsize 36e}$,
\AtlasOrcid[0000-0001-5329-6640]{H.~Abramowicz}$^\textrm{\scriptsize 155}$
% ...
\bigskip
\\

% ...
$^{17}$Department for Physics and Technology, University of Bergen, Bergen; Norway.\\
% ...
$^{30}$Physics Department, Brookhaven National Laboratory, Upton NY; United States of America.\\
% ...
$^{36}$$^{(a)}$Facult\'e des Sciences Ain Chock, Universit\'e Hassan II de Casablanca;$^{(b)}$Facult\'{e} des Sciences, Universit\'{e} Ibn-Tofail, K\'{e}nitra;$^{(c)}$Facult\'e des Sciences Semlalia, Universit\'e Cadi Ayyad, LPHEA-Marrakech;$^{(d)}$LPMR, Facult\'e des Sciences, Universit\'e Mohamed Premier, Oujda;$^{(e)}$Facult\'e des sciences, Universit\'e Mohammed V, Rabat;$^{(f)}$Institute of Applied Physics, Mohammed VI Polytechnic University, Ben Guerir; Morocco.\\
% ...
$^{46}$Physics Department, Southern Methodist University, Dallas TX; United States of America.\\
% ...
$^{50}$Fakult\"{a}t Physik , Technische Universit{\"a}t Dortmund, Dortmund; Germany.\\
% ...
$^{56}$II. Physikalisches Institut, Georg-August-Universit\"{a}t G\"ottingen, G\"ottingen; Germany.\\
% ...
$^{104}$CPPM, Aix-Marseille Universit\'e, CNRS/IN2P3, Marseille; France.\\
% ...
$^{119}$$^{(a)}$New York University Abu Dhabi, Abu Dhabi;$^{(b)}$United Arab Emirates University, Al Ain; United Arab Emirates.\\
% ...
$^{123}$Homer L. Dodge Department of Physics and Astronomy, University of Oklahoma, Norman OK; United States of America.\\
% ...
$^{155}$Raymond and Beverly Sackler School of Physics and Astronomy, Tel Aviv University, Tel Aviv; Israel.\\
% ...
$^{a}$ Also Affiliated with an institute covered by a cooperation agreement with CERN.\\
$^{b}$ Also at An-Najah National University, Nablus; Palestine.\\
$^{c}$ Also at Borough of Manhattan Community College, City University of New York, New York NY; United States of America.\\
$^{d}$ Also at Center for High Energy Physics, Peking University; China.\\
$^{e}$ Also at Center for Interdisciplinary Research and Innovation (CIRI-AUTH), Thessaloniki; Greece.\\
$^{f}$ Also at CERN, Geneva; Switzerland.\\
% ...

\end{flushleft}


%

%
%

%

%
%
%
%

\end{document}